\documentclass[vecphys]{svmult}

\usepackage{epsfig}
\usepackage{graphics}
\usepackage{psfrag}
\usepackage{amssymb}
\usepackage{amsmath}

\newcommand{\rf}[1]{(\ref{#1})}
\newcommand{\beq}{\begin{equation}}

\newcommand{\eeq}{\end{equation}}
\newcommand{\bea}{\begin{eqnarray}}
\newcommand{\eea}{\end{eqnarray}}

\newcommand{\e}{\mbox{e}}
\renewcommand{\d}{\mbox{d}}
\newcommand{\g}{\gamma}

\newcommand{\lam}{\lambda}
\newcommand{\La}{\Lambda}
\renewcommand{\b}{\beta}
\renewcommand{\a}{\alpha}
\newcommand{\n}{\nu}
\newcommand{\m}{\mu}

\newcommand{\ep}{\varepsilon}     

\newcommand{\om}{\omega}
\newcommand{\del}{\delta}
\newcommand{\Del}{\Delta}
\newcommand{\sg}{\sigma}
\renewcommand{\k}{\kappa}

\newcommand{\oh}{\frac{1}{2}}

\newcommand{\dg}{\dagger}

\newcommand{\tr}{\mathrm{tr}\,}
\newcommand{\Tr}{\tr}
\newcommand{\ra}{\rangle}
\newcommand{\la}{\langle}
\newcommand{\lan}{\left\la}
\newcommand{\ran}{\right\ra}
\newcommand{\ket}{\ran}
\newcommand{\bra}{\lan}
\newcommand{\prt}{\partial}
\newcommand{\mi}{\!-\!}
\newcommand{\equ}{\!=\!}

\newcommand{\cD}{{\cal D}}

\newcommand{\cT}{{\cal T}}

\newcommand{\cO}{{\cal O}}

\renewcommand{\ts}{{\tilde{s}}}
\newcommand{\tC}{{\tilde{C}}}
\newcommand{\tG}{{\tilde{G}}}
\newcommand{\tL}{{\tilde{\La}}}

\newcommand{\hH}{{\hat{H}}}

\newcommand{\hC}{{\hat{C}}}

\newcommand{\hI}{{\hat{I}}}

\newcommand{\bN}{{\bar{N}}}
\newcommand{\bg}{{\bar{g}}}
\newcommand{\bz}{{\bar{z}}}
\newcommand{\bV}{{\bar{V}}}

\newcommand{\no}{\nonumber}
\newcommand{\nn}{\no\\}

\newcommand{\cdtL}{\La_{{\rm cdt}}}
\newcommand{\cdtZ}{Z_{\rm cdt}}
\newcommand{\cdtW}{W_{\rm cdt}}

\usepackage{makeidx}         
\usepackage{multicol}        
\usepackage[bottom]{footmisc}

\makeindex             

\begin{document}

\title*{Quantum gravity as sum over spacetimes}
\author{J. Ambj\o rn
,  J. Jurkiewicz
 ~and R. Loll
}
\institute{
The Niels Bohr Institute, Copenhagen, Denmark\\
Institute of Physics, Jagellonian University, Krakow, Poland.\\
Institute for Theoretical Physics, Utrecht University, The Netherlands.\\
~\\
\texttt{~ambjorn@nbi.dk,~jurkiewicz@th.if.uj.edu.pl,~r.loll@uu.nl}
}
%
\maketitle


\section{Introduction}
\label{intro}

A major unsolved problem in theoretical physics is to reconcile
the classical theory of general relativity with quantum mechanics.
These lectures will deal with an attempt to describe 
quantum gravity as a path integral over geometries.
Such an approach has to be non-perturbative since
gravity is a non-renormalizable\index{non-renormalizable} 
quantum field theory\index{quantum field theory} when 
the dimension of spacetime is four. 
In that case the dimension of the gravitational coupling constant $G$ is -2 
in units where $\hbar =1$ and $c=1$ and the dimension of mass is one.
Thus conventional, perturbative quantum field theory is only expected
to be good for energies
\beq\label{i1}
E^2 \ll 1/{G}.
\eeq  
That is still perfectly good in all experimental situations
we can imagine in the laboratory, but an indication that 
something ``new'' has to happen at sufficiently large energy,
or equivalently, at sufficiently short distances. It is possible,
or maybe even likely, that a breakdown of perturbation theory 
when \rf{i1} is not satisfied indicates that new degrees of 
freedom should be present in a theory valid at higher energies.
Indeed, we have a well known example in the electroweak theory.
Originally the electroweak theory was described by a four-fermion
interaction. Such a theory is not renormalizable and perturbation
theory\index{perturbation theory} 
breaks down at sufficiently high energy. In fact it breaks
down unless the energy satisfies \rf{i1} with
the gravitational coupling constant $G$ replaced
the coupling constant $G_F$ in front of the four-Fermi interaction (since
$G_F$ also has mass dimension -2). The breakdown reflects the
appearance of new degrees of freedom, the $W$ and the $Z$ particles,
and the four-Fermi interaction is now just an approximation to 
the process where a fermion interacts via $W$ and $Z$ particles with
another fermion. The corresponding electroweak theory is renormalizable.

When it comes to gravity there seems to be no ``simple'' fix like
the one just described. However, string theory\index{string theory} 
is an example of 
a theory which tries to solve the problem by adding (infinitely many) new
degrees of freedom. Loop quantum gravity\index{loop quantum gravity} 
is another approach to 
quantum gravity which tries to circumvent the problem of 
non-renormalizability\index{non-renormalizable}  
by introducing rules of quantization which 
are unconventional from a perturbative point of view. The
point of view taken here in these lectures is much more mundane.
In a sum-over-histories\index{path integral} 
approach\index{sum-over-histories} we will attempt to 
define a non-perturbative\index{non-perturbative} quantum field theory which has
as its infrared limit ordinary classical general relativity 
and at the same time has a nontrivial 
ultra\-vio\-let limit. From this point of view it is in the 
spirit of the renormalization group approach, first advocated 
long ago by Weinberg \cite{weinberg}, 
and more recently substantiated by several groups of researchers 
\cite{reuteretc}. 

To understand the possibility of a nontrivial ultraviolet fixed 
point\index{ultraviolet fixed point} 
let us first apply ordinary perturbation theory\index{perturbation theory} 
to quantum gravity 
in a regime where \rf{i1} is satisfied. One can in a reliable way
calculate the lowest-order quantum correction to the 
gravitational potential of a point particle:
\beq\label{i2}
\frac{G}{r} \to  \frac{G(r)}{r},~~~G(r)=G 
\left(1-\om \frac{G}{r^2}+ \cdots\right),~~~\om = \frac{167}{30\pi}.
\eeq  
Thus the gravitational coupling constant becomes scale-dependent and
transferring from distance to energy we have 
\beq\label{i3}
G(E) = G (1-\om G E^2 +\cdots) \approx \frac{G}{1+\om G E^2}.
\eeq
It should be stressed that the scenario described in \rf{i2} and \rf{i3}
is completely standard in quantum field theory. Let us take the
simplest quantum field theory relevant in nature: quantum electrodynamics.
The electron as we observe it in low-energy scattering experiments
is screened by vacuum polarization: virtual electron-positron pairs,
created out of the vacuum and annihilated again so fast that 
one has consistency with the energy-time uncertainty relations, act
like dipoles and the observed charge becomes less than the ``bare'' charge.
To lowest order (one loop) \rf{i2} and \rf{i3} are replaced by
\beq\label{i1a}
\frac{e^2}{r} \to  \frac{e^2(r)}{r},~~~e(r)=e 
\left(1-\frac{e^2}{6\pi^2}\ln (m\,r)\right)+ \cdots),~~~m\,r \ll 1.
\eeq
\beq\label{i2a}
e^2(E) = e^2\left( 1 + \frac{e^2}{6\pi^2}\ln (E/m)\right) + \cdots) \approx 
 \frac{e^2}{1-\frac{e^2}{6\pi^2} \ln (E/m)}.
\eeq
The last expression in eq.\ \rf{i2a} is precisely the 
renormalization group\index{renormalization group}-improved 
formula for the running coupling 
constant in QED. In fact, it is easy to calculate the so-called 
(one-loop) $\b$-function\index{$\beta$-function} 
for QED from the first equation in 
\rf{i2a} and use this $\b$-function to obtain the expression on the
left-hand side of 
\rf{i2a}. Contrary to \rf{i3} it breaks down for sufficiently 
high energy: it has a so-called Landau pole and it  
reflects  that we expect the interactions of QED to 
be infinitely strong at short distances. We do not believe that 
such quantum field theories really exist as ``stand-alone'' theories.
They either require an explicit cut-off, which will then enter
in the observables at high energies, or they have to be embedded 
in a larger theory without a Landau pole\index{Landau pole}.

Assume that the last expression in \rf{i3} 
was exact for all $E$ (which it is not). 
One can argue that one should use $G(E)$ in eq.\ \rf{i1} rather than 
$G$, in which case one obtains
\beq\label{i4}
G(E) E^2 < \frac{1}{\om} = \frac{30\pi}{167} ~~~(< 1).
\eeq
Thus, assuming \rf{i3} it suddenly seems as if 
quantum gravity had become a reliable 
quantum theory at all energy scales, the reason being that the 
effective coupling constant $G(E)$ becomes weaker at high energies.
The behaviour \rf{i3} can be described in terms of a $\b$-function for 
quantum gravity. Introduce the dimensionless coupling constant $\tG(E)$:
\beq\label{i5}
\tG(E) = G(E)E^2.
\eeq
From \rf{i3} it follows that $\tG(E)$ satisfies the following equation:
\beq\label{i6}
E \frac{\d \tG}{\d E} = \b(\tG),~~~~\b(\tG) = 2 \tG - 2 \om \tG^2.
\eeq
The zeros of $\b(\tG)$ determine the fixed points of the running coupling 
constant $\tG(E)$. The zero at $\tG = 0$ is an infrared fixed point:
for $E \to 0$ the coupling constant $\tG(E) \to 0$ and correspondingly
the coupling constant $G(E) \to G$. The zero at $\tG = 1/\om$ is 
an ultraviolet fixed point
\index{ultraviolet fixed point}: for $E\to \infty$ the coupling
constant $\tG(E) \to 1/\om$ (and $G(E) \to 0$ as $1/(\om E^2)$).

\begin{figure}[t]
\centerline{\scalebox{0.8}{\rotatebox{0}{\includegraphics{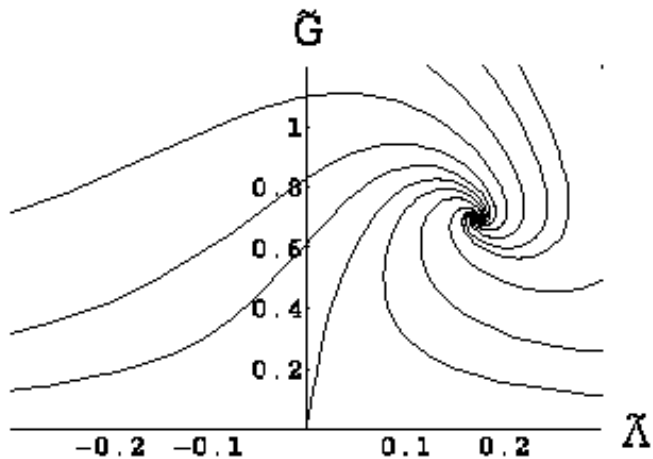}}}}
\caption{The flow of dimensionless gravitational and cosmological 
coupling constants $\tG$ and $\tL$ from a 
UV fixed point\index{ultraviolet fixed point}. 
One line flows to the 
infrared Gaussian fixed point where $\tG$ and $\tL$ are zero.}
\label{flow}
\end{figure}

While there is no compelling reason to take the above arguments
very seriously, since they are based on a one-loop calculation, the 
claim in \cite{reuteretc} is that a more careful analysis using 
the so-called exact renormalization group\index{renormalization group} 
equations or a systematic
$2+\ep$ expansion confirms the picture of a 
non-trivial UV fixed point\index{ultraviolet fixed point}.
A typical result of the renormalization group calculation is shown in
Fig.\ \ref{flow} in the case where the effective action has been
``truncated'' to just include the Einstein term and a cosmological term.
The figure has a UV fixed point\index{ultraviolet fixed point} 
from where the flow to low energies
starts and an infrared fixed point (the origin in the coupling constant 
coordinate system).

\begin{figure}[t]
\psfrag{B}{{\bf{\LARGE $\b(g)$}}}
\centerline{\scalebox{0.5}{\rotatebox{0}{\includegraphics{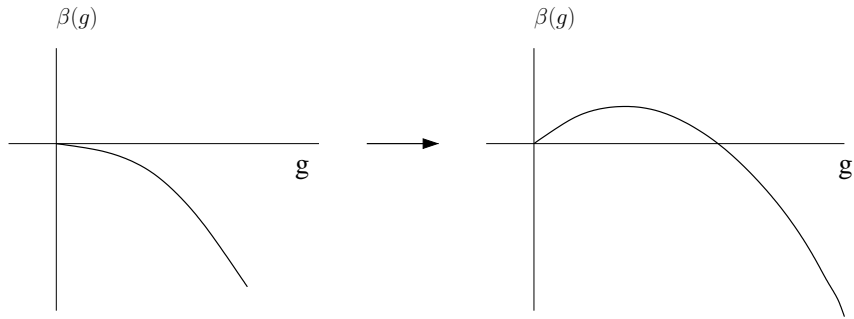}}}}
\caption{The change in the beta function $\b(g)$ in an asymptotically
free theory when the dimension changes from the critical dimension
$d$, where the coupling constant $g$ is dimensionless,
to $d+\ep$.}
\label{beta-f}
\end{figure}

Where do we meet such scenarios (IR fixed points at zero 
coupling and a nontrivial UV fixed point for a suitably defined
dimensionless coupling constant)? Assume one has an asymptotically 
free field theory\index{asymptotic freedom} 
in $d$ dimensions, i.e.\ $g =0$ is an 
UV fixed point\index{ultraviolet fixed point}.
The strong interactions in four-dimensional flat spacetime, 
quantum chromodynamics\index{quantum chromodynamics} (QCD), 
are such a theory. In high-energy 
scattering experiments the effective, running coupling constant 
goes to zero. This has been beautifully verified in high-energy 
experiments. The non-linear sigma model in two-dimensional 
spacetime is another model. It plays a very important role in string 
theory\index{string theory}, 
but even before that was extensively studied as a toy 
model of QCD since it is asymptotically free (i.e.\ the running, effective
coupling constant goes to zero at high energies). 
The asymptotically free theories
have a negative $\b$-function\index{$\beta$-function}. 
This is what makes the running coupling
constant go to zero at high energies. Change now (artificially) the 
dimension of spacetime infinitesimally from $d$ to $d+\ep$. Then
the $\b$-function to lowest order in $\ep$ will change as follows:
\beq\label{i7}
\b_{d}(g) \to \b_{d+\ep} = \ep g + \b_g(g),
\eeq   
and the situation is as shown in Fig.\ \ref{beta-f}: $g=0$ changes from 
an UV fixed point to an infrared fixed point while the 
new UV fixed point\index{ultraviolet fixed point} 
will be displaced to finite positive value $g_c(\ep)$
of $g$, a value which goes to zero when $\ep$ goes to zero.
In the case of gravity we have formally a renormalizable theory 
when $d=2$, the dimension where the gravitational coupling constant
$G$ is dimensionless. One can show that the two-dimensional 
theory can be viewed as asymptotically free (see eq.\ \rf{beta} below)
To apply eq.\ \rf{i7} to four-dimensional quantum gravity
starting with a renormalizable theory of gravity means that
$\ep$ has to be two, which is not very small.
Thus the considerations
above make little sense at a quantitative level, but the use
of the exact renormalization group indicates, as mentioned, that the
qualitative picture is correct.

The discussion above shows that there might be a chance that quantum
gravity can be defined as an ordinary quantum field theory at a 
non-trivial ultraviolet fixed point\index{ultraviolet fixed point}. 
It clearly requires non-perturbative
tools to address the question of the existence of such a fixed point 
and to analyze the properties of the 
field theory defined by approaching the fixed
point. One way to proceed is by using a lattice
regularization\index{lattice regularization} 
of the quantum field theory in question. The lattice 
provides an UV regularization of the quantum field theory, namely,
the inverse lattice spacing $1/a$. The task is then to define
a suitable ``continuum'' limit of this lattice theory. 
The procedure used is typically as follows:
let $\cO(x_n)$ be an observable, $x_n$ denoting a lattice point.
We write $x_n = a\, n$, $n$ measuring the position in integer lattice 
spacings. One can then obtain, either by computer simulations or by 
analytical calculations, the correlation length $\xi(g_0)$ in
lattice units, from
\beq\label{1.1}
-\log \la  \cO(x_n) \cO(y_m) \ra \sim |n-m|/ \xi(g_0) + o(|n-m|).
\eeq
A continuum limit of the lattice theory may then exist if it is possible to
fine-tune the bare coupling constant $g_0$ of the theory to a critical value
$g_0^c$ such that the correlation length\index{correlation length} 
goes to infinity, 
$\xi(g_0) \to \infty$. Knowing how $\xi(g_0)$ 
diverges for $g_0 \to g_0^c$ determines how the lattice spacing 
$a$ should be taken to zero as a function of the coupling constants, namely,
\beq\label{1.2}
\xi(g_0)\propto \frac{1}{|g_0-g_0^c|^\n},~~~~~a(g_0)\propto |g_0-g_0^c|^\n.
\eeq
This particular scaling of the lattice spacing ensures that one can 
define a physical mass $m_{ph}$ by 
$$
m_{ph} a(g_0) = 1/\xi(g_0)
$$ 
such that the correlator\index{correlation function} 
$\la  \cO(x_n) \cO(y_m) \ra$ falls off exponentially like
$\e^{-m_{ph} |x_n-y_m|}$ for $g_0 \to g_0^c$ when $|x_n-y_m|$, but not
$|n-m|$, is kept fixed in the limit $g_0\to g_0^c$.
Thus we have created a picture
where the underlying lattice spacing goes to zero while the physical mass
(or the correlation length measured in physical length units, not in 
lattice spacings) is kept fixed. This is the standard Wilsonian scenario
for obtaining the continuum (Euclidean) quantum 
field theory associated with the critical point $g_0^c$ of a {\it second
order} phase transition\index{phase transition} 
(for second-order phase transitions there
exists a correlation length which diverges, usually associated
with the order parameter characterizing the transition).

We would like to apply a similar approach to quantum gravity, and
thus obtain a new way to investigate if quantum
gravity can be defined non-perturbatively\index{non-perturbative} 
as a quantum field theory. The predictions from such a theory could 
then be compared with the renormalization 
group predictions related to the 
asymptotic safety\index{asymptotic safety} 
picture described above. It should be mentioned that the 
asymptotic safety picture is not the only suggestion for 
a {\it continuum} quantum theory of gravity using only ``conventional'' ideas
of quantum field theory. Very recently two other scenarios
have been suggested. One is called Lifshitz gravity\index{Lifshitz gravity} 
\cite{horava} and 
is a theory where the non-renormalizability of the Einstein-Hilbert
theory\index{Einstein-Hilbert action} 
is cured by adding higher-order spatial derivatives in a way 
somewhat similar to what Lifshitz did many years ago in statistical models.
In fact, the setup of the theory has some resemblance with the lattice-theory setup of 
``Causal Dynamical Triangulations (CDT)"
\index{CDT}\index{causal dynamical triangulation: see CDT}, to be described below,
since a time foliation is assumed and the infrared limit is that of GR.
However, contrary to Lifshitz gravity, we do not attempt to 
put in higher spatial derivatives in the lattice theory. However, when a 
continuum limit in the lattice theory is taken in a specific way which is not 
entirely symmetric in space and time one cannot rule out 
that higher spatial derivatives can play a role.
The other model goes by the name of
``scale-invariant gravity"\index{scale-invariant gravity} 
\cite{shaposh1,shaposh2}. It modifies
gravity into a renormalizable theory by introducing a scalar degree of 
freedom in addition to the transverse gravitational degrees of freedom.
Also this model has interesting features not incompatible with
the results of computer simulations using the CDT lattice model.

As already mentioned, we will use a lattice approach known as  
{\it causal dynamical triangulations} (CDT) as a regularization.
In Sec.\ \ref{CDT} we give a short description of the formalism, providing
the definitions which are needed later to describe the measurements.
CDT establishes a non-perturbative way of performing the sum over 
four-geometries\index{sum-over-histories} 
(for more extensive definitions, see \cite{ajl4d,blp}).
It sums over the class of piecewise linear 
four-geometries\index{piecewise linear geometries} which can be 
assembled from four-dimensional simplicial building blocks of link length $a$, 
such that only {\it causal} spacetime histories\index{spacetime history} 
are included. The challenge when searching for a {\it field theory} 
of quantum gravity is to find a theory
which behaves as described above, i.e.\ as in eq.\ \rf{1.2}. 
The challenge is three-fold: (i) to find a suitable
non-perturbative\index{non-perturbative} 
formulation of such a theory which satisfies a 
minimum of reasonable requirements, (ii) to find
observables which can be used to test relations like \rf{1.1}, and (iii) to
show that one can adjust the coupling constants of the
theory such that \rf{1.2} is satisfied. Although we will focus on (i) in what
follows, let us immediately mention that 
(ii) is notoriously difficult in a theory of 
quantum gravity, where one is faced with a number of 
questions originating in the dynamical nature of geometry.
What is the meaning of distance when integrating
over all geometries? How do we attach a meaning to local spacetime
points like $x_n$ and $y_n$? How can
we define at all local, diffeomorphism-invariant quantities in the continuum
which can then be translated to the regularized (lattice) theory? --
What we want to point out here is that although (i)-(iii) are
standard requirements when relating critical phenomena
and (Euclidean) quantum field theory, gravity {\it is} special and 
may require a reformulation of (part of) the standard scenario
sketched above. We will return to this issue later.

Our proposed non-perturbative\index{non-perturbative} formulation
of four-dimensional quantum gravity has a number of nice properties.

First, it sums over a class of piecewise 
linear geometries\index{piecewise linear geometries}.
The characteristic feature of piecewise linear geometries is 
that they admit a description without the use of coordinate systems.
In this way we perform the sum over geometries\index{sum over geometries} 
directly, avoiding the cumbersome procedure of first introducing 
a coordinate system and then getting rid of the ensuing gauge redundancy,
as one has to do in a continuum calculation. 
Our underlying assumptions are that 1) the class of piecewise
linear geometries is in a suitable sense dense in the set of all geometries 
relevant for the path integral\index{path integral} 
(probably a fairly mild assumption), and
2) that we are using a correct measure on the set of geometries. 
This is a more questionable assumption since we do not even 
know whether such a measure exists. Here one has to take a 
pragmatic attitude in order to make progress. 
We will simply examine the outcome of our construction and try to judge 
whether it is promising. 

Secondly, our scheme is background-independent\index{background-independent}. 
No distinguished geometry, accompanied by 
quantum fluctuations\index{quantum fluctuations}, 
is put in by hand. 
If the CDT-regularized theory\index{CDT} is to be taken seriously 
as a potential theory of quantum gravity, there has to be a
region in the space spanned by the bare coupling constants where the geometry 
of spacetime bears some resemblance with the kind of universe we
observe around us. That is, the theory should 
create dynamically an effective background 
geometry\index{background geometry} 
around which there are (small) quantum fluctuations.
This is a very nontrivial property of the theory and one we are going
to investigate in some detail. 
Computer simulations presented in these lectures
confirm in a much more direct
way the indirect evidence for such a scenario which we have known
for some time and first reported in \cite{emerge,semi}. 
They establish the de Sitter\index{de Sitter universe} nature of the 
background spacetime\index{background geometry}, 
quantify the fluctuations around it,
and set a physical scale for the universes we are dealing with.
The main results of these investigations, without the numerical details, 
were announced in \cite{agjl} and a detailed account 
of the results was presented in \cite{bigs4}.   

The remainder of these lecture notes is organized as follows:
in Sec.\ \ref{CDT} we describe the lattice formulation of 
four-dimensional quantum gravity. In Sec.\ \ref{numerical} 
the numerical results in 4d are summarized. We view these 
results as very important, but they also serve
as a motivation for moving to two dimensions. While there is no
propagating graviton in two-dimensional quantum gravity, it is 
a diffeomorphism-invariant theory and almost all of the conceptional 
problems mentioned above are present there.
Thus it is an important exercise to solve two-dimensional
quantum gravity. Surprisingly this can be done in the lattice
regularization known as ``dynamical triangulation''. An important 
corollary is that (1) one can explicitly construct the 
continuum limit\index{continuum limit} of the lattice theory
and (2) show that it agrees with the 
so-called Liouville 2d quantum gravity theory. This latter
theory is a continuum conformal field theory, explicit solvable,
and, when viewed in the correct way, a diffeomorphism-invariant theory.  
Thus there is indeed no problem having a 
lattice regularization\index{lattice regularization} 
of a diffeomorphism-invariant theory. In Sec.\ \ref{2d-euclid}
we solve what is known as two-dimensional 
Euclidean quantum gravity\index{Euclidean quantum gravity}.
In Sec.\ \ref{2d-CDT} we show how one can interpolate from Euclidean
2d quantum gravity to ``Lorentzian'' 2d 
quantum gravity\index{Lorentzian quantum gravity} which is a
two-dimensional version of the four-dimensional gravity theory we
have discussed in Sec.\ \ref{numerical}. 
Finally Sec.\ \ref{discuss} discusses the results obtained and 
outlines perspectives.

\section{CDT\index{CDT}}\label{CDT}

The use of so-called causal dynamical triangulations
(CDT) stands in the tradition of 
\cite{teitelboim}, which advocated that in a gravitational path 
integral\index{path integral} 
with the correct, Lorentzian signature\index{Lorentzian signature} 
of spacetime one should sum 
over causal geometries\index{sum over geometries} only. 
More specifically, we adopted this idea when it
became clear that attempts to formulate a {\it Euclidean} 
non-perturbative\index{non-perturbative} 
quantum gravity\index{Euclidean quantum gravity}
theory run into trouble in spacetime dimension $d$ larger than two
as will be described below.

This implies that we start from Lorentzian simplicial 
spacetimes with $d=4$ and insist
that only causally well-behaved geometries appear in the
(regularized) Lorentzian path integral\index{Lorentzian path integral}. 
A crucial property of our explicit construction is that 
each of the configurations allows for a rotation to 
Euclidean signature\index{Euclidean signature}.
We rotate to a Euclidean regime in order to perform the sum over geometries
(and rotate back again afterward if needed).
We stress here that although the sum is performed over geometries 
with Euclidean signature, it is different from what one would 
obtain in a theory of quantum gravity based ab initio on Euclidean spacetimes.
The reason is that not all Euclidean geometries with a given topology are 
included in the ``causal" sum since in general they have no correspondence 
to a causal Lorentzian geometry.
 
\begin{figure}[t]
\psfrag{a}{{\bf{\LARGE $t_i$}}}
\psfrag{b}{{\bf{\LARGE $t_f$}}}
\psfrag{c}{{\hspace{-20pt}\bf{\LARGE $x_i$}}}
\psfrag{d}{{\hspace{-20pt}\bf{\LARGE $x_f$}}}
\psfrag{t}{{\bf{\LARGE $t$}}}
\psfrag{x}{{\bf{\LARGE $x$}}}
\psfrag{u}{{\bf{\LARGE $a$}}}
\centerline{\scalebox{0.7}{\rotatebox{0}{\includegraphics{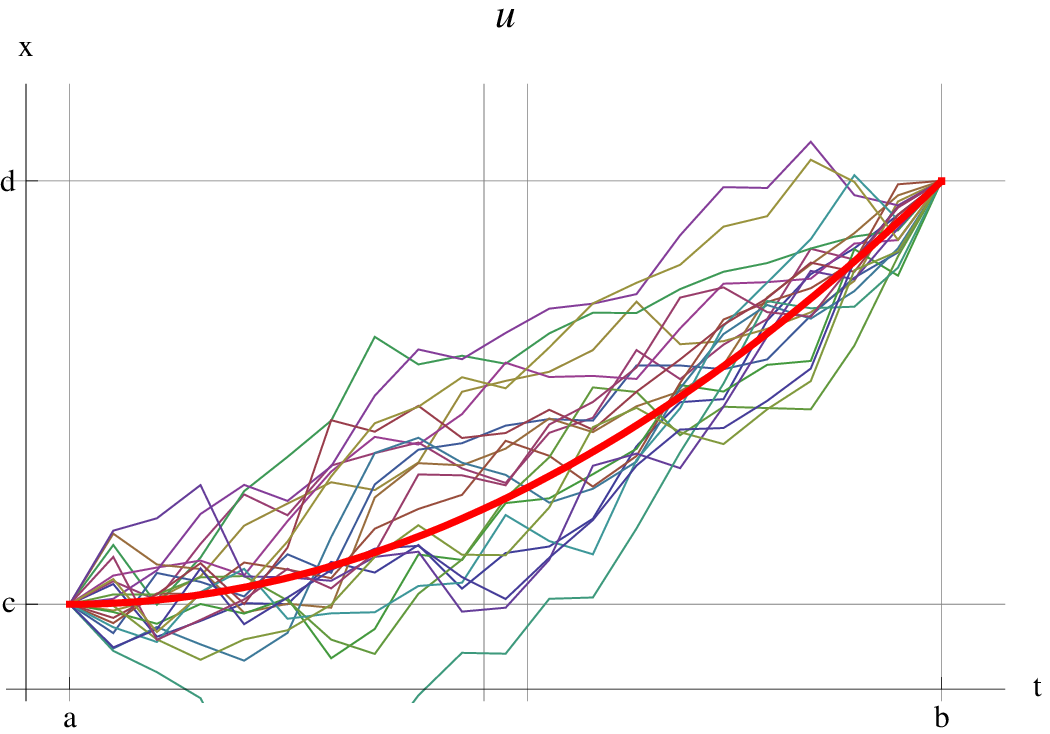}}}}
\vspace{12pt}
\centerline{\scalebox{0.3}{\rotatebox{90}{\includegraphics{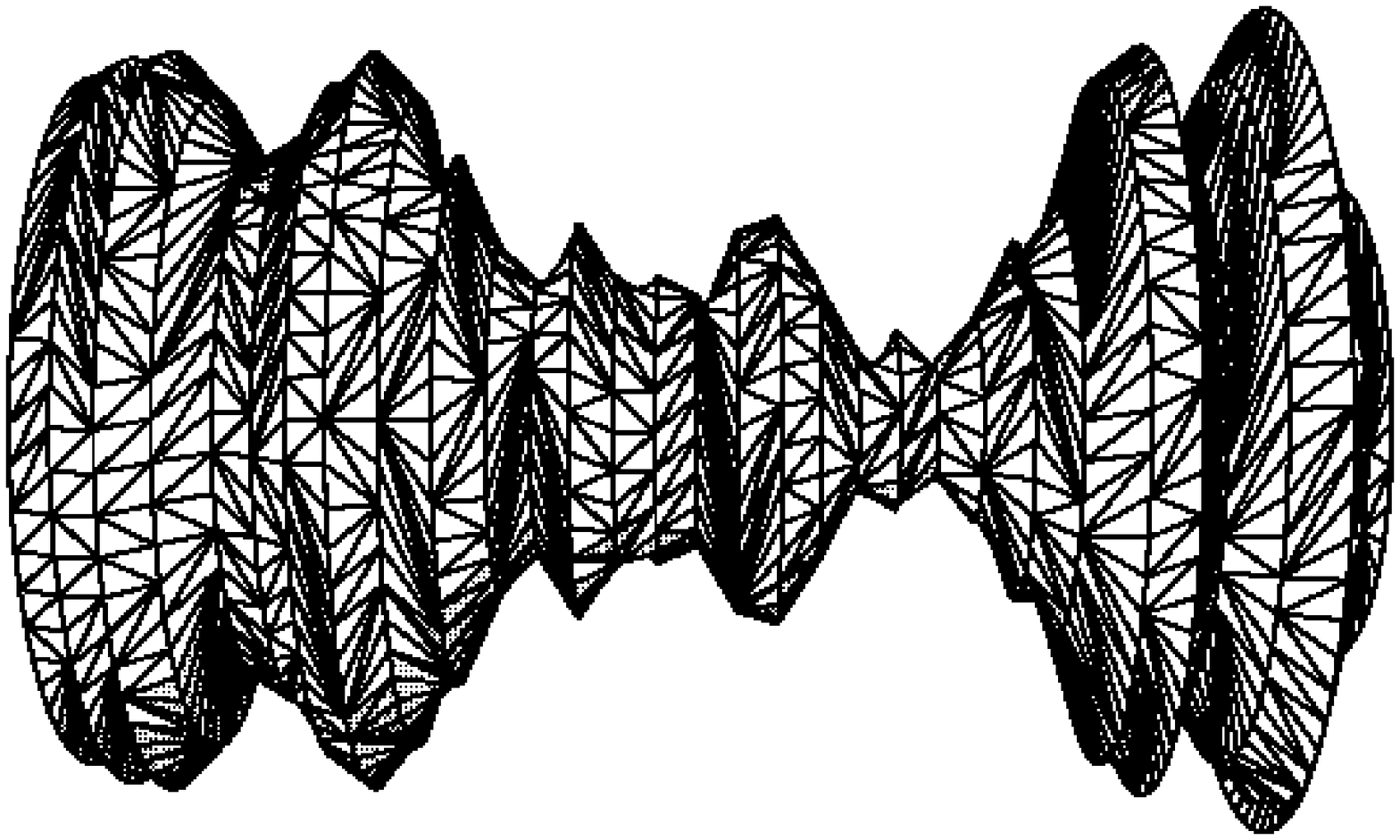}}}}
\caption{Piecewise linear space-time histories 
in quantum mechanics and in (1+1)-dimensional  quantum gravity. In 
the gravity case we show only a single space-time history, while 
in the quantum particle case we show many such histories as well
as the average path (thick line).}
\label{fig0}
\end{figure}

\begin{figure}[t]
\centerline{\scalebox{0.5}{\rotatebox{0}{\includegraphics{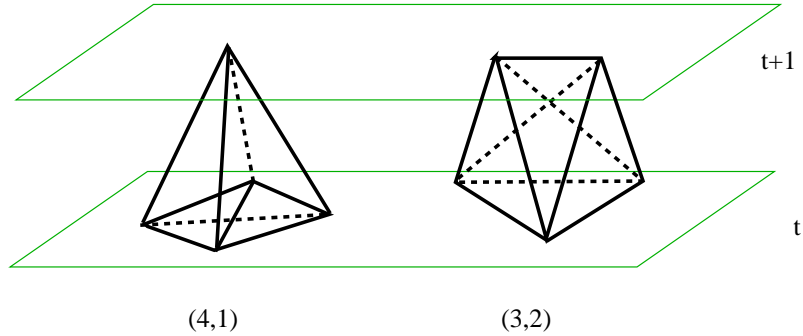}}}}
\caption{(4,1) and a (3,2) simplices connecting two neighbouring spatial 
slices. We also have symmetric (1,4) and (2,3) simplices with a vertex and
a line, respectively, at time $t$ and a tetrahedron and a triangle,
respectively, at time $t+1$. For simplicity we denote the 
total number of (4,1) and (1,4) simplices by $N^{(4,1)}_4$ and 
similarly the total number of (3,2) and (2,3) simplices by $N_4^{3,2}$. 
}
\label{connect}
\end{figure}

We refer to \cite{ajl4d} for a detailed description
of how to construct the class of 
piecewise linear geometries\index{piecewise linear geometries} used 
in the Lorentzian path integral. 
The most important assumption is the existence of 
a global proper-time foliation. This is symbolically illustrated
in Fig.\ \ref{fig0} where we compare the construction to the 
one of ordinary quantum mechanics: the path integral of ordinary 
quantum mechanics is regularized as a sum over piecewise linear 
paths\index{path integral} from point $x_i$ to point $x_f$ in time $t_f-t_i$. 
The time steps have length $a$ and the continuum limit\index{continuum limit}
is obtained when the length $a$ of these ``building blocks'' goes to 
zero. Similarly, in the quantum gravity case we have a sum over 
four-geometries, ``stretching'' between two three-geometries separated 
a proper time $t$ and constructed from four-dimensional building blocks,
as described below. On the figure we show for illustrational simplicity 
only a single space-time history and 
replace the three-dimensional spatial 
geometries with a one-dimensional one of $S^1$-topology. 
Moving from left to right we have a time foliation where
at each discrete time step space is represented 
by a circle. Neighbouring
circles are then connected by piecewise flat building blocks, usually triangles, as illustrated 
in Fig.\ \ref{2dminkowski} in Sec.\ \ref{2d-CDT}.
In the ``real'' four-dimensional
case, the spatial slices of topology $S^1$ will be replaced
by spatial slices of topology $S^3$, and neighbouring spatial $S^3$ 
slices are then connected by four-simplices as illustrated in 
Fig.\ \ref{connect} and described in detail below.
There is an  important difference between the quantum mechanical 
sum over paths and our sum over geometries with a time foliation:
the time $t$ in the quantum mechanical example is 
an external parameter, while the time $t$ in the case of quantum gravity is 
intrinsic. Also, again for the purpose of illustration, the two-dimensional
geometry has been drawn embedded in three-dimensional space, but in
the path integral\index{path integral} 
implementing the summation over geometries there is no such embedding present. 
Finally, we cannot refrain from mentioning that the paths shown for the 
quantum mechanical particle are in fact typical paths which appear for the 
(Euclideanized) path integral of a particle placed in an external potential.
They are picked out from an actual Monte Carlo simulation of such
a physical system. Similarly, the 2d surface is a surface picked out
from a Monte Carlo simulation of 2d quantum gravity and thus corresponds
to a typical 2d surface which appears in the path integral. 
This is the reason for the somewhat poor graphic representation: there is no 
natural length-preserving representation of the surface in 3d such
that the surface is not self-intersecting. For 
better graphic illustrations (animations) of the 2d surfaces which appear in
the 2d quantum-gravitational path integral we refer to the link \cite{link}.    

As mentioned above, we assume that the spacetime
topology is that of $S^3 \times R$, 
the spatial topology being that of $S^3$ merely for convenience.
The spatial geometry at each discrete proper-time step $t_n$ is represented
by a triangulation of $S^3$, made up of equilateral spatial tetrahedra with 
squared side-length $\ell_s^2\equiv a^2 >0$. 
In general, the number $N_3(t_n)$ of tetrahedra and how 
they are glued together to form a piecewise flat three-dimensional manifold
will vary with each time-step $t_n$. 
In order to obtain a four-dimensional triangulation,
the individual three-dimensional slices must 
still be connected in a causal way, 
preserving the $S^3$-topology at all intermediate times $t$ between 
$t_n$ and $t_{n+1}$ 
\footnote{This implies the absence of branching of the spatial
universe into several disconnected pieces, so-called {\it baby universes}, 
which (in Lorentzian signature\index{Lorentzian signature}) would 
inevitably be associated with causality violations 
in the form of degeneracies in
the light cone structure, as has been discussed elsewhere 
(see, for example, \cite{causality}).}. 
This is done as illustrated in Fig.\
\ref{connect}, introducing what we call (4,1)-simplices and (3,2)-simplices.
More precisely, a $(4,1)$-simplex is a four-simplex with four of its
vertices (i.e.\ a boundary tetrahedron) belonging to the 
triangulation of  $S^3(t_n)$, the time-slice corresponding to time $t_n$, 
and the fifth vertex belonging to the triangulation
of  $S^3(t_{n+1})$, the time-slice corresponding to time $t_{n+1}$. 
Similarly, a $(3,2)$
simplex has three vertices, i.e.\ a triangle, belonging to the 
triangulation of $S^3(t_n)$ and two vertices, i.e.\ a link, belonging to
the triangulation of $S^3(t_{n+1})$. We have also simplices
of type $(1,4)$ and $(2,3)$, which are defined in an obvious way,
interchanging the role of $S^3(t_n)$ and $S^3(t_{n+1})$. One can show
that two triangulations  of $S^3(t_n)$ and $S^3(t_{n+1})$ can 
be ``connected'' by these four building blocks glued together 
in a suitable way such that we have a four-dimensional triangulation
of $S^3\times [0,1]$. Also, two given triangulations  of 
$S^3(t_n)$ and $S^3(t_{n+1})$ can be connected in many ways compatible 
with the topology $S^3\times [0,1]$. 
In the path integral\index{path integral} 
we will be summing over all possible ways 
to connect a given triangulation $S^3(t_n)$ to a given
triangulation of $S^3(t_{n+1})$ compatible with the topology $S^3\times [0,1]$.
In addition we will sum over all 3d triangulations
of $S^3$ at all times $t_n$. 

We allow for an asymmetry between temporal and spatial lattice length assignments.
Denote by $\ell_t$ and $\ell_s$ the length of the time-like links and 
the space-like links, respectively. Then 
$\ell_t^2\equ -\alpha \ell_s^2$, $\alpha >0$.  
The explicit rotation to Euclidean 
signature is done by performing the rotation $\a \to -\a$ in the complex lower
half-plane, $|\a| > 7/12$, such that we have 
$\ell_t^2 = |\a| \ell_s^2$ (see \cite{ajl4d} for a discussion).

The Einstein-Hilbert action\index{Einstein-Hilbert action} 
$S^{\rm EH}$ has a natural geometric implementation
on piecewise linear geometries\index{piecewise linear geometries} 
in the form of the Regge action\index{Regge action}. 
This is given by the sum of the so-called deficit 
angles around the two-dimensional ``hinges" 
(subsimplices in the form of triangles), 
each multiplied with the volume
of the corresponding hinge. In view of the fact that we 
are dealing with piecewise
linear, and not smooth metrics, there is no unique ``approximation'' 
to the usual Einstein-Hilbert action\index{Einstein-Hilbert action}, 
and one could in principle work with a different form of the
gravitational action. We will stick with the Regge action, 
which takes on a very simple form in our case, where  
the piecewise linear manifold is constructed from just two different 
types of building blocks.
After rotation to Euclidean signature\index{Euclidean signature} 
one obtains for the action (see \cite{blp} for details)
\bea
S_E^{\rm EH}&=& \frac{1}{16\pi^2 G} \int d^4x \sqrt{g} (-R+2\La) ~~
\longrightarrow \\
 S_E^{\rm Regge}&=& 
-(\kappa_0+6\Delta) N_0+\kappa_4 (N_{4}^{(4,1)}+N_{4}^{(3,2)})+
\Delta (2 N_{4}^{(4,1)}+N_{4}^{(3,2)}),\nonumber
\label{actshort}
\eea 
where $N_0$ denotes the total number of vertices in the four-dimensional 
triangulation and $N_{4}^{(4,1)}$ and $N_{4}^{(3,2)}$ denote 
the total number of the four-simplices described above, so that the
total number $N_4$ of four-simplices is $N_4=N_{4}^{(4,1)}+N_{4}^{(3,2)}$. 
The dimensionless coupling constants $\k_0$ and $\k_4$ are 
related to the bare gravitational and bare cosmological coupling constants,
with appropriate powers of the lattice spacing $a$ already absorbed into 
$\k_0$ and $\k_4$. The {\it asymmetry parameter}
$\Del$ is related to the parameter $\a$ introduced above, which describes 
the relative scale
between the (squared) lengths of space- and time-like links. 
It is both convenient and natural to keep track of 
this parameter in our set-up, which from the
outset is not isotropic in time and space directions, see again \cite{blp} 
for a detailed discussion. 
Since we will in the following work with the path integral\index{path integral}
after Wick rotation, let us redefine $\tilde \a:=-\a$ \cite{blp}, 
which is positive in the Euclidean
domain.\footnote{The most symmetric choice is $\tilde\a=1$, 
corresponding to vanishing asymmetry, $\Delta=0$.} 
For future reference, the Euclidean four-volume of our universe 
for a given choice of $\tilde\a$ is given by 
\beq\label{vol2}
V_4 = \tilde{C}_4(\xi)\, a^4\, N_4^{(4,1)} =  
\tilde{C}_4(\xi)\, a^4 \, N_4/(1+\xi),
\eeq
where $\xi$ is the ratio
\beq\label{vol3}
\xi = N_4^{(3,2)}/N_4^{(4,1)},
\eeq
and $\tilde{C}_4(\xi)\, a^4$ is a measure of the ``effective four-volume''
of an ``average'' four-simplex. $\xi$ will depend
on the choice of coupling constants in a rather complicated way
(for a detailed discussion we refer to \cite{ajl4d,bigs4}).

The path integral\index{CDT path integral} 
or partition function for the CDT version of quantum gravity is now
\beq\label{2.1}
Z(G,\La) = \int \cD [g] \; \e^{-S_E^{\rm EH}[g]} ~~~\to~~~ 
Z(\k_0,\k_4,\Del) = 
\sum_{\cT} \frac{1}{C_\cT} \; \e^{-S_E(\cT)},
\eeq
where the summation is over all causal triangulations $\cT$ of the kind 
described above, and we have dropped the superscript ``Regge" on 
the discretized action. 
The factor $1/C_\cT$ is a symmetry factor, given by the order of 
the automorphism group of the triangulation $\cT$. 
The actual set-up for the simulations is as
follows. We choose a fixed number $N$ of spatial slices at proper times $t_1$,
$t_2= t_1 + a_t$, up to $t_N = t_1 + (N\mi 1) a_t$, 
where $\Delta t\equiv a_t$ is the discrete lattice
spacing in temporal direction and $T = N a_t$ the 
total extension of the universe in proper time. 
For convenience we identify $t_{N+1}$
with $t_1$, in this way imposing the topology $S^1\times S^3$ rather
than $I\times S^3$. This choice does not affect physical results,
as will become clear in due course.

Our next task is to {\it evaluate} the non-perturbative\index{non-perturbative}
sum in \rf{2.1}, if possible, analytically.
This can be done in spacetime dimension $d=2$ 
(\cite{al,alwz} (and we discuss this in detail below) 
and at least partially in $d=3$ 
\cite{worm,blz}, but presently an analytic solution in
four dimensions is out of reach. 
However, we are in the fortunate situation that
$Z(\kappa_0,\kappa_4,\Delta)$ can be studied quantitatively with the 
help of Monte Carlo simulations\index{Monte Carlo simulations}. 
The type of algorithm needed to update 
the piecewise linear geometries\index{piecewise linear geometries} 
has been around for a while,
starting from the use of dynamical triangulations in bosonic string
theory (two-dimensional Euclidean triangulations) \cite{adf,david,migdal} 
and was later extended to their application in Euclidean 
four-dimensional quantum gravity\index{Euclidean quantum gravity} 
\cite{aj,migdal1}.
In \cite{ajl4d} the algorithm was modified to 
accommodate the geometries of the CDT set-up.
The algorithm is such that it takes the 
symmetry factor $C_\cT$ into account automatically. 

\begin{figure}[t]
\centerline{\scalebox{0.6}{\rotatebox{0}{\includegraphics{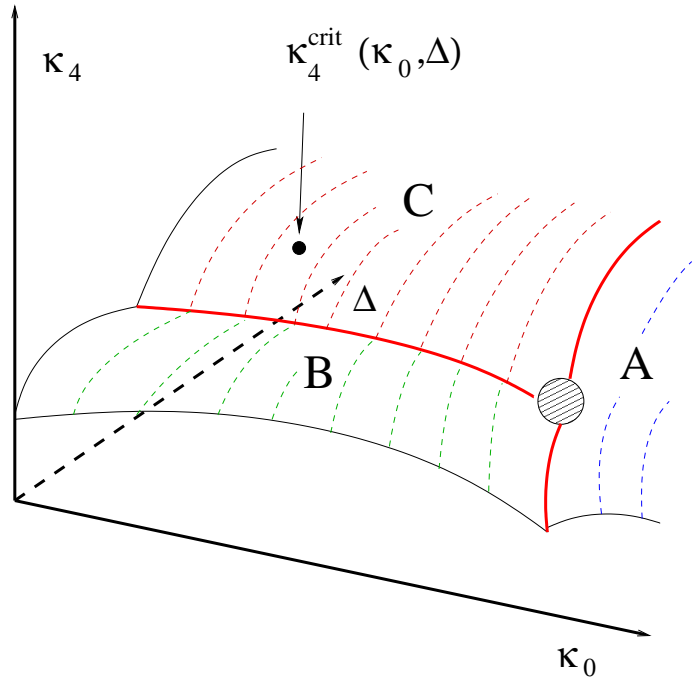}}}}
\caption{The phases A, B and C in the coupling constant space 
$(\k_0,\Del,\k_4)$. Phase C is the one where extended 
four-dimensional geometries emerge. }
\label{ph-diagram}
\end{figure}

We have performed extensive 
Monte Carlo simulations\index{Monte Carlo simulations} of the 
partition function $Z$ for a number of values of the bare coupling constants.
As reported in \cite{blp}, there are regions of the coupling 
constant space which do not appear relevant for continuum physics in
that they seem to suffer from problems similar to the ones found
earlier in {\it Euclidean} quantum gravity\index{Euclidean quantum gravity} 
constructed in terms of dynamical
triangulations, which essentially led to its abandonment in $d>2$.  
What is observed in Euclidean four-dimensional quantum 
gravity is the following:  when the (inverse, bare) 
gravitational coupling $\k_0$ is sufficiently large one sees
so-called branched polymers\index{branched polymers}, 
i.e.\ not really a four-dimensional universe,
but a universe which branches out like a tree with so many branches that 
it becomes truly fractal when the number of four-simplices 
becomes infinite, and its Hausdorff dimension\index{Hausdorff dimension} is 2. 
Such triangulations represent the most extended triangulations
one can construct unless one explicitly forbids branching.
When the (inverse, bare) 
gravitational coupling $\k_0$ is sufficiently small one observes a 
totally crumpled universe\index{crumpled phase} 
with almost no extension. In this phase there
exist vertices of very high order and the connectivity 
of the triangulation is such that it is possible to move from any four-simplex 
to any other crossing only a few neighbouring
four-simplices. The Hausdorff dimension\index{Hausdorff dimension} 
of such a triangulation
is infinite in the limit where the number of four-simplices goes
to infinity. These two phases, the crumpled and the 
branched polymer\index{branched polymers} phase,
are separated by a phase transition line along which there is a 
first-order transition. It was originally hoped that one could find 
a point on the critical line where the first-order transition 
becomes second order and which could then be used
as a fixed point where a continuum theory of quantum gravity
could be defined along the lines suggested by eqs.\ \rf{1.1} and \rf{1.2}.
However, such a second-order transition was not found\cite{firstorder}, 
and eventuallythe idea of a theory of four-dimensional 
Euclidean quantum gravity\index{Euclidean quantum gravity}
was abandoned. A new principle for selecting the class of geometries
one should use in the path integral was needed and this led to the
suggestion to include only {\it causal} triangulations in the sum over 
spacetime histories\index{sum-over-histories}\index{spacetime history}.

\begin{figure}[t]
\centerline{{\scalebox{0.5}{\rotatebox{90}{\includegraphics{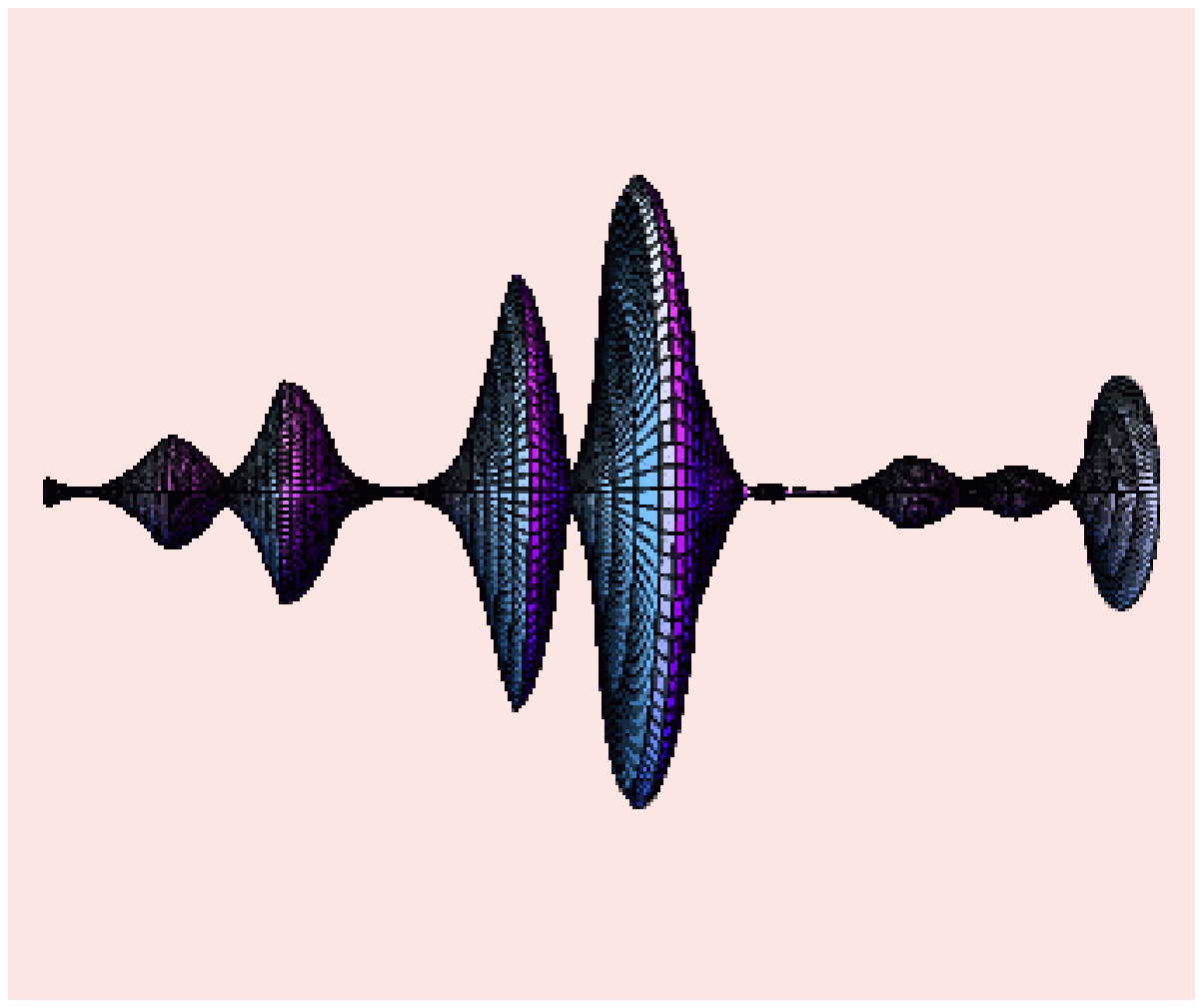}}}}
{\scalebox{0.5}{\rotatebox{90}{\includegraphics{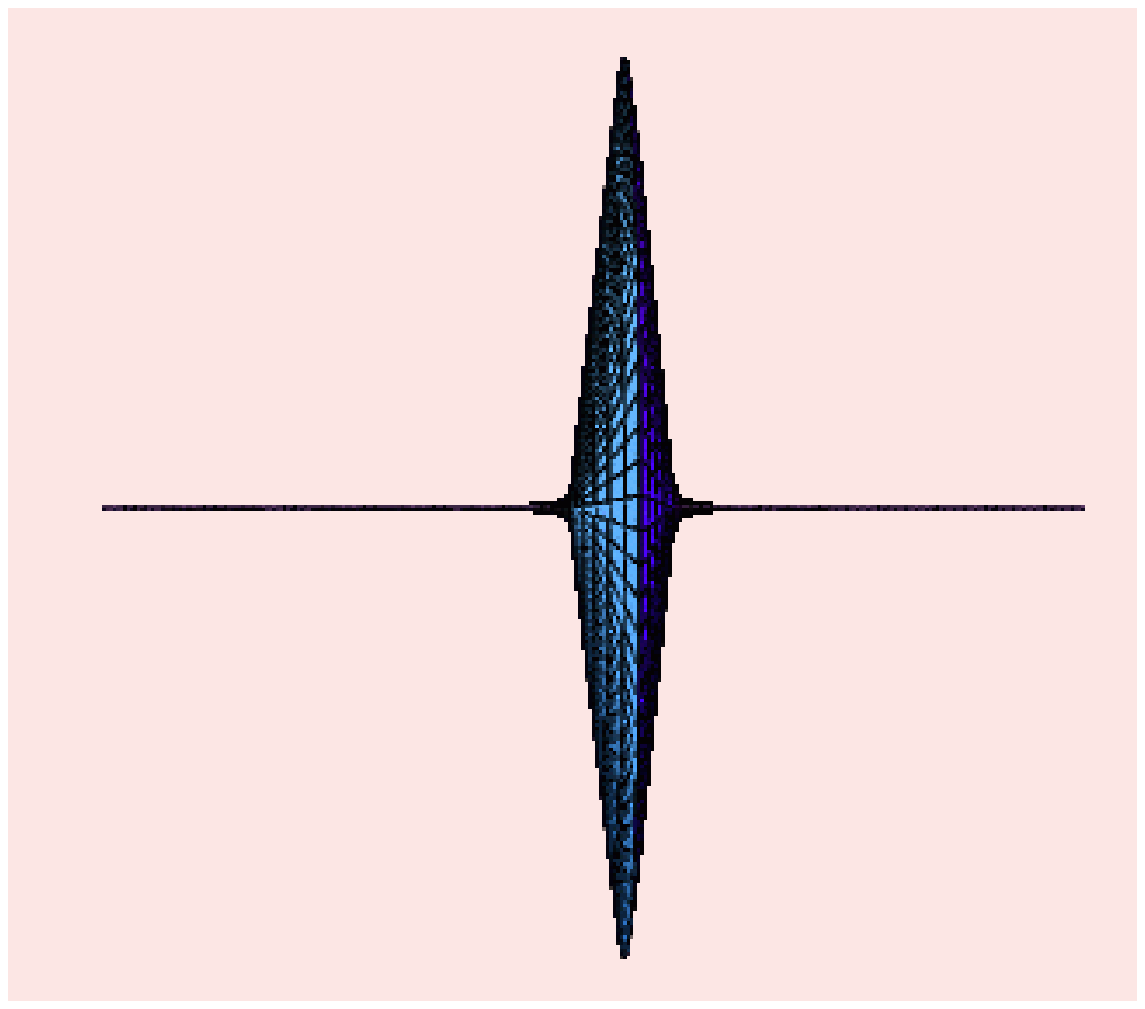}}}}}
\centerline{{\scalebox{1.03}{\rotatebox{90}{\includegraphics{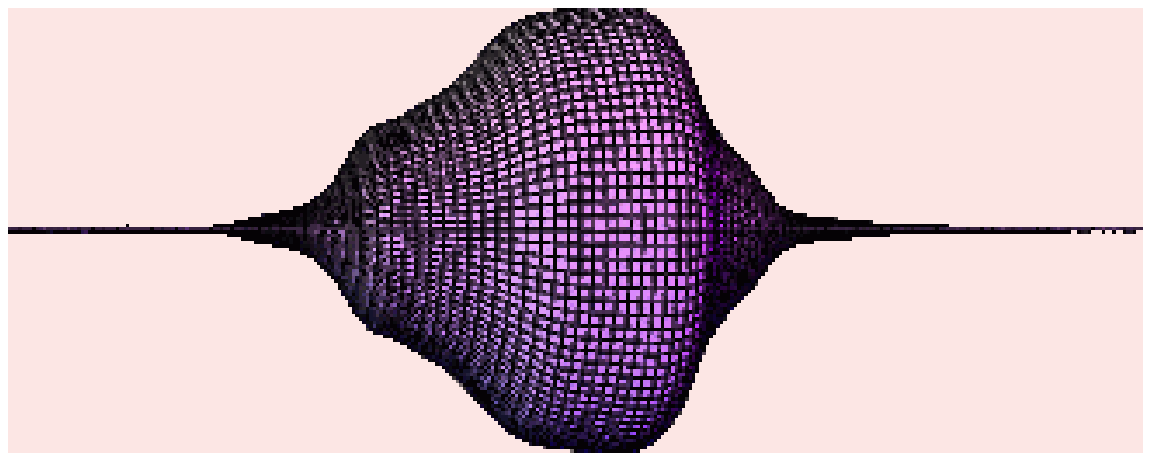}}}}}
\caption{Typical configurations in the phases A, B and C (lowest figure).
Phase C is the one where extended four-dimensional geometries emerge.}
\label{blobs}
\end{figure}

When we include only the causal triangulations in the 
path integral\index{CDT path integral},
we still see a remnant of the Euclidean structure just described,
namely, when the (inverse, bare) gravitational coupling $\k_0$ is sufficiently 
large, the Monte Carlo simulations\index{Monte Carlo simulations} exhibit
a sequence in time direction of small, disconnected universes, none of 
them showing any sign of the scaling one would 
expect from a macroscopic universe. We denote this phase by A. 
We believe that this phase of the system is a Lorentzian version of the 
branched-polymer phase\index{branched polymers} of Euclidean 
quantum gravity\index{Euclidean quantum gravity}. 
By contrast, when $\Del$ is sufficiently small, the simulations
reveal a universe with a vanishing temporal extension of only a few 
lattice spacings, ending both in past 
and future in a vertex of very high order, 
connected to a large fraction of all vertices. 
This phase is most likely related
to the so-called crumpled phase\index{branched polymers} 
of Euclidean quantum gravity. We denote this phase by B.
The crucial and new feature of the quantum superposition in terms of 
{\it causal} dynamical triangulations is the appearance of a region 
in coupling constant space which is different and interesting and 
where continuum physics may emerge. 
It is in this region that we have performed the 
simulations discussed here 
and where work up to now has already uncovered a number of intriguing
physical results \cite{emerge,semi,blp,spectral}. 
In Fig.\ \ref{blobs} we have shown how different configurations 
look in the three phases discussed above, and in
Fig.\ \ref{ph-diagram} we have shown the tentative phase diagram 
in the coupling constant space of $\k_0,\k_4$ and $\Del$.
A ``critical'' surface is shown in the figure. Keeping $\k_0$ and
$\Del$ fixed, $\k_4$ acts as a chemical potential for $N_4$;
the smaller $\k_4$, the larger $\la N_4\ra$. At some critical 
value $\k_4(\k_0,\Del)$, depending on the choice of $\k_0$ and $\Del$,
$\la N_4\ra \to \infty$. For $\k_4 < \k_4(\k_0,\Del)$ the partition
function is plainly divergent and not defined. When we talk about 
phase transitions\index{phase transition} 
we are always at the ``critical'' surface
\beq\label{cs}
\k_4 = \k_4(\k_0,\Del),
\eeq 
simply because we cannot have a phase transition\index{phase transition} 
unless $N_4 = \infty$.
We put ``critical'' into quotation marks since it only means
that we probe infinite four-volume. No 
continuum limit\index{continuum limit} is 
necessarily associated with a point on this surface. To decide
this issue requires additional investigation. A good analogy is
the Ising model on a finite lattice. To have a genuine phase
transition for the Ising model we have to take the lattice volume
to infinity since there are no genuine 
phase transitions\index{phase transition} for finite
systems. However, just taking the lattice volume to infinity is not
sufficient to ensure critical behaviour of the Ising model. We also 
have to tune the coupling constant to its critical value.
Being on the ``critical'' surface, or rather ``infinite-volume'' surface
\rf{cs}, we can discuss various phases, and these are the ones indicated
in the figure.  
The different phases are separated by phase transitions, which
might be first-order. However, we have not yet 
conducted a systematic investigation of the order of the transitions. 
Looking at Fig.\ \ref{ph-diagram}, we have two lines of 
phase transitions, separating phase A and phase C and separating phase
B and phase C respectively. They meet in the point indicated on the 
figure. It is tempting to speculate that this point might be associated with 
a higher-order transition, as is common for statistical systems in 
such a situation. We will return to this point later.

In the Euclideanized setting the value of the cosmological constant 
determines the spacetime volume $V_4$ since the two appear in the action as
conjugate variables. We therefore have $\la V_4\ra \sim G/\La$ in a continuum 
notation, where 
$G$ is the gravitational coupling constant and $\La$ the cosmological 
constant. In the computer simulations it is more 
convenient to keep the four-volume fixed or partially fixed. We will
implement this by fixing the total number of 
four-simplices of type $N_4^{(4,1)}$ or, equivalently, the total number $N_3$
of tetrahedra making up the spatial $S^3$ triangulations at
times $t_i$, $i= 1,\dots,N$,
\beq\label{2.3}
N_3 =\sum_{i=1}^N N_3(t_i) = \oh N_4^{(4,1)}.
\eeq 
We know from the simulations
that in the phase of interest $\la N_4^{(4,1)} \ra 
\propto \la N_4^{(3,2)} \ra$ as the 
total volume is varied \cite{blp}. This
effectively implies that we only have two bare coupling constants 
$\k_0,\Del$ in \rf{2.1}, while we compensate by hand for the coupling
constant $\k_4$ by studying the partition function $Z(\k_0,\Del;N_4^{(4,1)})$
for various $N_4^{(4,1)}$. 
To keep track of the ratio $\xi(\k_0,\Del)$ between the expectation value 
$\la N_4^{(3,2)} \ra$ and $N_4^{(4,1)}$, which depends 
weakly on the coupling constants, we write (c.f. eq.\ \rf{vol3})
\beq\label{ratio}
 \la N_4 \ra = N_4^{(4,1)} + \la N_4^{(3,2)}\ra = 
N_4^{(4,1)}(1+\xi(\k_0,\Del)).
\eeq
For all practical purposes we can regard $N_4$ in
a Monte Carlo simulation\index{Monte Carlo simulations} as fixed. 
The relation between the partition function 
we use and the partition function with variable four-volume is 
given by the Laplace transformation
\beq\label{2.3x}
Z(\k_0,\k_4,\Del) = \int_0^\infty \d N_4 \; \e^{-\k_4 N_4} \; 
Z(\k_0,N_4,\Del), 
\eeq
where strictly speaking the integration over $N_4$ 
should be replaced by a summation over the discrete values $N_4$ can take.
Returning to Fig.\ \ref{ph-diagram}, keeping $N_4$ fixed rather than
fine-tuning $\k_4$ to the critical value $\k_4^c$ implies that one 
is already on the ``critical'' surface drawn in Fig.\ \ref{ph-diagram},
assuming that $N_4$ is sufficiently large (in principle infinite).
Whether $N_4$ {\it is} sufficiently large to qualify as ``infinite''
can be investigated by performing the computer simulations for 
different $N_4$'s and comparing the results. This is a technique
we will use over and over again in the following.

\section{Numerical results}\label{numerical}

The Monte Carlo simulations\index{Monte Carlo simulations} 
referred to above will generate a sequence of 
spacetime histories\index{spacetime history}. 
An individual spacetime history is not an
observable, in the same way as a path $x(t)$ of a particle in the 
quantum-mechanical path integral is not. However, 
it is perfectly legitimate to talk about the 
{\it expectation value} $\la x(t) \ra$ as 
well as the {\it fluctuations around} $\la x(t) \ra$.
Both of these quantities are in principle calculable in quantum mechanics.
Let us make a slight digression and discuss this in some detail since 
it illustrates well the picture we also hope emerges in a theory 
of quantum gravity. Consider the particle example shown in Fig.\ \ref{fig0}.
We have a particle moving from $x_i$ at $t_i$ to $x_f$ at $t_f$.
In general there will be a classical motion of the particle satisfying
these boundary conditions (we will assume that for simplicity). If $\hbar$
can be considered small compared to the other parameters entering into 
the description of the system, the classical path will be a good 
approximation to $\la x(t) \ra$ according to Ehrenfest's theorem.
In Fig.\ \ref{fig0} the smooth curve represents $\la x(t) \ra$. 
In the path integral we sum over all continuous paths from $(x_i,t_i)$ to
$(x_f,t_f)$ as illustrated in Fig.\ \ref{fig0}. However, when all 
other parameters in the problem are large compared to $\hbar$ we expect 
a ``typical'' path to be close to $\la x(t) \ra$ which also will be 
close to the classical path. Let us make this explicit in the simple
case of the harmonic oscillator. Let $x_{cl}(t)$ denote the solution to
the classical equations of motion such that $x_{cl}(t_i)=x_i$ and 
$x_{cl}(t_f)=x_f$. For the harmonic oscillator the decomposition 
$$
x(t) = x_{cl}(t) + y(t),~~~~ y(t_i)=y(t_f)=0
$$
leads to an exact factorization of the path integral thanks 
to the quadratic nature of the action. The part involving $x_{cl}(t)$ 
gives precisely the classical action and the part involving $y(t)$ 
the contributions from the fluctuations, 
independent of the classical part. Taking 
the classical path to be macroscopic gives a picture of 
a macroscopic path dressed with small 
quantum fluctuations\index{quantum fluctuations}, small
because they are independent of the classical motion. Explicitly
we have for the fluctuations (Euclidean calculation):
$$ 
\bra\int_{t_i}^{t_f} \d t\; y^2(t)\ket = 
\frac{\hbar}{2m\omega^2} \;
\left(\frac{\omega (t_f-t_i)}{\tanh(\omega (t_f-t_i))}-1\right).
$$
Thus the harmonic oscillator is a simple example of what we 
hope for in quantum gravity: 
Let the size of the system be macroscopic, i.e.\ $x_{cl}(t)$ is 
macroscopic (put in by hand): then the quantum fluctuations 
around this path are small and of the order
$$ 
\la |y|\ra \propto \sqrt{\frac{\hbar}{m \omega^2 (t_f-t_i)}}.
$$
We hope this translates into the description of our universe: 
the macroscopic size of the  
universe dictated by the (inverse) cosmological constant in any 
Euclidean description (trivial to 
show in the model by simply differentiating the partition 
function with respect to the cosmological constant 
and in the simulations thus put in by hand) and the small 
quantum fluctuations\index{quantum fluctuations} dictated by the 
other coupling constant, namely, the gravitational coupling constant.  

\subsection{The emergent\index{emergent universe} de 
Sitter background\index{background geometry}\index{de Sitter universe}}

Obviously, there are many more dynamical variables in quantum gravity than 
there are in the particle case. 
We can still imitate the quantum-mechanical situation 
by picking out a particular one, for example, 
the spatial three-volume $V_3(t)$ at proper time $t$. We 
can measure both its expectation value $\la V_3(t)\ra $ 
as well as fluctuations around it. The former 
gives us information about the large-scale  
``shape'' of the universe we have created in the computer. First
we will describe the measurements of $\la V_3(t)\ra $, keeping a more
detailed discussion of the fluctuations to Sec.\ \ref{fluctuations} below.

A ``measurement'' of $V_3(t)$ consists of a table $N_3(i)$, where
$i=1,\ldots,N$ denotes the number of time-slices. Recall from Sec.\ \ref{CDT}
that the sum over slices $\sum_{i=1}^N N_3(i)$ is kept constant. The 
time axis has a total length of $N$ time steps, where $N=80$ in the 
actual simulations, and we have cyclically identified time-slice $N+1$ with
time-slice 1. 

What we observe in the simulations is that for the range of discrete volumes
$N_4$ under study the universe does {\it not} extend
(i.e. has appreciable three-volume) over the entire time axis, but rather is
localized in a region much shorter than 80 time slices. 
Outside this region the spatial extension $N_3(i)$ will be minimal, 
consisting of the minimal number (five) of tetrahedra needed to 
form a three-sphere $S^3$, plus occasionally a few more 
tetrahedra.\footnote{This
kinematic constraint ensures that the triangulation remains a {\it simplicial
manifold} in which, for example, 
two $d$-simplices are not allowed to have more than 
one $(d-1)$-simplex in common.} This thin ``stalk" therefore
carries little four-volume and in a given 
simulation we can for most practical purposes
consider the total four-volume of the remainder, 
the extended universe, as fixed. 

\begin{figure}[t]
\centerline{\scalebox{0.75}{\rotatebox{0}{\includegraphics{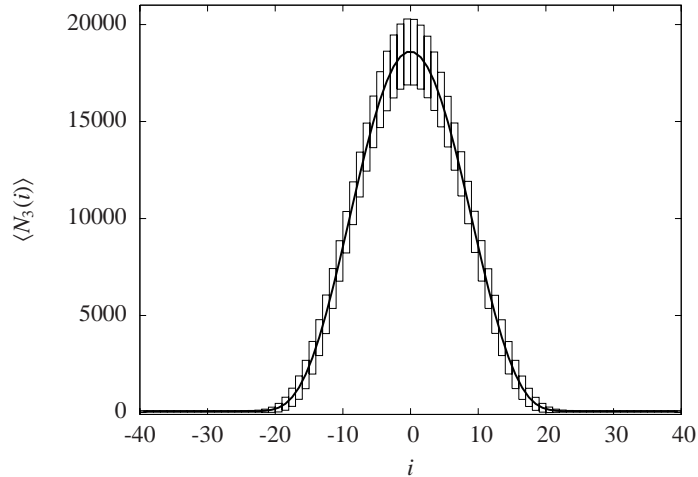}}}}
\caption{\label{fig1} Background geometry $\langle N_3(i)\rangle$: 
MC measurements for fixed $N_4^{(4,1)}= 160.000$ ($N_4=362.000$)
and best fit \rf{n1} yield indistinguishable curves at given plot resolution. 
The bars indicate the average size of 
quantum fluctuations\index{quantum fluctuations}.}
\end{figure}

In order to perform a meaningful average over geometries 
which explicitly refers to the extended part of the universe, 
we have to remove the translational zero mode which is present. 
We refer to \cite{bigs4} for a discussion of the procedure.
Having defined the centre of volume along the 
time-direction of our spacetime configurations we can now 
perform superpositions of configurations and 
define the average $\la N_3(i)\ra $ as a function of the discrete time $i$.
The results of measuring the average discrete 
spatial size of the universe at various discrete times $i$ are illustrated 
in Fig.\ \ref{fig1} and can be succinctly summarized by the formula
\beq\label{n1}
N_3^{cl}(i):= \la N_3(i)\ra  = 
\frac{N_4}{2(1+\xi)}\;\frac{3}{4} \frac{1}{s_0 N_4^{1/4}}  
\cos^3 \left(\frac{i}{s_0 N_4^{1/4}}\right),~~~s_0\approx 0.59,
\eeq
where $N_3(i)$ denotes the number of three-simplices in the spatial slice 
at discretized time $i$ and $N_4$ the 
total number of four-simplices in the entire universe. Since we are 
keeping $N_4^{(4,1)}$ fixed in the simulations and since $\xi$ changes
with the choice of bare coupling constants, it is sometimes 
convenient to rewrite \rf{n1} as
\beq\label{n1x}
N_3^{cl}(i) = \oh N_4^{(4,1)}\;\frac{3}{4} 
\frac{1}{\tilde{s}_0 (N_4^{(4,1)})^{1/4}}  
\cos^3 \left(\frac{i}{\tilde{s}_0 (N_4^{(4,1)})^{1/4}}\right),
\eeq
where $\tilde{s}_0$ is defined by
$\tilde{s}_0 (N_4^{(4,1)})^{1/4} = s_0 N_4^{1/4}$.
Of course, formula \rf{n1} is only valid in the extended part of the universe 
where the spatial three-volumes are larger than the minimal cut-off size.

\begin{figure}[t]
\psfrag{t}{{\bf{\large $\sigma$}}}
\psfrag{v}{{\bf{\large $P(\sigma)$}}}
\centerline{\scalebox{1.1}{\rotatebox{0}{\includegraphics{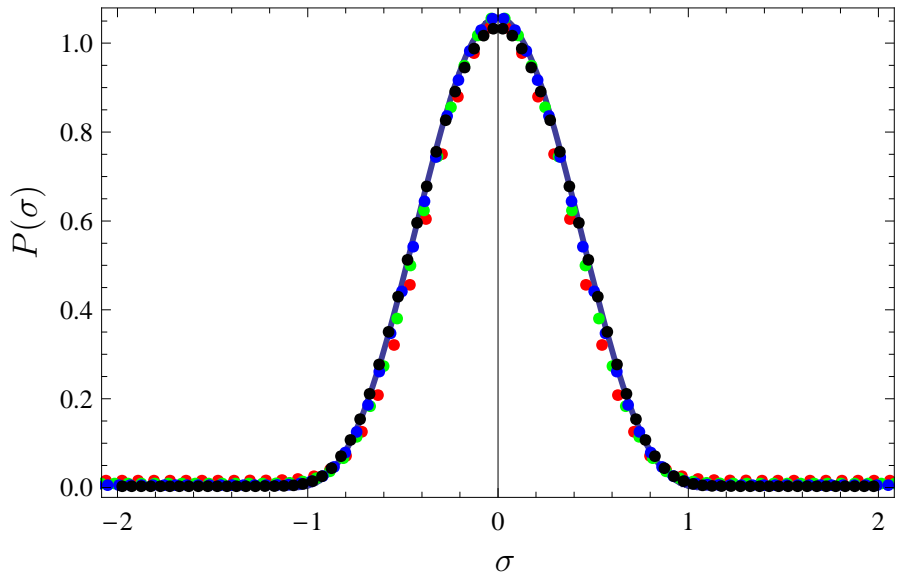}}}}
\caption{Rescaling of time and volume variables according to relation
\rf{n1} for $N_4 =$ 45.500, 91.000, 181.000 and 362.000. The plot 
also include the curve \rf{n1}. More precisely: $\sg \propto 
i/N_4^{1/4}$ and $P(\sg) \propto N_3(i)/N_4^{3/4}$.}
\label{fig1a}
\end{figure}

The data shown in Fig.\ \ref{fig1} have been 
collected at the particular values  
$(\k_0,\Del) = (2.2,0.6)$ of the bare coupling constants and for 
$N_4= 362.000$ (corresponding to $N_4^{(4,1)} = 160.000$). 
For this value of $(\k_0,\Del)$
we have verified relation \rf{n1} for $N_4$ ranging from 45.500 to 362.000
building blocks (45.500, 91.000, 181.000 and 362.000). 
After rescaling the time and volume variables by suitable powers of $N_4$ 
according to relation (\ref{n1}), and plotting them in the same way as 
in Fig.\ \ref{fig1}, one finds almost total
agreement between the curves for different spacetime volumes.
This is illustrated in Fig.\ \ref{fig1a}. Thus we have here 
a beautiful example of finite-size scaling, and at least when
we discuss the average three-volume $V_3(t)$ all our discretized
volumes $N_4$ are large enough that we can treat them as infinite, in the
sense that no further change will occur for larger $N_4$.

By contrast, the quantum fluctuations\index{quantum fluctuations} indicated in 
Fig.\ \ref{fig1} as vertical bars {\it are} 
volume-dependent and will be the larger the
smaller the total four-volume, see Sec.\ \ref{fluctuations} below for details.
eq.\ \rf{n1} shows that
spatial volumes scale according to $N_4^{3/4}$ and time intervals
according to $N_4^{1/4}$, as one would expect for
a genuinely {\it four}-dimensional spacetime and this is exactly 
the scaling we have used in Fig.\ \ref{fig1a}. This strongly suggests
a translation of \rf{n1} to a continuum notation.
The most natural identification is given by  
\beq\label{n2}
\sqrt{g_{tt}}\; V_3^{cl}(t) = V_4 \;
\frac{3}{4 B} \cos^3 \left(\frac{t}{B} \right),
\eeq
where we have made the identifications
\beq\label{n3}
\frac{t_i}{B} = \frac{i}{s_0 N_4^{1/4}}, ~~~~
\Del t_i \sqrt{g_{tt}}\;V_3(t_i) = 2 \tC_4 N_3(i) a^4,
\eeq
such that we have
\beq\label{n3z}
\int dt \sqrt{g_{tt}} \; V_3(t) = V_4.
\eeq
In \rf{n3}, $\sqrt{g_{tt}}$ is the constant proportionality 
factor between the time
$t$ and genuine continuum proper time $\tau$, $\tau=\sqrt{g_{tt}}\; t$.
(The combination $\Del t_i\sqrt{g_{tt}}V_3$ contains $\tC_4$, related to the 
four-volume of a four-simplex rather than the three-volume corresponding
to a tetrahedron, because its time integral must equal $V_4$).
Writing $V_4 =8\pi^2 R^4/{3}$, and $\sqrt{g_{tt}}=R/B$,
eq.\ \rf{n2} is seen to describe 
a Euclidean {\it de Sitter universe}\index{de Sitter universe} 
(a four-sphere, the maximally symmetric space for 
positive cosmological constant)
as our searched-for, dynamically generated 
background geometry!\index{background geometry}
In the parametrization of \rf{n2} 
this is the classical solution to the action
\beq\label{n5}
S= \frac{1}{24\pi G} \int d t \sqrt{g_{tt}}
\left( \frac{ g^{tt}\dot{V_3}^2(t)}{V_3(t)}+k_2 V_3^{1/3}(t)
-\lam V_3(t)\right),
\eeq
where $k_2= 9(2\pi^2)^{2/3}$ and  $\lam$ is a Lagrange multiplier,
fixed by requiring that the total four-volume 
be $V_4$, $\int d t \sqrt{g_{tt}} \;V_3(t) = V_4$. 
Up to an overall sign, this is precisely 
the Einstein-Hilbert action\index{Einstein-Hilbert action} for the scale 
factor $a(t)$ of a homogeneous, isotropic universe
(rewritten in terms of the spatial three-volume $V_3(t) =2\pi^2 a(t)^3$), 
although we of course never put any such
simplifying symmetry assumptions into the CDT model\index{CDT}. 

A discretized, dimensionless version of \rf{n5} is
\beq\label{n7b}
S_{discr} =
k_1 \sum_i \left(\frac{(N_3(i+1)-N_3(i))^2}{N_3(i)}+
\tilde{k}_2 N_3^{1/3}(i)\right),
\eeq
where $\tilde{k}_2\propto k_2$.
This can be seen by applying the scaling \rf{n1}, 
namely, $N_3(i) = N_4^{3/4} n_3(s_i)$ and $s_i = i/N_4^{1/4}$. 
This enables us to finally conclude that the identifications 
\rf{n3} when used in the action \rf{n7b} lead 
na\"ively to the continuum expression \rf{n5} under the identification
\beq\label{n7c}
G = \frac{a^2}{k_1} \frac{\sqrt{\tC_4}\; \ts_0^2}{3\sqrt{6} }.
\eeq

Next, let us comment on the universality of these results. 
First, we have checked that they are not
dependent on the particular definition of time-slicing we have 
been using, in the following sense. By 
construction\index{piecewise linear geometries} of the
piecewise linear CDT-geometries\index{CDT} we have at each integer 
time step $t_i= i\, a_t $
a spatial surface consisting of $N_3(i)$ tetrahedra. Alternatively, one
can choose as reference slices for the measurements of the spatial volume
non-integer values of time, for example, 
all time slices at discrete times $i-1/2$,
$i=1,2,...$ . In this case the ``triangulation" of the 
spatial three-spheres consists of
tetrahedra -- from cutting a (4,1)- or a (1,4)-simplex half-way -- 
and ``boxes", obtained by cutting a (2,3)- or (3,2)-simplex 
(the geometry of this is worked out
in \cite{dl}). We again find a relation
like \rf{n1} if we use the total number of spatial building blocks
in the intermediate slices (tetrahedra+boxes) instead of just the 
tetrahedra. 

Second, we have repeated the measurements for other values of the bare 
coupling constants. As long as we stay in the phase where 
an extended universe is observed, the phase C in Fig.\ \ref{ph-diagram}, 
a relation like \rf{n1} remains valid. In addition, the value of $s_0$, 
defined in eq.\ \rf{n1}, is almost unchanged 
until we get close to the phase transition\index{phase transition}
lines beyond which the extended universe disappears.
Only for the values of $\k_0$ around 3.6 and larger will the
measured $\la N_3(t)\ra $ differ significantly from the value at 2.2.
For values larger than 3.8 (at $\Del = 0.6$), the
universe will disintegrate into a number of small and disconnected components 
distributed randomly along the time axis, 
and one can no longer fit the distribution 
$\la N_3(t)\ra $ to the formula \rf{n1}. 
Later we will show that while $s_0$ is almost unchanged, the constant 
$k_1$ in \rf{n7b}, which governs the 
quantum fluctuations\index{quantum fluctuations} around the 
mean value $\la N_3(t)\ra $, is more sensitive to a change
of the bare coupling constants, in particular, in the case where we
change $\k_0$ (while leaving $\Del$ fixed).

\subsection{Fluctuations around de 
Sitter space\index{de Sitter universe}\label{fluctuations}}

In the following we will test in more detail 
how well the actions \rf{n5} and \rf{n7b} 
describe the computer data. A crucial test is
how well it describes the quantum fluctuations\index{quantum fluctuations} 
around the emergent de Sitter background\index{emergent universe}.

The correlation function\index{correlation function} 
(the covariance matrix $\hC$) is  defined  by
\beq\label{3h.1}
C_{N_4}(i,i') = \la \delta N_3(i) \delta N_3 (i')\ra,
~~~~\del N_3(i) \equiv N_3(i) -\bN_3(i),
\eeq 
where we have included an additional subscript $N_4$ to emphasize that 
$N_4$ is kept constant in a given simulation. 

The first observation extracted from
the Monte Carlo simulations\index{Monte Carlo simulations} 
is that under a change in the four-volume
$C_{N_4}(i,i')$ scales as\footnote{We stress again that 
the form \rf{n7f} is only valid in that part of the universe whose
spatial extension is considerably 
larger than the minimal $S^3$ constructed from 5 tetrahedra. 
(The spatial volume of the stalk typically fluctuates between 
5 and 15 tetrahedra.)} 
\beq
C_{N_4}(i,i') =
N_4 \; F\Big({i}/N^{1/4}_4,{i'}/N_4^{1/4}\Big), \label{n7f}
\eeq
where $F$ is a universal scaling function.
\begin{figure}[t]
\centerline{{\scalebox{1.0}{\rotatebox{0}{\includegraphics{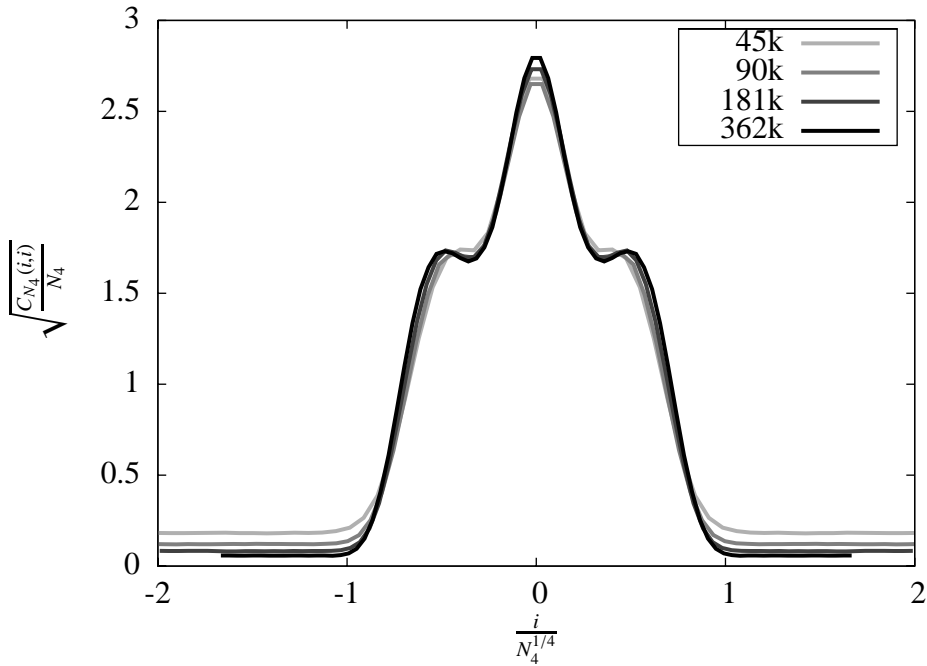}}}}}
\caption{\label{fig7} Analyzing the quantum fluctuations of Fig.\ \ref{fig1}: 
diagonal entries $F(t,t)^{1/2}$ of the universal scaling function 
$F$ from \rf{n7f}, for $N_4^{(4,1)}=$ 20.000,
40.000, 80.000 and 160.000.}
\end{figure} 
This is illustrated by Fig.\ \ref{fig7} for the rescaled version
of the diagonal part $C_{N_4}^{1/2}(i,i)$, 
corresponding precisely to the quantum fluctuations 
$\la (\del N_3(i))^2\ra^{1/2}$ 
of Fig.\ \ref{fig1}. While the height of the curve in Fig.\ \ref{fig1}
will grow as $N_4^{3/4}$, the superimposed fluctuations
will only grow as $N_4^{1/2}$. We conclude that {\it for fixed bare 
coupling constants} the relative fluctuations  
will go to zero in the infinite-volume limit . 

Let us rewrite the minisuperspace action \rf{n5} for a fixed, finite
four-volume $V_4$ in terms of dimensionless variables 
by introducing $s=t/V_4^{1/4}$ and $V_3(t) = V_4^{3/4} v_3(s)$:
\beq\label{n5a}
S =  \frac{1}{24\pi}\; \frac{\sqrt{V_4}}{G} \int d s \sqrt{g_{ss}}
\left( \frac{ g^{ss}\dot{v_3}^2(s)}{v_3(s)}+k_2 v_3^{1/3}(s) \right),
\eeq
now assuming that $\int ds \sqrt{g_{ss}}\; v_3(s) = 1$, 
and with $g_{ss}\equiv g_{tt}$. The same rewriting can be 
done to \rf{n7b} which becomes
\beq\label{n7bb}
 S_{discr} =
k_1 \sqrt{N_4} \sum_i  \Delta s  \left(\frac{1}{n_3(s_i)}     
\left(\frac{n_3(s_{i+1})-n_3(s_i)}{\Delta s}\right)^2+
\tilde{k}_2 n_3^{1/3}(s_i)\right),
\eeq
where $N_3(i) = N_4^{3/4} n_3(s_i)$ and $s_i = i/N_4^{1/4}$.

From the way the factor $\sqrt{N_4}$ appears as an overall scale in
eq.\ \rf{n7bb} it is clear that to the extent 
a quadratic expansion around the effective background 
geometry\index{background geometry} is valid one will have a scaling
\beq\label{sc1}
\la \del N_3(i) \del N_3(i')\ra = 
N_4^{3/2} \la \del n_3(t_i) \del n_3(t_{i'})\ra = 
N_4 F(t_i,t_{i'}),
\eeq
where $t_i = i/N_4^{1/4}$. This implies that \rf{n7f} provides additional 
evidence for the validity of the quadratic approximation 
and the fact that our choice of action 
(\ref{n7b}), with $k_1$ independent of $N_4$ is indeed consistent.

To demonstrate in detail that the full
function $F(t,t')$ and not only its diagonal part is described by the
effective actions \rf{n5}, \rf{n7b}, let us for convenience adopt a 
continuum language and compute its expected behaviour.
Expanding \rf{n5} around the classical solution 
according to $V_3(t) = V_3^{cl}(t) + x(t)$,
the quadratic fluctuations are given by
\bea
\la x(t) x(t')\ra &\! =\! & 
\int \cD x(s)\; x(t)x(t')\; 
e^{ -\frac{1}{2}\int\!\!\int d s d s' x(s) M(s,s') x(s')} \nonumber\\
&\! =\! &  M^{-1}(t,t'),\label{n7a}
\eea
where $\cD x(s)$ is the normalized measure and 
the quadratic form $M(t,t')$ is determined by expanding the 
effective action $S$ to second order in $x(t)$, 
\beq\label{n8}
S(V_3) = S(V_3^{cl}) + \frac{1}{18\pi G}\frac{B}{V_4}  \int \d t \; x(t) \hH 
 x(t).
\eeq
In expression \rf{n8}, $\hH$ denotes the Hermitian operator
\beq\label{n9}
\hH= -\frac{\d }{\d t} \frac{1}{\cos^3 (t/B)}\frac{\d }{\d t} -
\frac{4}{B^2\cos^5(t/B)},
\eeq
which must be diagonalized 
under the constraint that $\int dt \sqrt{g_{tt}}\; x(t) =0$, since
$V_4$ is kept constant. 

Let $e^{(n)}(t)$ be the eigenfunctions of the quadratic form 
given by \rf{n8} with the volume constraint enforced, 
ordered according to increasing eigenvalues $\lam_n$.
As we will discuss shortly, the lowest eigenvalue is $\lam_1 =0$,
associated with translational invariance in time direction, 
and should be left out
when we invert $M(t,t')$, because we precisely fix the centre of volume
when making our measurements. Its dynamics is therefore not accounted for
in the correlator $C(t,t')$\index{correlation function}.

\begin{figure}[t]
\centerline{{\scalebox{0.8}{\rotatebox{0}{\includegraphics{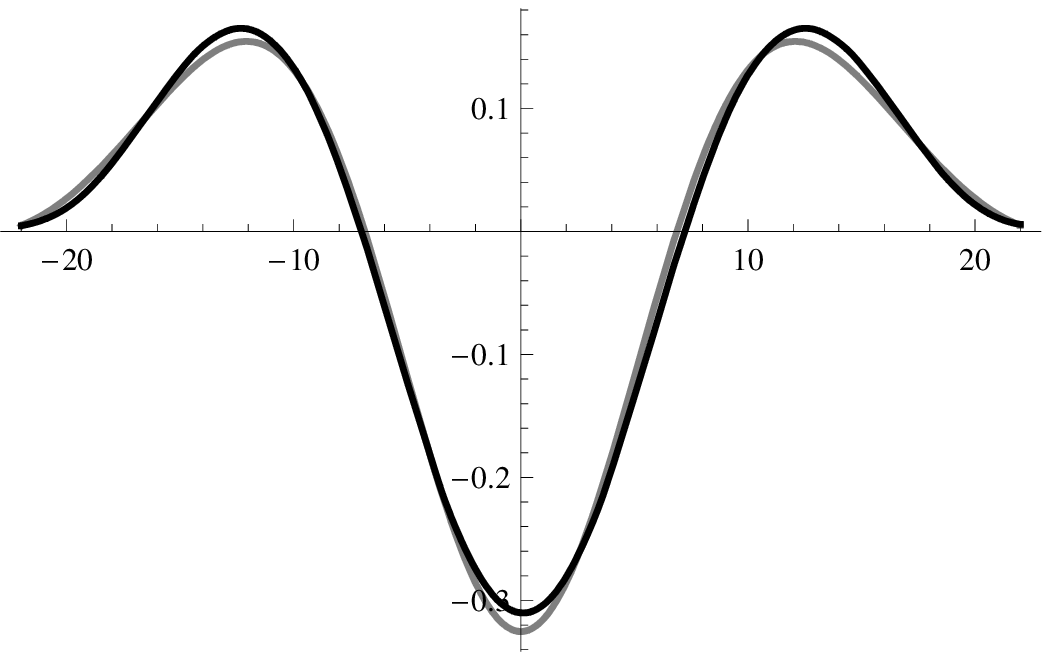}}}}}
\centerline{{\scalebox{0.8}{\rotatebox{0}{\includegraphics{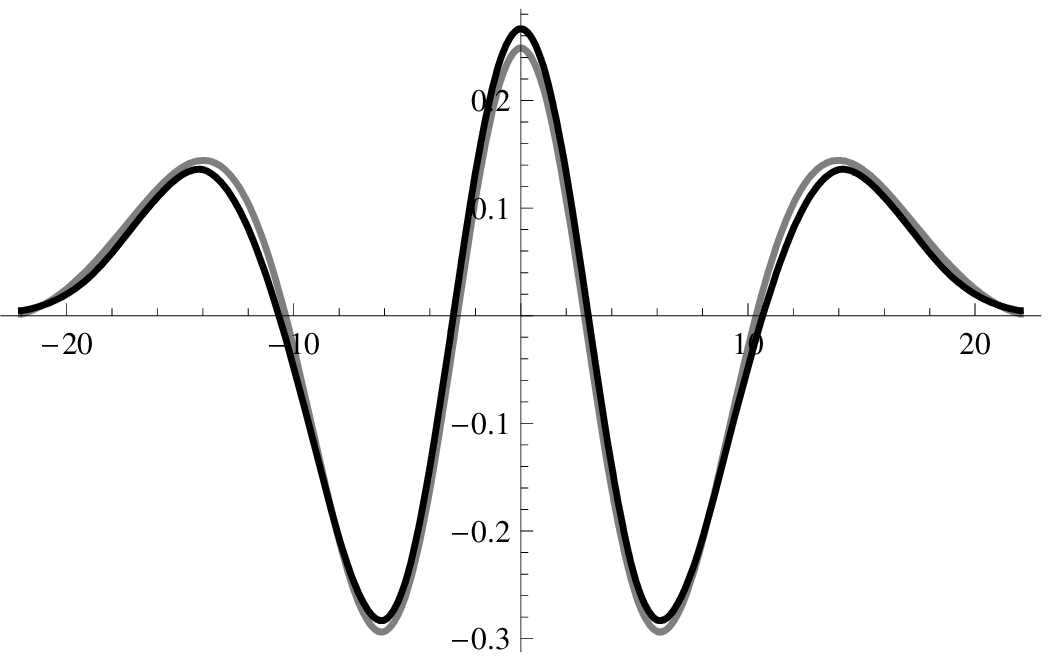}}}}}
\caption{\label{fig8}Comparing the two highest even eigenvector of the 
covariance matrix $C(t,t')$ measured directly
(gray curves) with the two lowest even 
eigenvectors of $M^{-1}(t,t')$, calculated semi-classically (black curves). }
\end{figure}
If this cosmological continuum model were to give the correct 
description of the computer-generated universe, the matrix 
\beq\label{n12}
M^{-1}(t,t') = \sum_{n=2}^\infty \frac{e^{(n)}(t)e^{(n)}(t')}{\lam_n}
\eeq 
should be proportional to the measured correlator $C(t,t')$.
\index{correlation function}
Fig.\ \ref{fig8} shows the eigenfunctions $e^{(2)}(t)$ and $e^{(4)}(t)$
(with two and four zeros respectively),
calculated from $\hH$ with the constraint $\int dt \sqrt{g_{tt}}\; x(t) =0$
imposed. Simultaneously we show the corresponding eigenfunctions  
calculated from the data, i.e.\ from the matrix $C(t,t')$, which
correspond to the (normalizable) eigenfunctions with the 
highest and third-highest eigenvalues. The agreement is very good, 
in particular, when taking into consideration that
no parameter has been adjusted in the action (we simply take 
$B=s_0 N_4^{1/4}\Del t$ in \rf{n2} and \rf{n8}, which gives $B=14.47a_t$
for $N_4 = 362.000$).  

The reader may wonder why the first eigenfunction exhibited
has two zeros. As one would expect, the 
ground state eigenfunction $e^{(0)}(t)$
of the Hamiltonian \rf{n9}, corresponding to the lowest eigenvalue, 
has no zeros, but it does not satisfy the volume
constraint $\int dt \sqrt{g_{tt}}\; x(t) =0$. 
The eigenfunction $e^{(1)}(t)$ of $\hH$ with next-lowest 
eigenvalue has one zero and is given by the simple analytic function
\beq\label{e1}
e^{(1)}(t) = \frac{4}{\sqrt{\pi B}}  \sin \Big(\frac{t}{B}\Big) \, 
\cos^2 \Big(\frac{t}{B}\Big) = c^{-1} \frac{d V_3^{cl} (t)}{dt},
\eeq 
where $c$ is a constant.
One realizes immediately that $e^{(1)}$ is the translational zero mode of the 
classical solution $V_3^{cl}(t)$ ($\propto \cos^3 t/B$). Since the action is 
invariant under time translations, we have
\beq\label{cl1}
S(V_3^{cl}(t+\Del t)) = S(V_3^{cl}(t)),
\eeq
and since $V_3^{cl}(t)$ is a solution to the 
classical equations of motion we find to second order (using the 
definition \rf{e1})
\beq\label{cl2}
S(V_3^{cl}(t+\Del t)) = S(V_3^{cl}(t)) + 
\frac{c^2 (\Del t)^2 }{18\pi G}\frac{B}{V_4} 
\int dt \;e^{(1)}(t) \hH e^{(1)}(t),
\eeq 
consistent with $e^{(1)}(t)$ having eigenvalue zero. 

It is clear from Fig.\ \ref{fig8} that some of the eigenfunctions of $\hH$
(with the volume constraint imposed) agree very well 
with the measured eigenfunctions.
All even eigenfunctions (those symmetric with respect to reflection about the
symmetry axis located at the centre of volume) turn out to agree very well. 
The odd eigenfunctions of $\hH$ agree less well with the 
eigenfunctions calculated from the measured $C(t,t')$.
The reason seems to be that we have not managed to eliminate the motion
of the centre of volume completely from our measurements. 
There is an inherent ambiguity in fixing the centre of volume of 
one lattice spacing, which 
turns out to be sufficient to reintroduce the zero mode in the data.
Suppose we had by mistake misplaced the centre of volume by a small distance 
$\Del t$. This would introduce a modification 
\beq\label{mod}
\Del V_3 = \frac{dV_3^{cl}(t)}{dt} \; \Del t
\eeq
proportional to the zero mode of the potential $V_3^{cl}(t)$.
It follows that the zero mode can re-enter whenever we have an ambiguity 
in the position of the centre of volume. 
In fact, we have found that the first odd eigenfunction
extracted from the data can be perfectly described by 
a linear combination of $e^{(1)}(t)$ and $e^{(3)}(t)$.  
It may be surprising at first that an ambiguity of one lattice
spacing can introduce a significant mixing. However, if we translate 
$\Del V_3$ from eq.\ \rf{mod} to ``discretized'' dimensionless units using 
$V_3(i) \sim N_4^{3/4} \cos (i/N_4^{1/4})$, 
we find that $\Del V_3 \sim \sqrt{N_4}$,
which because of $\la (\del N_3(i))^2\ra \sim N_4$ 
is of the same order of magnitude as the fluctuations 
themselves. In our case, this apparently does affect the odd eigenfunctions. 
\begin{figure}[t]
\centerline{{\scalebox{0.7}{\rotatebox{90}{\includegraphics{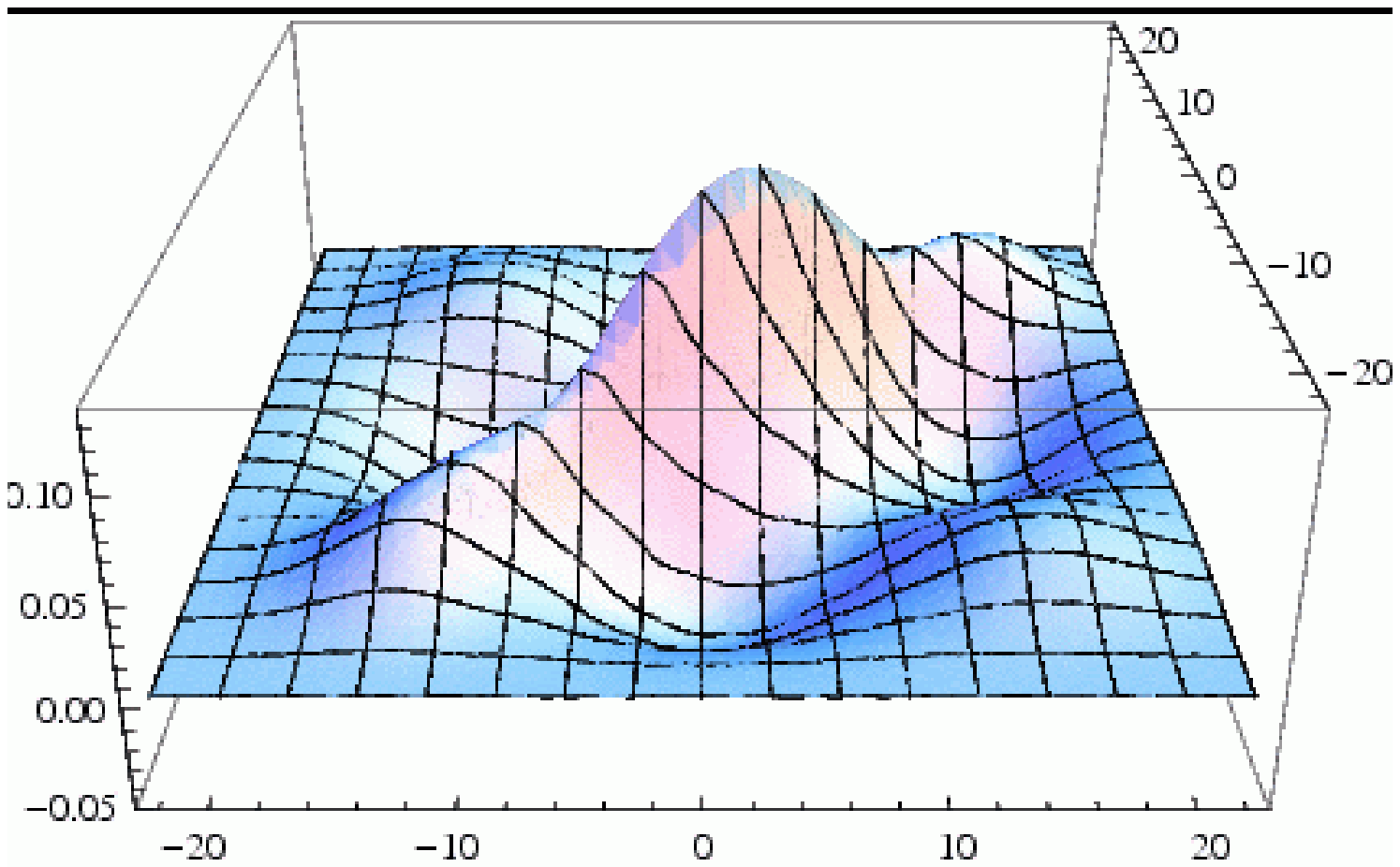}}}}}
\centerline{{\scalebox{0.7}{\rotatebox{90}{\includegraphics{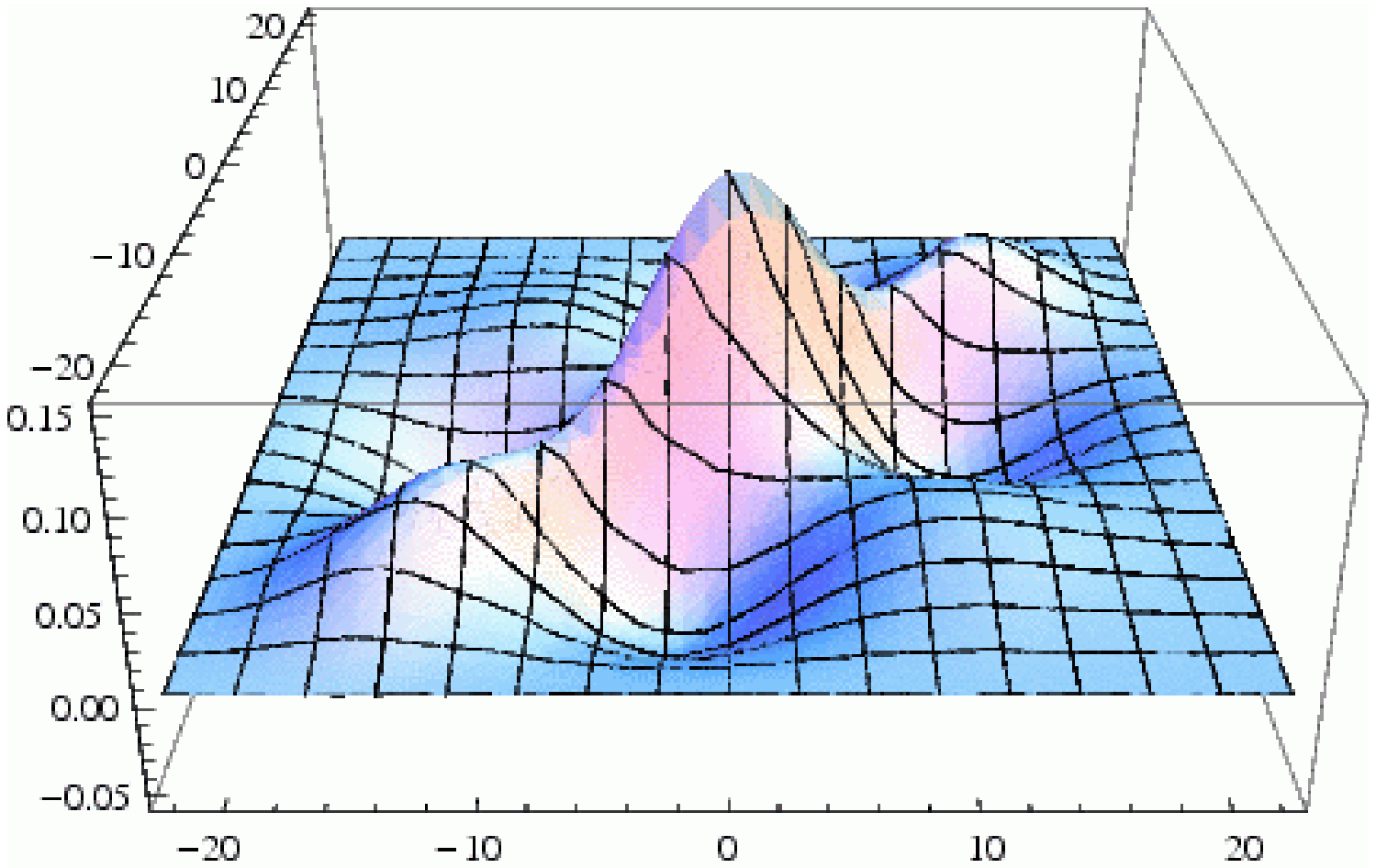}}}}}
\caption{\label{fig9}Comparing data for the extended part of the universe: 
measured $C(t,t')$ (above)
versus $M^{-1}(t,t')$ obtained from analytical calculation (below). 
The agreement is good, and
would have been even better had we included only the even modes.}
\end{figure}

One can also compare the data and 
the matrix $M^{-1}(t,t')$ calculated from \rf{n12} directly. 
This is illustrated in Fig.\ \ref{fig9}, where we have restricted ourselves to
data from inside the extended part of the universe. We imitate the
construction \rf{n12} for $M^{-1}$, using the data to 
calculate the eigenfunctions, rather than $\hH$. One 
could also have used $C(t,t')$ directly, but the use of the 
eigenfunctions makes it somewhat easier to perform the restriction to the bulk.
The agreement is again good (better than 15\% at any point
on the plot), although less spectacular than 
in Fig.\ \ref{fig8} because of the 
contribution of the odd eigenfunctions to the data.

\subsection{The size of the universe and the flow of $G$\label{size}}

It is natural to view the coupling constant $G$ in front of 
the effective action for the scale factor as the gravitational
coupling constant $G$. The effective action which described our 
computer-generated data was given by eq.\ \rf{n5} and its
dimensionless lattice version by \rf{n7b}. The computer data 
allows us to extract $k_1 \propto a^2/G$, $a$ being the 
spatial lattice spacing, the precise constant 
of proportionality being given by eq.\ \rf{n7c}: 
\beq\label{n7cc}
{G} = \frac{a^2}{k_1} \frac{\sqrt{\tC_4}\; \ts_0^2}{3\sqrt{6} }.
\eeq
For the bare coupling constants
$(\k_0,\Del)= (2.2,0.6)$ we have high-statistics measurements
for $N_4$ ranging from 45.500 to 362.000 four-simplices 
(equivalently, $N_4^{(4,1)}$
ranging from 20.000 to 160.000 four-simplices). The choice of 
$\Del$ determines the asymmetry parameter $\a$, and the 
choice of $(\k_0,\Del)$ determines the ratio $\xi$ 
between $N_4^{(3,2)}$ and $N_4^{(4,1)}$. This in turn determines 
the ``effective'' four-volume $\tC_4$ of an average four-simplex, which
also appears in \rf{n7cc}. The number $\ts_0$ in \rf{n7cc} 
is determined directly from the time 
extension $T_{\rm univ}$ of the extended universe according to
\beq\label{width}
T_{\rm univ}=\pi\; \ts_0 \Big(N_4^{(4,1)}\Big)^{1/4}.
\eeq
Finally, from our measurements we have determined $k_1= 0.038$. 
Taking everything together according to \rf{n7cc}, 
we obtain $G\approx 0.23 a^2$, or 
$\ell_{Pl}\approx 0.48 a$, where $\ell_{Pl} = \sqrt{G}$ is the 
Planck length\index{Planck length}.

From the identification of the volume of the four-sphere, 
$V_4 =  8\pi^2 R^4/{3} = \tC_4 N_4^{(4,1)} a^4$,
we obtain that $R=3.1 a$. In other words, 
{\it the linear size $\pi R$ of 
the quantum de Sitter universes\index{de Sitter universe} 
studied here lies in the range of 12-21 Planck lengths for $N_4$ in the 
range mentioned above and for the bare 
coupling constants chosen as $(\k_0,\Del)=(2.2,0.6)$}.

Our dynamically generated universes are therefore not very big, and the 
quantum fluctuations\index{quantum fluctuations} 
around their average shape are large as is apparent from 
Fig.\ \ref{fig1}. It is rather 
surprising that the semi-classical minisuperspace formulation is applicable
for universes of such a small size, a fact that should be 
welcome news to anyone
performing semi-classical calculations to describe the 
behaviour of the early universe.
However, in a certain sense our lattices are still coarse compared
to the Planck scale $\ell_{Pl}$ because the Planck length\index{Planck length} 
is roughly half a lattice
spacing. If we are after a theory of quantum gravity valid on all scales, 
we are in particular interested in uncovering phenomena 
associated with Planck-scale
physics. In order to collect data free from unphysical 
short-distance lattice artifacts at this
scale, we would ideally like to work with a
lattice spacing much smaller than the Planck length,
while still being able to set by hand the physical volume of the  
universe studied on the computer.

The way to achieve this, under the assumption that the coupling 
constant $G$ of formula \rf{n7cc} is indeed a true 
measure of the gravitational coupling constant, is as follows.
We are free to vary the discrete four-volume $N_4$ and the bare coupling 
constants $(\k_0, \Del)$ of the Regge action (see \cite{blp}
for further details on the latter). Assuming for the
moment that the semi-classical minisuperspace action is valid, 
the effective coupling constant $k_1$ in front of it
will be a function of the bare coupling constants $(\k_0, \Del)$,
and can in principle be determined as described above for the case 
$(\k_0, \Del)=(2.2,0.6)$. If we adjusted the bare coupling 
constants such that in the limit as $N_4 \to \infty$ both
\beq\label{n100}
V_4 \sim N_4 a^4~~~{\rm and}~~~G\sim a^2/k_1(\k_0,\Del)
\eeq
remained constant (i.e.\ $k_1(\k_0,\Del) \sim 1/\sqrt{N_4}$), 
we would eventually reach a region where the 
Planck length\index{Planck length} was 
significantly smaller than the lattice spacing $a$, in which event the lattice
could be used to approximate spacetime structures of Planckian size
and we could initiate a genuine study of the sub-Planckian regime.
\begin{figure}[th]
\centerline{{\scalebox{0.62}{\rotatebox{0}{\includegraphics{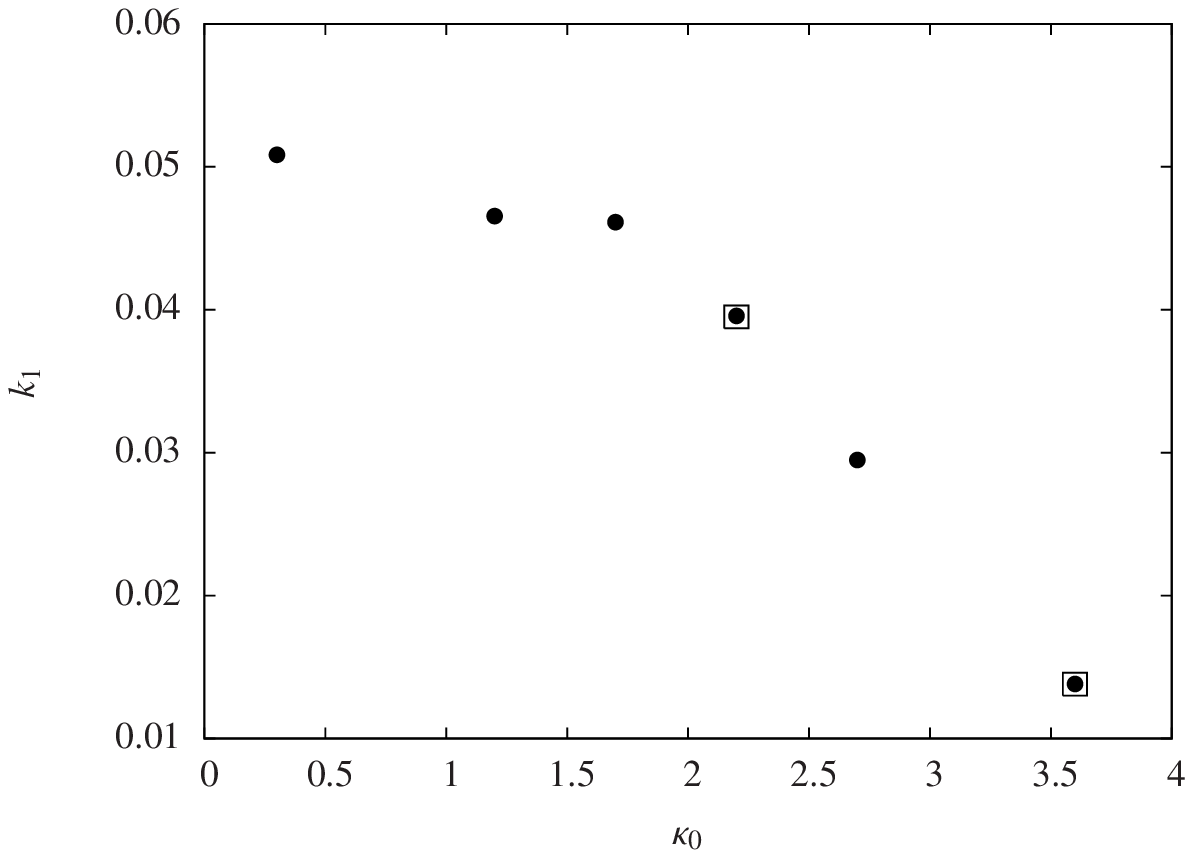}}}}}
\centerline{{\scalebox{0.62}{\rotatebox{0}{\includegraphics{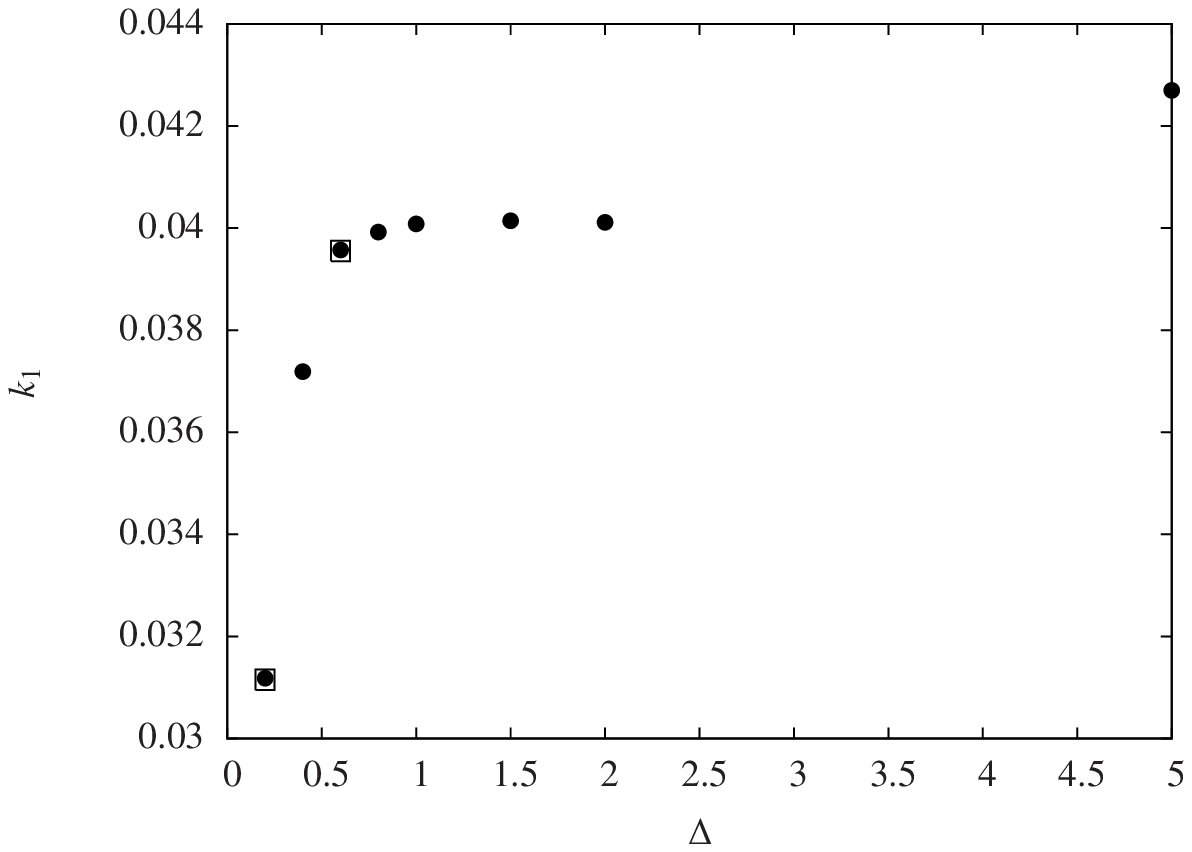}}}}}
\caption{\label{fig12}The measured effective 
coupling constant $k_1$ as function
of the bare $\k_0$ (top, $\Del = 0.6$ fixed) and the asymmetry $\Del$ (bottom, 
$\k_0=2.2$ fixed). The marked point near the middle of the data points sampled
is the point $(\k_0,\Del)=(2.2,0.6)$ where most measurements 
in the remainder of the
paper were taken. The other marked points are those closest to the two phase 
transitions, to the ``branched-polymer phase"\index{branched polymers} (top), 
and the ``crumpled phase"\index{branched polymers} (bottom).}
\end{figure} 
Since we have no control over the effective coupling constant 
$k_1$, the first obvious question which arises is whether we can at all
adjust the bare coupling constants in such a way that at large 
scales we still see a four-dimensional universe, with
$k_1$ going to zero at the same time. 
The answer seems to be in the affirmative, as we will go on to explain.

Fig.\ \ref{fig12} shows the results of extracting 
$k_1$ for a range of bare coupling constants for which we still
observe an extended universe. In the top figure 
$\Del = 0.6$ is kept constant while $\k_0$ is varied. 
For $\k_0$ sufficiently large
we eventually reach a point where a 
phase transition\index{phase transition} takes place (the point
in the square in the bottom right-hand corner is the measurement closest to
the transition we have looked at).    
For even larger values of $\k_0$, beyond this transition, 
the universe disintegrates into a number of small universes, 
in a CDT-analogue\index{CDT} of the 
branched-polymer phase\index{branched polymers} of
Euclidean quantum gravity\index{Euclidean quantum gravity}. 
The plot shows that 
the effective coupling constant $k_1$ becomes smaller and 
possibly goes to zero as the phase transition point is approached, although
our current data do not yet allow us to conclude that $k_1$ 
does indeed vanish at the transition point.

Conversely, the bottom figure of 
Fig.\ \ref{fig12} shows the effect of varying  
$\Del$, while keeping $\k_0=2.2$ fixed.
As $\Del$ is decreased towards 0, we 
eventually hit another phase transition\index{phase transition}, 
separating the physical phase of 
extended universes from the CDT-equivalent of the 
crumpled phase\index{branched polymers} 
of Euclidean quantum gravity, where the 
entire universe will be concentrated within a few time steps,
as already mentioned above. 
(The point closest to the transition where
we have taken measurements is the one in the bottom left-hand corner.)
Also when approaching this phase transition
the effective coupling constant $k_1$ goes to 0, leading to the tentative 
conclusion that $k_1 \to 0$ along the entire phase boundary.

However, to extract the coupling constant $G$ 
from \rf{n7cc} we not only have to take into 
account the change in $k_1$, but also
that in $\ts_0$ (the width of the distribution $N_3(i)$) and 
in the effective four-volume $\tC_4$ as a 
function of the bare coupling constants.
Combining these changes, we arrive at
a slightly different picture. Approaching the boundary where spacetime
collapses in time direction (by lowering $\Del$), the
gravitational coupling constant $G$ {\it decreases},
despite the fact that $1/k_1$ increases. This is a consequence of $\ts_0$ 
decreasing considerably. 
On the other hand, when (by increasing $\k_0$) we approach the 
region where the universe breaks up into several independent components, 
the effective gravitational coupling constant $G$ increases, more or 
less like $1/k_1$, where the behaviour of $k_1$ is shown in 
Fig.\ \ref{fig12} (top). This implies that  
the Planck length\index{Planck length} $\ell_{Pl} = \sqrt{G}$ increases 
from approximately $0.48 a$ to $0.83 a$ when $\k_0$ changes from
2.2 to 3.6. Most likely we can make it even bigger in terms of Planck units
by moving closer to the phase boundary.       

On the basis of these arguments, it seems likely that the non-perturbative
CDT-formulation\index{non-perturbative} of quantum gravity does allow us to 
penetrate into the sub-Planckian regime 
and probe the physics there explicitly. 
Work in this direction is currently ongoing. 
One interesting issue under investigation is whether and to what extent 
the simple minisuperspace description remains valid as we go to shorter scales.
We have already seen deviations from classicality at short scales
when measuring the spectral dimension \cite{spectral,blp}, and one would expect
them to be related to additional terms in the 
effective action \rf{n5} and/or a nontrivial scaling behaviour of $k_1$.
This raises the interesting possibility of being able to test explicitly the
scaling violations of $G$ predicted by renormalization group methods
in the context of asymptotic safety\index{asymptotic safety} \cite{reuteretc}.

\section{Two-dimensional Euclidean 
quantum gravity\index{Euclidean quantum gravity}} \label{2d-euclid}  

The results described above are of course interesting and suggest
that there might exist a field theory of quantum gravity in four 
dimensions (three space and one time dimension). However, the
results are based on numerical simulations. As already mentioned
it is of great conceptional interest that we have a toy 
model, two-dimensional quantum gravity, where both the lattice theory 
and the continuum quantum gravity theory can be solved analytically
and agree.  Of course we can still be in the situation that there 
exists no description of quantum gravity as a field theory in four 
dimensions (although we have presented some evidence in favour of 
such a scenario above), but we can then not blame 
the underlying formalism for being inadequate.

\subsection{Continuum formulation}

Let $M^h$ denote a closed, compact, connected and
orientable surface of genus $h$ and Euler
characteristic $\chi(h) = 2-2h$. The partition function of two-dimensional
Euclidean quantum gravity\index{Euclidean quantum gravity} 
is formally given by
\beq\label{4.1a}
Z(\La,G)= \sum_{h=0}^\infty\int
\cD [g] \, e^{-S(g;\La,G)},
\eeq
where $\La$ denotes the
cosmological constant, $G$ is the gravitational
coupling constant and $S$ is the continuum
Einstein--Hilbert action\index{Einstein-Hilbert action} defined by
\beq\label{4.2}
S(g;\La,G) = \La\int_{M^h} d^2 \xi \sqrt{g} -
\frac{1}{2\pi G} \int_{M^h} d^2\xi \sqrt{g}\, R.
\eeq
In eq.\ \rf{4.1a}, we take the sum to include all possible topologies of
two-dimensional manifolds (i.e.\ over all genera $h$),
and in eq.\ \rf{4.2} $R$ denotes the scalar curvature
of the metric $g$ on the manifold $M^h$.
The functional integration is over all
{\it diffeomorphism equivalence classes} $[g]$ of metrics on $M^h$.

In two dimensions the curvature part of the 
Einstein--Hilbert action\index{Einstein-Hilbert action} 
is a topological invariant according to the
Gauss--Bonnet theorem, which allows us to write
\beq\label{4.1y}
Z(\La,G) = \sum_{h=0}^\infty e^{\chi(h)/G} Z_h(\La),
\eeq
where
\beq\label{4.1z}
Z_h(\La) = \int\cD [g]\, e^{-S(g;\La)},
\eeq
and
\beq\label{4.1w}
S(g;\La) = \La\, V_g,
\eeq
where $V_g =  \int d^2 \xi \,\sqrt{g}$ is the volume of the
universe for a given diffeomorphism class of metrics. In the remainder of this
section we will, for simplicity, restrict our attention to manifolds
homeomorphic to $S^2$ or $S^2$ with a fixed number of holes unless explicitly stated
otherwise. In this case we disregard the topological
term in the action since it is
a constant. The sphere $S^2$ with $b$ boundary components
will be denoted $S^2_b$ and
we denote the partition function for the sphere,
$Z_0(\La)$ in eq.\ \rf{4.1z}, by $Z (\La)$.

In the presence of a boundary it is natural to add to the
action a boundary term
\beq\label{4.z30}
S(g;\La,Z_1,...,Z_b) = \La V_g +\! \sum_{i=1}^b Z_i L_{i,g},
\eeq
where $L_{i,g}$ denotes the length of the $i$th boundary component with
respect to the metric $g$. We refer to the $Z_i$'s as the
cosmological constants of the boundary components.
The partition function is in this case given by
\beq\label{4.z31}
W(\La;Z_1,...,Z_b) = \int \cD[g]\, e^{-S(g;\La,Z_1,\ldots ,Z_b)}.
\eeq
Since the lengths of the boundary components are invariant under
diffeomorphisms, it makes sense to fix them
to values $L_1,\ldots ,L_b$ and define the {\em
Hartle--Hawking wave functionals}\index{Hartle-Hawking wavefunction} by
\beq\label{4.z33}
W(\La;L_1,...,L_b) = \int \cD [g] \, e^{-S(g;\La)}
\,\prod_{i=1}^b \del(L_i - L_{i,g}),
\eeq
where $S(g;\La)$ is given by eq.\ \rf{4.1w}.
Since eq.\ \rf{4.z31} is the Laplace
transform of eq.\ \rf{4.z33}, i.e.
\beq\label{4.z32}
W(\La;Z_1,...,Z_b) = \int_0^\infty \prod_{i=1}^b dL_i \,e^{-Z_i L_i}\,
W(\La;L_1,...,L_b),
\eeq
we denote them by the same symbol. We distinguish between the two by
the names of the arguments.

\subsection{The lattice regularization\index{lattice regularization}}

At the outset we restrict the
topology of surfaces to be that of $S^2$ with a fixed number of holes.
We view abstract triangulations of $S^2_b$ as defining a grid in
the space of diffeomorphism equivalence classes of metrics on $S^2_b$.
Each triangle is a ``building block'' with side lengths $a$. 
This $a$ will be a UV cut-off which we will relate to the bare 
coupling constants on the lattice. However, presently it is convenient 
to view $a$ as being 1 (length unit).

Let $T$ denote a triangulation of $S^2_b$.
The regularized theory of gravity will be
defined by replacing the action $S_g(\La,Z_1,\ldots,Z_b)$ 
in eq.\ \rf{4.z30} by
\beq\label{5x.8}
S_T(\m,\lam_1,\ldots,\lam_b) =\m N_t + \sum_{i=1}^b  \lam_i l_i,
\eeq
where $N_t$ denotes the number of triangles in $T$ and $l_i$ is the number of
links in the $i$th boundary component.
The parameter $\m$ is the bare cosmological constant and the $\lam_i$'s
are the bare cosmological constants of the boundary components.
The integration over diffeomorphism equivalence classes of metrics in eq.\ \ \rf{4.z31}
becomes a summation over non-isomorphic triangulations. We
define the loop functions\index{loop correlator} 
(discretized versions of $W(\La,Z_1,\ldots,Z_b)$)
by summing over all triangulations of $S^2_b$:
\beq\label{5x.9}
w(\m ,\lam_1,\ldots,\lam_b) =   \sum_{l_1,\ldots ,l_b}\,
\sum_{T \in \cT (l_1,\ldots ,l_b)} \,
e^{-S_T(\m,\lam_1,\ldots,\lam_b)}.
\eeq
Analogously, we define the partition function
for closed surfaces by
\beq\label{5x.9zz}
Z(\m) =   \sum_{T \in \cT} \frac{1}{C_T}\,
e^{-S_T(\m)},
\eeq
where $C_T$ is the symmetry factor of $T$ and $S_T(\m)=\m N_t$.
Since we consider surfaces of
a fixed topology we have left out the curvature term in the action.
It will be introduced later, when the restriction on topology is lifted.

Next we write down the regularized version of the
Hartle--Hawking wave functionals\index{Hartle-Hawking wave function}
$W(\La,L_1,\ldots,L_b)$:
\beq\label{5x.10}
w(\m,l_1,\ldots,l_b)  = \sum_{T \in \cT (l_1,\ldots,l_b)} e^{-S_T(\m)}
\eeq
with an abuse of notation similar to the one in the previous section.
This can also be written in the form
\beq\label{5x.11b}
w(\m,l_1,\ldots,l_b) = \sum_k e^{-\m k} w_{k,l_1,\ldots,l_b}\; ,
\eeq
where we have introduced the notation
$$
w_{k,l_1,\ldots,l_b}
$$
for the number of triangulations in $\cT (l_1,\ldots ,l_b)$ with $k$
triangles.

The discretized analogues of the Laplace transformations which
relate $W(\La,Z_1,\ldots ,Z_b)$ and $W(\La,L_1,\ldots ,L_b)$ are
\beq\label{5x.11bb}
w(\m,\lam_1,\ldots,\lam_b) = \sum_{l_1,\ldots ,l_b}
e^{-\sum _i\lam_i l_i} w(\m,l_1,\ldots,l_b)
\eeq
and eq.\ \rf{5x.11b}.
Similarly, we have for the partition functions
\bea
Z(\m) &=&  \sum_k e^{-\m k} Z (k),\label{5x.11} \\
Z(k) &=& \sum_{T \in \cT, N_t=k} \frac{1}{C_T}. \label{5x.new11}
\eea
It follows from the definitions \rf{5x.11b}--\rf{5x.11bb}
that $w(\m,\lam_1,\ldots,\lam_b)$ is the generating function
for the numbers $w_{k,l_1,\ldots,l_b}$, the arguments of the
generating function being $e^{-\m}$ and $e^{-\lam_i}$.
{\it In this way the evaluation of the 
loop functions\index{loop correlator} of two-dimensional
quantum gravity is reduced to the purely combinatorial problem of
finding the number of non-isomorphic triangulations of
$S^2$ or $S^2_b$ with a given number of triangles and boundary components of
given lengths.}

We use the notation
\beq\label{5x.gen}
w(g,z_1,\ldots,z_b) = \sum_{k,l_1,\ldots ,l_b}
w_{k,l_1,\ldots,l_b} \,g^k z_1^{-l_1-1} \cdots z_b^{-l_b-1}
\eeq
for the generating function with an extra factor $z_1^{-1}\ldots z_b^{-1}$,
i.e.\ we make the identifications
\beq\label{5x.gen1}
g = e^{-\m},~~~~z_i = e^{\lam_i}.
\eeq
The reason for this particular choice of variables in the generating
function is motivated by its analytic structure, which will be revealed
below.

In the following we consider a particular class of triangulations
\begin{figure}[t]
\centerline{\hbox{\psfig{figure=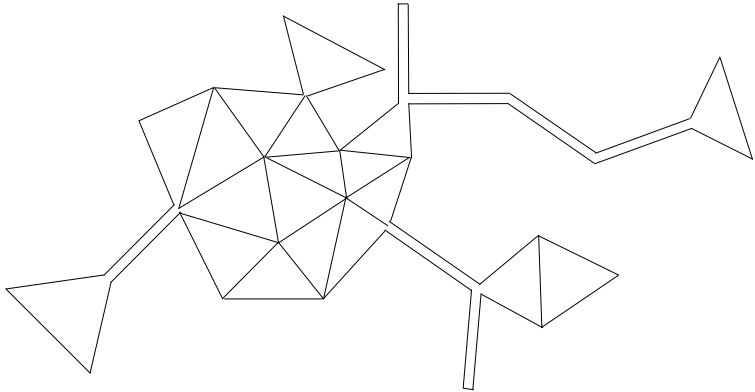,width=10cm}}}
\caption[fig4.3]{A typical unrestricted ``triangulation''.}
\label{fig4.3}
\end{figure}
which includes degenerate boundaries. It may be defined as the class of
complexes homeomorphic to the sphere with a number of holes that one obtains
by successively gluing together a collection of triangles and a collection of
double-links which we consider as
(infinitesimally narrow) strips, where links, as
well as triangles, can be glued onto the boundary of a
complex both at vertices
and along links. Gluing a double link along a link makes no change in the
complex. An example of such a complex
is shown in Fig.\ \ref{fig4.3}. The reason we use this class of 
triangulations is that they match the ``triangulations'' we obtain
from the so-called matrix models\index{matrix model} 
to be considered below. We call
this class of complexes ``unrestricted triangulations''.   

One could have chosen a more regular class of triangulations,
corresponding more closely to our intuitive
notion of a surface. However the degenerate structures present in the
unrestricted triangulations appear on a slightly larger
scale in the regular triangulations in the form of narrow strips
consisting of triangles. Since we want to take the lattice
side $a$ of a triangle to zero in the continuum limit\index{continuum limit},
there should be no difference in that limit between various classes
of triangulations, unless more severe constraints are 
introduced. We say that ``the continuum limit is universal". 
But at some point the constraint can be so strong that the 
continuum limit is changed. We will meet precisely such a 
change below, leading from (Euclidean) dynamical triangulations (DT) to 
causal dynamical triangulations (CDT)\index{CDT}.

Let $w(g,z)$ denote the generating function
for the (unrestricted) triangulations with one boundary component. 
Then we have
\beq\label{5x.11c}
w(g,z) = \sum_{k=0}^\infty \sum_{l=0}^\infty w_{k,l}\, {g^k}\,z^{-(l+1)} \equiv
\sum_{l=0}^\infty \frac{w_l(g)}{z^{l+1}}
\eeq
We have included the triangulation consisting of one point. It
gives rise to the term $1/z$ and we have $w_0(g) =1$. The function
$w_1(g)$ starts with the term $g$,
which corresponds to an unrestricted triangulation
with a boundary consisting
of one (closed) link with one vertex and containing one triangle.
The coefficients $w_{k,1}$ in the expansion
$w_{1,1}g+w_{3,1} g^3+\cdots\;$ of $w_1(g)$ are the numbers of unrestricted
triangulations with a boundary consisting of one link.

\begin{figure}
\centerline{\scalebox{0.7}{\rotatebox{0}{\includegraphics{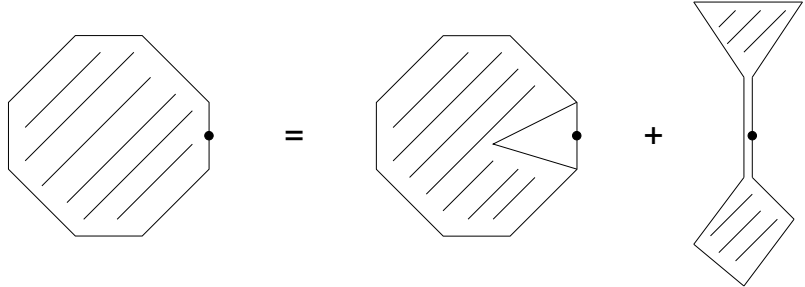}}}}
\caption{{\small 
Graphical representation of relation \rf{5x.11e}: The boundary contains 
one marked link which is part of a triangle or a double link. 
Associated to each
triangle is a weight $g$, and to each double link a weight 1. 
}}
\label{fig4.4}
\end{figure}

The coefficients of $w(g,z)$ fulfill a recursion
relation which has the simple graphical
representation shown in Fig.\ \ref{fig4.4}.
The diagrams indicate two operations that one can perform on a marked
link on the boundary to produce a triangulation which
has either fewer triangles or fewer boundary links.
The first term on the right-hand side of 
Fig.\ \ref{fig4.4} corresponds to the
removal of a triangle. The second term corresponds to the
removal of a double-link.
Note that removing a triangle creates a new double-link
if the triangle has two boundary links. In addition, note that we
count triangulations {\it with one marked link on each boundary
component} and adopt the notation introduced above for the corresponding
quantities.

The equation associated with the diagrams is
\beq\label{5x.11e}
\left[w(g,z)\right]_{k,l} = \left[g z w(g,z)\right]_{k,l} +
\left[\frac{1}{z} w^2(g,z)\right]_{k,l}.
\eeq
The subscripts $k,l$ indicate the coefficient of $g^k/z^{l+1}$.
Let us explain the equation in some detail.
The factor $g z$ in eq.\ \rf{5x.11e} is present since the triangulation
corresponding to the first term
on the right-hand side of Fig.\ \ref{fig4.4} has one
triangle less and one boundary link
more than the triangulation on the left-hand side. The function $w^2(g,z)$
in the last term in eq.\  \rf{5x.11e} arises from the two blobs
connected by the double-link in Fig.\ \ref{fig4.4} and the $1/z$ in front of
$w^2(g,z)$ is inserted to make up for the decrease by two in the length of the
boundary when removing the double link.

As the reader may have discovered, eqs.\ \rf{5x.11e} 
is not correct for the smallest values
of $l$. Consider Fig.\ \ref{fig4.4}. The first term on the
left-hand side of eq.\ \rf{5x.11e} (a single vertex) 
has no representation on the diagram. In order for eq.\ \rf{5x.11e}
to be valid for $k=l=0$ we have to add the
term $1/z$ on the right-hand side of eq.\ \rf{5x.11e}. 
Furthermore, it is clear from Fig.\ \ref{fig4.4} that the first term 
on the right-hand side has at least
two boundary links. Consequently, the term $gz w(g,z)$ on the
right-hand side of eq.\ \rf{5x.11e} should be replaced by
$gz (w(g,z)-1/z-w_1(g)/z^2)$
such that all terms corresponding to triangulations with
boundaries of length 0 and 1 are subtracted.
It follows that the correct equation is
\beq\label{5x.11g}
(z-gz^2)w(g,z)-1+g(w_1(g) +z)=w^2(g,z).
\eeq

We will refer to eq.\ \rf{5x.11g} as the 
{\it loop equation}\index{loop equation}. It is
a second-order equation in $w(g,z)$.
As will be clear in the following this algebraic
feature allows us to extract asymptotic formulas for the number of
triangulations with $k$ triangles in the limit
$k \to \infty$. 

\subsection{Counting graphs}\label{counting}

\begin{figure}
\centerline{\hbox{\psfig{figure=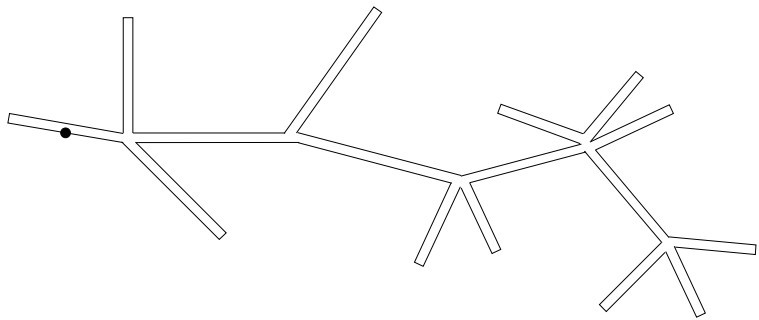,width=10cm}}}
\caption[fig4.5a]{Rooted branched polymers created by gluing of a boundary
with one marked link.}
\label{fig4.5a}
\end{figure}

Let us begin by solving  eq.\ \rf{5x.11g} in the
limit $g =0$. In this
case there are no internal triangles and the triangulations are in 
one-to-one correspondence with rooted 
branched polymers\index{branched polymers} \footnote{One might think
that such polymers are not relevant at all for studying real surfaces
made of triangles, not to mention higher piecewise linear manifolds,
but in fact the branched polymer structure is quite generic. 
Surfaces or higher dimensional manifolds can ``pinch'', 
such that two parts of the 
triangulation are only connected by a minimal ``neck''. If this 
happens in many places one can effectively obtain a branched polymer
structure even for higher dimensional piecewise linear manifolds.
Such minimal necks have been used to measure critical exponents 
of various ensembles of piecewise linear manifolds \cite{jain} and
in four-dimensional Euclidean quantum gravity one has indeed, as mentioned
above, observed a phase where the four-dimensional piecewise linear
manifolds degenerate to branched polymers \cite{aj}. The same is the 
case for bosonic strings with central charge $c > 1$ \cite{ad}. }.
The double-links correspond to the links of the branched polymers and the
root is the marked link, see Fig.\ \ref{fig4.5a}.
If $g=0$ then eq.\ \,\rf{5x.11g} reads
\beq\label{5x.11x}
w^2(z)-zw(z) +1 = 0.
\eeq
The above equation has two solutions. The one that corresponds to the
counting problem has a Taylor expansion in $z^{-1}$  whose first
term is $z^{-1}$ (recall that $w_{0,0} =1$). This solution is given by
\beq\label{5x.11xa}
 w(z) = \oh\left(z - \sqrt{z^2-4}\right).
\eeq
Expanding in powers of $1/z$ yields
\beq\label{5x.11y}
w(z) = \sum_{l=0}^\infty \frac{w_{2l}}{z^{2l+1}},
\eeq
where
\beq\label{5x.11y2}
w_{2l} = \frac{ (2l) !}{(l+1)!\, l !} = \frac{1}{\sqrt{\pi}}\,
l^{-3/2}\, 4^l\, (1+O(1/l)),
\eeq
and $w_{2l}$ is the number of rooted polymers with $l$ links.
Note that $w_{2l}$ are the Catalan numbers, 
known from many combinatorial problems.

The generating function $w(z)$ is analytic in the complex plane $C$
with a cut on the real axis along the interval $[-2,2]$.
The endpoints of the
cut determine the radius of convergence of $w(z)$ as a function of $1/z$
or, equivalently, the exponential growth of $w_{2l}$.

We can solve the second-order equation \rf{5x.11g} and obtain
\beq\label{loopx1}
w(g,z) = \oh\left( V'(z)- \sqrt{(V'(z))^2 - 4Q(z)} \right) ,
\eeq
where, anticipating generalizations, we have introduced the notation
\beq\label{loopx2}
V'(z) = z-gz^2,~~~~Q(z) = 1-gw_1(g)-gz.
\eeq
The sign of the square root is determined as in eq.\ \rf{5x.11xa}
by the requirement that $w(g,z) = 1/z+ O(1/z^2)$ for large
$z$ (since $w_{0,0} =1$).
If $g=0$ then $V'(z)^2-4Q(z)=z^2-4$. For $g > 0$, on the other hand,
$V'(z)^2-4Q(z)$  is a fourth-order polynomial of the form
\bea
\lefteqn{V'(z)^2-4Q(z)=\{z-(2+2g)+O(g^2)\}}\nonumber\\
&&~~~\times\{z+(2-2g)+O(g^2)\}\{gz-(1-2g^2)+O(g^3)\}^2\label{E3}
\eea
in a neighbourhood of $g=0$ since
the analytic structure of $w(g,z)$ as a function of $z$ cannot
change discontinuously at $g=0$. We can therefore write
\beq\label{pot1}
V'(z)^2-4Q(z)=(z-c_+(g))(z-c_-(g))(c_2(g)-gz)^2,
\eeq
and, by eq.\ \rf{loopx1},
\beq\label{loopx3}
w(g,z) = \oh\left(z-gz^2+(gz-{c}_2)\sqrt{(z-c_+)(z-c_-)} \right),
\eeq
where $c_-,c_+$ and ${c}_2$ are functions of $g$, analytic
in a neighbourhood of $g=0$.
We label the roots so that $c_-\leq c_+$.
{\it The numbers $c_-$, $c_+$ and $c_2$ are uniquely determined by the requirement that
$w(g,z) = 1/z+O(1/z^2)$, again originating from $w_{0,0}=1$.}
This requirement gives three equations for the
coefficients of $z,z^0,z^{-1}$.

\begin{figure}[t]
\centerline{\scalebox{0.6}{\rotatebox{0}{\includegraphics{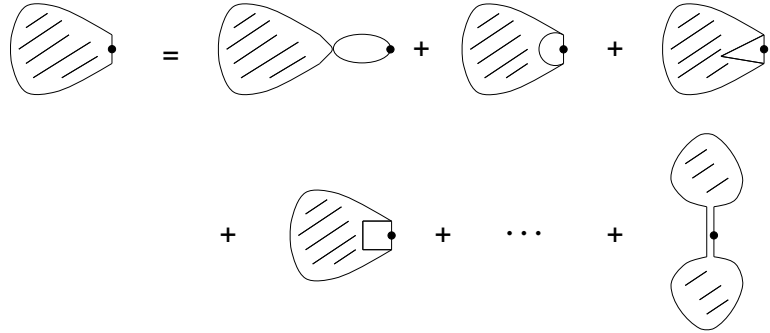}}}}
\caption{{\small 
Graphical representation of relation \rf{*4.44x}: the marked link
of the boundary belongs either to an $i$-gon (associated weight $g\, t_i$) 
or a double link (associated weight 1).
It is also a graphical representation of eq.\ \rf{3.18a}
if instead of weight 1 we associate a weight $g_s$ to the marked double link.
}}
\label{fig4.6}
\end{figure}

We can generalize the above counting problem to planar complexes
made up of polygons with an arbitrary number $j\leq n$ of sides,
including ``one-sided'' and ``two-sided'' polygons.
If we attribute a weight $gt_j$ to each
$j$-sided polygon and a weight $z$ to each boundary link,
and adopt the notation
\beq\label{loopx26}
V'(z)=  z -g(t_1 +t_2z+t_3 z^2+\cdots + t_n z^{n-1}),
\eeq 
\beq\label{*4.43b}
Q(z)= 1-g\sum_{j=2}^n t_j
\sum_{l=0}^{j-2} z^l w_{j-2-l}(g),~~~~w_0(g)=1,
\eeq
the analogue of eq.\ \rf{5x.11g} is
\beq\label{*4.44a}
w(g,z)^2 = V'(z)w(g,z) -Q(z),
\eeq
or
\bea
w(g,z) &=& g\left(t_1\frac{1}{z}+t_2+ t_3 z 
+ \cdots + t_nz^{n-2}\right)w(g,z)
\nonumber\\
&& +\frac{1}{z}Q(z) + \frac{1}{z}w^2(g,z). \label{*4.44x}
\eea
The graphical representation of \rf{*4.44x} is shown in Fig.\ \ref{fig4.6}.
The subtraction of the polynomial $Q(z)$ in eq.\ \rf{*4.44a} reflects
the fact that the term with a $j$-sided polygon
in Fig.\ \ref{fig4.6} must have
a boundary of length at least $j-1$ for $j >1$.
The constant term $1$ in $Q$ corresponds to
the complex consisting of a single vertex.

The solution can be written
\beq\label{loopx11}
w(g,z) =\oh \left(V'(z) - M(z) \sqrt{(z-c_+(g))(z-c_-(g)} \right),
\eeq
where $M(z)$ is a polynomial of a degree which is one less than that
of $V'(z)$. Again, the polynomial $M$ is uniquely determined by the
requirement that $w(g,z)$ falls off at infinity as before, i.e.\
$w(g,z)=1/z+O(1/z^2)$, and the additional requirement that $w(z)$ 
has a single cut. It is sometimes convenient to write \rf{loopx11} as
\beq\label{loopx12}
w(g,z) = \oh \left(V'(z) - \sum_{k=1}^{n-1}
M_k(g) (z-c_+)^{k-1} \sqrt{(z-c_+(g))(z-c_-(g)} \right).
\eeq
One can show the following: for $t_1,\ldots,t_{n-1} \geq 0$ and 
$t_n > 0$ one has 
\beq\label{crit1}
M_k(g) < 0,~~~~~k < n,
\eeq
while 
\beq\label{crit2}
M_1(g) > 0
\eeq
in a neighbourhood of $g=0$. When we increase $g$ we first
reach a point $g_c$, where \rf{crit1} is still satisfied but
\beq\label{crit3}
M_1(g_c) = 0.
\eeq
The coupling constant point $g_c$ is thus the point
where the analytical structure of $w(g,z)$ changes from
being identical to that of the branched 
polymer\index{branched polymers}, i.e.\
it behaves like $(z-c_+)^{1/2}(z-c_-)^{1/2}$, to  
$(z-c_+)^{3/2}(z-c_-)^{1/2}$. The function $w(g,z)$ is 
an analytic function around the point $g=0$ ($w(g=0,z)$ is 
the branched-polymer partition function discussed above). The radius 
of convergence is precisely $g_c$. If we return to the expansion 
in \rf{5x.11e} each term $w_l(g)$ has this radius of convergence.
It is the generating function for triangulations with one 
boundary consisting of $l$ links. The singularity of $w_l(g)$
for $g \to g_c$ determines the asymptotic behaviour of the
number of such triangulations for a large number of triangles, i.e.\
the leading behaviour of the numbers $w_{k,l}$ for large $k$
(see \rf{5lo.30} and \rf{5y.7} below).

Let us introduce $g_j= gt_j$ as new variables. We have 
\beq\label{loopx100}
w(g_i,z)
= \sum_{l,k_1,\ldots,k_n} w_{\{k_j\},l} \,z^{-(l+1)} \prod_{j=1}^{n}
g_{j}^{k_j},
\eeq
where $w_{\{k_j\},l}$ is the number of planar graphs with
$k_j$ $j$-sided polygons, $j=1,\ldots,n$, and a boundary of length $l$.

From $w(g_i ,z)$ we can derive the generating function
for planar graphs with two boundary components by applying the
{\it loop insertion operator}\index{loop insertion operator}
\beq\label{loopx101}
\frac{d}{dV(z)} = \sum_{j=1}^{\infty} \frac{j}{z^{j+1}} \, \frac{d}{d g_j}.
\eeq
One should think of this operator as acting on formal power series
in an arbitrary number of variables $g_j$.
The action of $d/d V(z_2)$ on $w(g_i,z_1)$ has in each term of the
power series the effect of
reducing the power $k_j$ of a specific coupling constant $g_j$
by one and adding a factor $j k_j/z_2^{j+1}$. The geometrical
interpretation is that a $j$-sided polygon is removed, leaving
a marked boundary of length $j$ to which the new indeterminate
$z_2$ is associated. The factor $k_j$ is due to the
possibility to make the replacement at any of the $k_j$ $j$-sided
polygons present in the planar graph, while $j$ is the number of
possibilities to choose the marked link on the new boundary component.
The generating function for planar graphs
with $b$ boundary components can therefore be expressed as
\beq\label{loopx102}
w(g_i ,z_1,\ldots,z_b) = \frac{d}{d V(z_b)}\cdots \frac{d}{d V(z_2)}
w(g_i,z_1).
\eeq

A most remarkable result is the following: for any potential $V(g_i,z)$
the two-loop function\index{loop correlator} $w(g_i,z_1,z_2)$ has the form
\bea
\lefteqn{w(g_i,z_1,z_2)} \label{loopx107}  \\
&&= \frac{1}{2(z_1-z_2)^2}
\left(\frac{z_1z_2 -\oh(z_1+z_2)(c_++c_-) + c_+c_-}{
\sqrt{[(z_1-c_+)(z_1-c_-)][(z_2-c_+)(z_2-c_-)]}}-1
\right). \nonumber
\eea
Note that there is no explicit reference to the potential $V(g_i,z)$,
but of course $c_+$ and $c_-$ depend on the potential.

From this formula one can in principle construct the 
multi-loop function\index{loop correlator}
$w(g,z_1,\ldots,z_b)$ by applying the loop insertion
operator\index{loop insertion operator} $b-2$ times. 
One can use this formula 
to find the leading singularity of $w(g,z_1,\ldots,z_b)$
when $g \to g_c$, the critical value of the coupling 
constant $g$ and the value where $M_1(g) = 0$. One finds 
\beq\label{5lo.30s}
w(g,z_1,\ldots,z_b) \sim \left(\frac{1}{\sqrt{g_c-g}}\right)^{2b-5},
\eeq
as $g \to g_c$.
This implies that the generating function $w(g,l_1,\ldots,l_b)$ for
the number of triangulations, $w_{k,l_1,\ldots,l_b}$,
constructed from $k$ triangles with $b$ boundary components
of length $l_1,\ldots,l_b$, has a singularity as $g\to g_c$ that is
independent of the length of the boundary components and is given by
\beq\label{5lo.30}
w_{h}(g,l_1,\ldots,l_b) \sim \left(\frac{1}{\sqrt{g_c-g}}\right)^{2b-5}.
\eeq
Finally, we obtain from eq.\ \rf{5lo.30} the asymptotic behaviour of
$w_{k,l_1,\ldots,l_b}$ as $k \to \infty$:
\beq\label{5y.7}
w_{k,l_1,\ldots,l_b} \sim
\left( \frac{1}{g_c} \right)^{k} \,k^{-\frac{5}{2}+b-1}.
\eeq
We note that these results can be generalized to triangulations
which have $h$ handles:
\beq\label{5y.9}
w_{h}(g,l_1,\ldots,l_b) \sim \left(\frac{1}{\sqrt{g_c-g}}\right)^{2b+(h-1)5}.
\eeq
\beq\label{5y.8}
w^{(h)}_{k,l_1,\ldots,l_b} \sim
\left( \frac{1}{g_c} \right)^{k} \,k^{(h-1)\frac{5}{2}+b-1}.
\eeq
For future applications it is important to note that the
position of the leading singularity $g_c$ in \rf{5y.9} or, alternatively,
the exponential growth of the number of triangles in \rf{5y.8}, is 
independent of the number of handles or the number of boundaries.

\subsection{The continuum limit\index{continuum limit}}\label{continuum}

We now show how continuum physics is related  to
the asymptotic behaviour of $w^{(h)}_{k,l_1,\ldots,l_b}$ for $k \to \infty$
{\it and} $l_1,\ldots,l_b \to \infty$ in a specific way, and we
use the results for the generating functions 
$w_h(g,z_1,\ldots,z_b)$
derived in the previous sections to study this limit.

Before discussing details it is useful to
clarify how we expect the continuum wave functionals
$W(\La,Z_1,\ldots,Z_b)$ to renormalize. Since the cosmological constants
$\La$ and $Z_i$ have dimensions $1/a^2$ and $1/a$, respectively, $a$ being
the length of the lattice cutoff, it is natural to expect that
they are subject to an additive renormalization
\beq\label{conz1}
\La_c = \frac{\m_c}{a^2} + \La,~~~~Z_{i,c} = \frac{\lam_{i,c}}{a} +Z_i,
\eeq
where $\La_c$ and $Z_{i,c}$
are the {\it bare} cosmological coupling constants.
Since our regularization is represented
in terms of discretized two-dimensional manifolds,
the bare cosmological constants should be related to the dimensionless
coupling constants $\m,\lam_i$ by
\beq\label{conz100}
\La_c =\frac{\m}{a^2},~~~~~Z_{i,c} = \frac{\lam_i}{a},
\eeq
so that eq.\ \rf{conz1} can be written
\beq\label{conz101}
\m-\m_c = a^2 \La,~~~~~~\lam_i -\lam_{i,c} = aZ_i.
\eeq
In the following we assume for simplicity that all the $\lam_{i,c}$s are
equal to $\lam_c$.
We identify the constants $\mu_c$ and $\lam_c$
with the critical couplings $g_c$ and $c_+(g_c)$ via the relations
\beq\label{conz2}
\frac{1}{g_c} = e^{\m_c},~~~~~~c_+(g_c) = e^{\lam_c}.
\eeq
Recalling the relation \rf{5x.gen1} between
$\m,g$ and $z,\lam$, it follows that the $a\to 0$ limit
of the functions $w(\mu ,\lambda_1,\ldots ,\lam_b)$ is determined
by their singular behaviour at $g_c$. The renormalization \rf{conz1}
has the effect of cancelling the exponential entropy factor for the
triangulations, see eq.\  \rf{5y.7}. Note that since we have
the same exponential factors for all genera, we expect
the renormalization of the cosmological constants to be
independent of genus.

We begin by studying the continuum limit\index{continuum limit} for
planar surfaces. Then we will discuss how to take higher genera into account,
thereby reintroducing the gravitational coupling constant
$G$ and also discussing its renormalization. 
This will lead us to the so-called
{\it double-scaling limit}\index{double-scaling limit}.
%


We are interested in a limit of the discretized models where the
length $a$ of the links goes to zero
while the number $k$ of triangles and the lengths $l_i$ of the boundary components
go to infinity in such a way that
\beq\label{conz4}
V = k a^2~~~~\mbox{\rm and}~~~~L_i= l_i a
\eeq
remain finite. The asymptotic behaviour of $w_{k,l_1,\ldots,l_b}$
is given by eq.\ \rf{5y.7} if the  $l_1,\ldots,l_b$ remain bounded.
In this case the leading term is of the form
$$
e^{\m_c V/a^2} (V/a^2)^\b ,
$$
where $\b$ is a critical
exponent. If the boundary lengths $l_1,\ldots,l_b$ diverge according
to eq.\ \rf{conz4}, we expect a corresponding factor
$$
e^{\lam_c L_i/a} (L_i/a)^\a ,
$$
where $\a$ is another critical exponent.
This form of the entropy was encountered for
branched polymers\index{branched polymers} 
in eq.\ \rf{5x.11y2}. We can therefore express the
expected asymptotic behaviour of
the coefficients $w_{k,l_1,\ldots,l_b}$ as
\beq\label{conz5}
w_{k,l_1,\ldots,l_b} \sim  e^{\frac{\m_c}{a^2} V} \;
e^{\lam_c \sum_i l_i} \; a^{-\a b-2\b}\, W(V,L_1,\ldots,L_b),
\eeq
as $a \to 0$, with $V$ and $L_i$ defined by eq.\ \rf{conz4} fixed.
The factor $a^{-\a b-2\b}$ may be thought of as
a
wave-function renormalization.

From eq.\ \rf{conz5} we deduce that the scaling behaviour of
the discretized wave functional $w(\m,\lam_1,\ldots.,\lam_b)$
is given by
\bea
\lefteqn{w(\m,\lam_1,\ldots,\lam_b) =
\sum _{k,l_1,\ldots,l_b} e^{-\m k}\, 
e^{-\sum_i \lam_i l_i}\,w _{k,l_1,\ldots,l_b}}
~~~~~~~ \nn
&& \sim \frac{1}{a^{\a b+ 2\b}} \sum_{k,l_1,\ldots,l_b} e^{-(\m-\m_0)k} \,
 e^{-\sum_i(\lam_i-\lam_0) l_i} W(V,L_1,\ldots,L_b) \nn
&& \sim \frac{1}{a^{(\a+1) b+(2\b+2)}}\, W(\La,Z_1,\ldots,Z_b), \label{conz6}
\eea
where we have used the relation
\beq
W(\La,Z_1,\ldots,Z_b) = \int_0^\infty dV \prod_{i=1}^b dL_i \;
e^{-\La V-\sum_i Z_i L_i}\; W(V,L_1,\ldots,L_b).
\eeq

Our next goal is to show that we can take a limit as suggested by
eqs.\ \rf{conz101} and \rf{conz6}.
In terms of the variables $g,z_i$ we have
\beq\label{conz9}
g=g_c(1-\La a^2),~~~~z_i=z_c(1+a Z_i),
\eeq
where we have introduced the notation
\beq\label{conz10}
z_c =c_+(g_c) = e^{\lam_c}
\eeq
for the critical
value $c_+(g_c)$ of $z$ corresponding to the 
largest allowed value of $g$.
Inserting \rf{conz9} and \rf{conz10} in the expression \rf{loopx107}
one obtains
\beq\label{5s.13}
w(g,z_1,z_2) \sim a^{-2}W(\La,Z_1,Z_2)
\eeq
where
\beq\label{5s.13a} 
W(\La,Z_1,Z_2)= \oh \frac{1}{(Z_1-Z_2)^2}\left(\frac{\oh(Z_1+Z_2)
+ \sqrt{\La}}{\sqrt{(Z_1+\sqrt{\La})(Z_2+\sqrt{\La})}}-1\right),
\eeq
Similarly one can show, using the 
loop insertion operator\index{loop insertion operator},
that when the number of boundaries is larger than two one has
\beq\label{5s.4}
w(g,z_1,\ldots,z_b)\sim \frac{1}{a^{7b/2-5}}
\left(-\frac{d}{d \La}\right)^{b-3} \,\left[\frac{1}{\sqrt{\La}}\,
\prod^{b}_{i=1} \frac{1}{(Z_i+\sqrt{\La})^{3/2}} \right] ,
\eeq
i.e.\
\beq\label{5s.5}
W(\La,Z_1,\ldots,Z_b) \sim 
\left(-\frac{d}{d \La}\right)^{b-3} \,\left[\frac{1}{\sqrt{\La}}\,
\prod^{b}_{i=1} \frac{1}{(Z_i+\sqrt{\La})^{3/2}} \right]\nonumber .
\eeq
The continuum expressions for the n-loop 
functions\index{loop correlator} are all 
universal and independent of the explicit form of the potential
$V(z)$ as long as the weights $t_i \geq 0$. For the one-loop function
the situation is different. As is seen by formally applying the 
counting of powers $a$ in \rf{5s.4} to the case $b=1$ one obtains
the power $a^{3/2}$, i.e.\ a positive power of $a$. The important 
point in \rf{5s.4} and \rf{5s.13} is that the power is negative: in
the scaling limit $a \to 0$ these terms will dominate. This is 
how the formulas should be understood: there are other terms 
too, but they will be subdominant when $a \to 0$, i.e.\ when 
$g\to g_c$ and $z \to z_c$ as dictated by \rf{conz9} and \rf{conz10}.  
For the one-loop function the term associated with $a^{3/2}$ will vanish
when $a \to 0$ and we will be left with a non-universal term explicitly
dependent on the potential $V$. However, the term associated with 
$a^{3/2}$ is still the leading term which is non-analytic in the 
coupling constant $g$, so if we differentiate a number of times
with respect to $g$ and {\it then} take the limit $a \to 0$ it will be 
dominant. No continuum physics is associated with 
the analytic terms since they contain $g$ only to some finite
positive power, and are thus associated with only a finite number
of triangles (of which the lattice length $a \to 0$ when 
$g \to g_c$) if we recall the interpretation of $w(g,z)$ as the generating
function of the number of triangulations. Inserting \rf{conz9} and 
\rf{conz10} in the expression \rf{loopx12} one obtains 
\beq\label{5s.16a}
w(g,z) = \oh (V'(z)+ a^{3/2} W(\La,Z) + O(a^5/2))
\eeq
where
\beq\label{5s.16}
W(\La,Z) =( Z -\oh \sqrt{\La})\sqrt{ Z+\sqrt{\La}}.
\eeq

This ends the calculation of the loop-loop correlation\index{loop correlator} 
functions\index{correlation function} 
for manifolds with topology $S^2_b$, the sphere with $b$ boundaries.
The results agree with continuum calculations using quantum 
Liouville theory and it follows that one can obtain a continuum,
diffeomorphism-invariant theory starting out with a suitable 
lattice theory, where the lattice link length acts as 
a diffeomorphism-invariant UV cut-off and simply taking the 
lattice spacing $a \to 0$, while renormalizing
the bare couplings in a standard way, namely, 
the cosmological and the boundary cosmological
coupling constants.

Let us end this description of 
Euclidean quantum gravity\index{Euclidean quantum gravity} 
by mentioning the corresponding results for higher-genus surfaces. 
The generalization of \rf{5s.4} is
\beq\label{5lo.28}
w_{h}(g,z_1,\ldots,z_b) \sim  \frac{1}{a^{7b/2+(5h-5)}}
W_{h}(\La,Z_1,\ldots,Z_b).
\eeq
In particular, taking $b=0$ leads to the expression
\beq\label{5lo.h3}
Z_{h}(g)  \sim \frac{\tau_{h}}{(a^2\La)^{5(h-1)/2}},
\eeq
for the singular part of the partition function,
where the constants $\tau_h$ can in principle be (and have been) computed.

In \rf{5lo.h3} we have actually completed the task of calculating
\rf{4.1z}. We can now reintroduce the gravitational coupling constant 
$G$ and try to calculate the complete partition function
\beq\label{5y.new1}
Z(G,\La) = \sum_{h=0}^\infty \tau_h\,\, e^{\frac{2-2h}{G}} \,a^{5(1-h)}
 \La^{\frac{5(1-h)}{2}}.
\eeq
The factor $a^{-5}$ present for each genus can be absorbed
in a renormalization of the gravitational coupling constant
\beq\label{5y.13}
\frac{1}{G_{ ren}} = \frac{1}{G(a)} - \frac{5}{4}\log \frac{\La}{a^2},
\eeq
where $G_{ ren}$ denotes the
{\it renormalized} gravitational coupling.
A continuum limit\index{continuum limit} 
of \rf{5y.new1} only exists in the limit $a \to 0$ if
we allow $G$ to be a function of the lattice spacing $a$ determined 
by \rf{5y.13} for fixed $G_{ren}$ and $\La$. The continuum limit is 
then given by 
 \beq\label{5y.13b}
Z(G,\La) =
\sum_{h=0}^\infty \tau_h \;\left(\e^{-1/G}\La^{-\frac{5}{4}}\right)^{2h-2},
\eeq 
and depends only on the variable $x= \La e^{4/(5G)}$. To actually calculate
$Z(G,\La)$ we have to perform the summation over the number of handles
$h$ in \rf{5y.13b}, an interesting task which we will not address here.
Rather, we will focus on eq.\ \rf{5y.13}, since we can use this 
equation to calculate the $\b$-function\index{$\beta$-function} for $G$ using
\beq\label{beta}
\b(G) \equiv \left. -a \frac{\d G(a)}{\d a}\right|_{\La,G_{ren}}  
= -\;{5}{2} G^2.
\eeq
Two-dimensional Euclidean quantum gravity\index{Euclidean quantum gravity} is 
asymptotically free\index{asymptotic freedom} 
as already mentioned in the introduction.

\section{Two-dimensional Lorentzian quantum 
gravity\index{Lorentzian quantum gravity}}\label{2d-CDT}

As already mentioned above, Euclidean quantum gravity does not really
work in more than two dimensions (in the sense of leading to a continuum
theory of higher-dimensional geometry). By contrast, the formalism
called CDT\index{CDT}, based on causal dynamical triangulations, seems
to lead to very interesting results. It is based on the idea
that there exists a globally defined (proper-)time variable, 
which can be used to describe the evolution of the universe. In addition, one 
assumes that the topology of space is unchanged with respect to the foliation
defined by this global time. 

These requirements are definitely not satisfied in two-dimensional {\it Euclidean} 
quantum gravity\index{Euclidean quantum gravity}. In principle one can also
``superimpose" a proper time on Euclidean quantum universes and follow
their evolution as first described in the 
seminal work by Kawai and collaborators. Starting out 
with a spatial universe of topology $S^1$, it
will immediately split up into many disconnected spatial ,
one-dimensional universes as a function of proper time.
It turns out that the structure is fractal, in the sense that 
an infinity of spatial universes, most of them of infinitesimal
spatial extension, will be created as a function of proper time.

Since two-dimensional Euclidean quantum gravity is explicitly
solvable, even on a lattice before the continuum limit
is taken, as described above, it is
of interest to understand the transition from the Euclidean
lattice gravity theory to the CDT lattice gravity theory\index{CDT}.
Clearly one has to suppress the splitting of a spatial 
universe into two or more disconnected spatial universes if one 
wants to move from the spacetime configurations which characterize
the Euclidean path integral to the configurations present
in the CDT path integral\index{CDT path integral}.
It makes sense to talk about the splitting of a 
spatial universe into two if the universe has  
Lorentzian signature\index{Lorentzian signature}, 
since such a splitting (in the simplest case)  
is associated with an isolated point where the 
metric and its associated light-cone
structure are degenerate, which has a diffeomorphism-invariant meaning.
This was the motivation for imposing such a constraint in the original
CDT model. By working in Lorentzian signature initially, and only later
rotating to Euclidean signature\index{Euclidean signature}, 
this constraint survives also in (the Euclidean version of) CDT.  
Going back to Fig.\ \ref{fig4.6}, this suggests that one should associate
a factor $g_s$ instead of a factor 1 with the graph with the double
line. Geometrically this figure can be viewed as a process where a triangle is 
removed at a marked link (and a new link is marked at the new boundary),
except in the case where the marked link does not belong to a
triangle, but is part of a double-link, in which case the double link is
removed and the triangulation is separated into two. 
If one thinks of the recursion process in Fig.\ \ref{fig4.6}
as a ``peeling away" of the triangulation as proper time advances, the
presence of a double link represents the ``acausal" splitting point beyond
which the triangulation splits into two discs with two separate boundary
components (i.e. two separate one-dimensional spatial universes). 
The interpretation of this process, 
advocated in \cite{fractal}, is that it represents a split 
of the spatial boundary with respect to (Euclidean) proper time.  
Associating an explicit weight $g_s$ with this situation and  
letting $g_s \to 0$ suppress this process compared to processes 
where we simply remove an $i$-gon from the triangulation.
Nevertheless, we will see below that there exists an
interesting scaling of $g_s$ with $a$ such that the 
process survives when we let $a\to 0$, but with a result different from the Euclidean 
quantum gravity theory\index{Euclidean quantum gravity}. 
We call this new limit 
{\it generalized} CDT\index{CDT}\index{generalized CDT model} \cite{alwz}.

Let us introduce the new coupling constant $g_s$ in eq.\ \rf{*4.44x}.
The equation is then changed to 
\beq\label{3.18a}
w(z)= g\left(\sum_{i=1}^n t_i z^{i-2}\right) w(z) + \frac{g_s}{z}\, w^2(z)
+\frac{1}{z}\, Q(z,g).
\eeq

In the analysis it will be
convenient to keep the coupling constant $t_1 > 0$, although
we are usually not so interested in situations with one-gons.
It can be motivated as follows.
Consider a ``triangulation'' consisting of $T_1$ one-gons,
$T_2$ two-gons, $T_3$ triangles, $T_4$ squares etc., up to $T_n$ n-gons. 
The total coupling-constant factor associated with the 
triangulation is given by
\beq\label{3.6}
g^{T_1+\cdots +T_n} g_s^{-T_1/2 +T_3/2 +\cdots + (n/2-1)T_n}.
\eeq
We observe that in the limit $g_s \to 0$, a necessary condition for obtaining 
a finite critical value $g_c(g_s)$ is $T_1 > 0$. 
We should emphasize that the analysis described below can be 
carried out also if we suppress the appearance of any one-gons
(by setting $t_1 =0$) in our triangulations, but it is 
slightly more cumbersome since then $g_c (g_s)\to \infty$ as $g_s \to 0$, 
requiring further rescalings.

For simplicity we will consider the simplest 
nontrivial model with potential\footnote{the rationale for calling $V$ a ``potential"
will become clear below}
\beq\label{3.7}
V(z)= \frac{1}{g_s}\left( -g z + \oh z^2 -\frac{g}{3} z^3\right)
\eeq
and analyze its behaviour in the limit $g_s \to 0$.
The disk amplitude\index{disk amplitude} \rf{loopx11} now has the form
\beq\label{3.9}
w(z) = \frac{1}{2g_s} \left(-g +z -gz^2 + g(z-c_2)\sqrt{(z-c_+)(z-c_-)}\right),
\eeq
and the constants $c_2$, $c_+$ and $c_-$ are determined by the requirement 
that $w(z) \to 1/z$ for $z \to \infty$. 
Compared with the analysis of the previous
section, the algebraic condition fixing the
coefficient of $1/z$ to be unity will now enforce a 
completely different scaling behaviour as $g_s \to 0$.

For the time being, we will think of $g_s$ as small and fixed, and perform
the scaling analysis for $g_c(g_s)$.
As already mentioned above, the 
critical point $g_c$ is determined by the additional 
requirement $M_1=0$ in the representation \rf{loopx12}, i.e.\
that $c_2(g_c)=c_+(g_c)$, which presently
leads to the equation 
\beq\label{3.10}
\left(1-4g_c^2\right)^{3/2} = 12 \sqrt{3} \;g_c^2 \,g_s.
\eeq
Anticipating that we will be
interested in the limit $g_s \to 0$, we write the critical points as
\beq\label{3.11}
g_c(g_s)= \oh (1-\Del g_c(g_s)),~~~\Del g_c(g_s) = 
\frac{3}{2} g_s^{2/3} + O(g_s^{4/3}),
\eeq
and 
\beq\label{3.12}
z_c (g_s) = c_+(g_c,g_s)= 
\frac{1}{2g_c(g_s)}\left(1+\sqrt{\frac{1-4g_c(g_s)^2}{3}}\right) =
1 + g_s^{1/3} + O(g_s^{2/3}),
\eeq
while the size of the cut in \rf{3.7}, $c_+(g_c)-c_-(g_c)$,  behaves as 
\beq\label{3.12a}
c_+(g_c)-c_-(g_c) = 4 g_s^{1/3} + 0(g^{2/3}_s).
\eeq
Thus the cut shrinks to zero as $g_s \to 0$.

Expanding around the critical point given by \rf{3.11}-\rf{3.12} a 
nontrivial limit can be obtained if we insist that 
in the limit $a\rightarrow 0$, $g_s$ scales according to
\beq\label{3.13}
g_s = G_s a^3,
\eeq
where $a$ is the lattice cut-off introduced earlier.
With this scaling the size of the cut scales to zero as $4\,a\, G_s^{1/3}$.
In addition $\sqrt{(z-c_+)(z-c_-)} \propto a$ 
if we introduce the standard identification \rf{conz9}: 
$z=c_+(g_c)+a\, Z$. This scaling is different from the conventional 
scaling in Euclidean quantum gravity
\index{Euclidean quantum gravity} where $\sqrt{(z-c_+)(z-c_-)}
\propto a^{1/2}$ since in that case
$(z-c_+)$ scales while $(z-c_-)$ does not scale.

We can now write
\beq\label{3.14}
g = g_c(g_s)(1-a^2 \La) = \bg(1 - a^2 \cdtL+O(a^4)),
\eeq
with the identifications
\beq\label{3.14a}
\cdtL \equiv \La + \frac{3}{2}G_s^{2/3},~~~~\bg = \oh,
\eeq
as well as
\beq\label{3.15}
z= z_c +a Z = \bz+a \cdtZ +O(a^2),
\eeq
with the identifications
\beq\label{3.15a}
\cdtZ\equiv Z+G_s^{1/3},~~~~\bz=1.
\eeq 
Using these definitions one computes in the limit $a \to 0$ that
\beq\label{3.16}
w(z) = \frac{1}{a} \; \frac{\cdtL -\oh \cdtZ^2 + 
\oh(\cdtZ-H)\sqrt{(\cdtZ+H)^2 -\frac{4G_s}{H}}}{2G_s}.
\eeq
In \rf{3.16}, the constant $H$ (or rather, its rescaled version 
$h=H/\sqrt{2\cdtL}\;$) satisfies the third-order equation
\beq\label{3.17}
h^3 -h + \frac{2G_s}{(2\cdtL)^{3/2}} =0,
\eeq
which follows from the consistency equations for the constants 
$c_2$, $c_+$ and $c_-$
in the limit $a \to 0$.
We thus define
\beq\label{3.17a}
w(z) = \frac{1}{a}\, \cdtW(\cdtZ,\cdtL,G_s) \equiv \frac{1}{a}\, W(Z,\La,G_s)
\eeq  
in terms of the continuum 
Hartle-Hawking wave functions\index{Hartle-Hawking wave function} 
$\cdtW(\cdtZ,\cdtL,G_s)$ and $W(Z,\La,G_s)$.

Notice that while the cut of $\sqrt{(z-c_+)(z-c_-)}$ goes to zero as the 
lattice spacing $a$, it nevertheless survives in the scaling limit
when expressed in terms of renormalized ``continuum'' variables,
as is clear from eq.\ \rf{3.16}. Only in the limit $G_s \to 0$ it
disappears and we have 
\beq\label{3.18}
w(z) = \frac{1}{a}\, \cdtW(\cdtZ,\cdtL,G_s)~~
\underset{G_s\to 0}{\longrightarrow}~~  
\frac{1}{a} \; \frac{1}{\cdtZ +\sqrt{2\cdtL}},
\eeq  
which is the original CDT disk amplitude\index{disk amplitude}
introduced in \cite{al}.

\vspace{6pt}

\begin{figure}[t]
\centerline{\scalebox{0.6}{\rotatebox{0}{\includegraphics{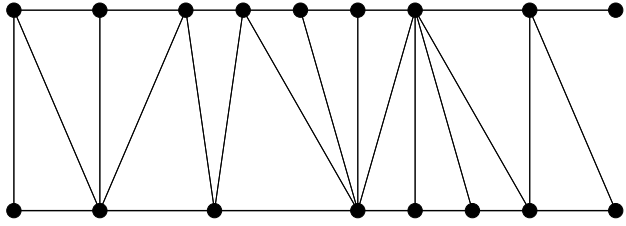}}}}
\caption{The propagation of a spatial slice from time $t$ to time $t+1$.
The end of the strip should be joined to form a band with topology $S^1\times
[0,1]$.}
\label{2dminkowski}
\end{figure}

\noindent
Let us make some  comments:

\vspace{6pt}

\noindent
{\bf (1)} We have dealt here directly with a ``generalized'' 
CDT model\index{CDT}\index{disk amplitude}
which in the limit $G_s \to 0$ reproduces the ``original'' CDT disk
amplitude\index{disk amplitude} \rf{3.18}. The original two-dimensional 
CDT model was defined according to 
the principles already outlined in our discussion of four-dimensional
quantum gravity. Thus the interpolation between two spatial slices
separated by one lattice spacing of proper time is shown in Fig.\ 
\ref{2dminkowski}. This figure is the two-dimensional analogue of 
Fig.\ \ref{connect}. The combinatorial problem of summing over
all such surfaces connecting two spatial boundaries separated by a
certain number of time steps can be solved \cite{al}, and the continuum 
limit can be taken. The corresponding continuum ``propagator'' is found to be 
\bea
\lefteqn{G(X,Y;T) = \frac{4\cdtL \e^{-2\sqrt{\cdtL}\,T}}{
(\sqrt{\cdtL}+X)+\e^{-2\sqrt{\cdtL}\,T}(\sqrt{\cdtL}-X)}}\label{26}\\
&&\times \, \frac{1}{(\sqrt{\cdtL}+X)(\sqrt{\cdtL}+Y)-\e^{-2\sqrt{\cdtL}\,T}
(\sqrt{\cdtL}-X)(\sqrt{\cdtL}-Y)},
\nonumber
\eea
where $X,Y$ are the boundary cosmological constants associated with
the two boundaries of the cylinder and $T$ is the proper time separating the
two boundaries. The propagator has an asymmetry 
between $X$ and $Y$ because we have marked a point (a vertex in the 
discretized model). By an inverse Laplace transform one can calculate
the propagator $G(X,L;T)$ as a function of the length of the unmarked boundary.
In particular, we have 
\beq\label{27}
 G(X,L=0;T) = \int_{-i\infty}^{i\infty} \d Y \;G(X,Y;T),
\eeq
and we define the CDT disk amplitude\index{disk amplitude} as 
\beq\label{28}
W^{(0)}(X) = \int_0^\infty  \d T \; G(X,L=0;T),
\eeq
and it is given by the expression on the far right in \rf{3.18}.

\begin{figure}[t]
\centerline{\scalebox{0.5}{\rotatebox{0}{\includegraphics{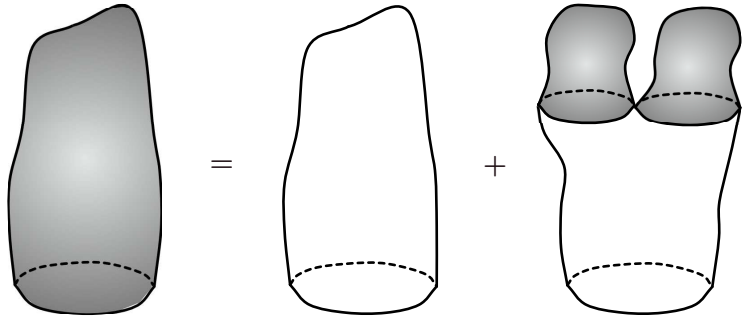}}}}
\caption{Graphical illustration of eq.\ \rf{29}. Shaded parts represent
the generalized CDT disk amplitude\index{disk amplitude}, 
unshaded parts the original CDT disk
amplitude and the original CDT propagator \rf{26}.}
\label{generaldisk}
\end{figure}

The generalized CDT model\index{CDT}\index{disk amplitude} 
allows branching of the spatial universes
as a function of proper time $T$, the branching being controlled 
by the coupling constant $G_s$. This results in the  
graphical equation for the generalized CDT disk amplitude\index{disk amplitude}
shown in Fig.\ \ref{generaldisk}. The corresponding equation is:
\bea\label{29}
W(X)& =& W^{(0)}(X) +G_s \int_0^\infty \d T \int_0^\infty \d L_1 \d L_2 
\\
 &&(L_1+L_2)G(X,L_1+L_2;T) W(L_1)W(L_2),
\nonumber
\eea
where $W(L)$ is the disk amplitude\index{disk amplitude} 
corresponding to boundary length $L$.
It can be solved \cite{alwz} for $W(X)$ and the solution is 
$\cdtW (X,\cdtL,G_s)$ found above.

\vspace{6pt}

\noindent
{\bf (2)} Using eq.\ \rf{3.17} we can expand $w(z)$ into a power series
in $G_s/(2\cdtL)^{3/2}$ whose radius of convergence is $1/3\sqrt{3}$. 
For fixed values of $\cdtL$, this value corresponds to the largest 
value of $G_s$ where \rf{3.17} has a positive solution for $h$. 
The existence of such a bound on $G_s$ for fixed $\cdtL$
was already observed in \cite{alwz}.
This bound can be re-expressed more transparently in the present
Euclidean context, where it is more natural to keep the ``Euclidean"
cosmological constant $\La$ fixed, rather than $\cdtL$. We have
\beq\label{3.19}
\frac{G_s}{(2\cdtL)^{3/2}}\leq  \frac{1}{3\sqrt{3}}~ \Rightarrow ~ 
\frac{{3} G_s^{2/3}}{2\La +{3} G_s^{2/3}} \leq 1,
\eeq
which for fixed $\La > 0$ is obviously satisfied for all positive
$G_s$. In order to see that the usual Euclidean 2d quantum gravity 
(characterized by some {\it finite} value for $g_s$) can be re-derived
from the disk amplitude\index{disk amplitude} \rf{3.16}, let us 
expand eq.\ \rf{3.16} for large $G_s$.
The square root part becomes
\beq\label{3.20}
a^{-1}\;G_s^{-5/6}\; 
\left(Z- \sqrt{2\La/3}\right)\sqrt{Z+2\sqrt{2\La/3}} , 
 \eeq
which coincides with the generic expression $a^{3/2} W(Z,\La)$  
in Euclidean 2d quantum gravity\index{Euclidean quantum gravity} 
(c.f. eqs.\ \rf{5s.16a} and \rf{5s.16})
if we take $G_s$ to infinity as $g_s/a^3$ (and take into account 
a trivial rescaling of the cosmological constant). However, if we reintroduce
the same scaling in the $V'(z)$-part of $w(z)$, it does not scale with $a$
but simply goes to a constant. This term would dominate $w(z)$
in the limit $a \to 0$ if one did not remove it by hand, as is usually done
in the Euclidean model.

\vspace{6pt}

\noindent
{\bf (3)} Why does the potential $V'(z)$ (and therefore the entire
disk amplitude\index{disk amplitude} $w(z)$) scale (like $1/a$) in the 
new continuum limit\index{continuum limit} with 
$g_s = G_s a^3$, $a\to 0$, contrary
to the situation in ordinary Euclidean 
quantum gravity?\index{Euclidean quantum gravity}
This is most clearly seen by looking again at 
the definitions \rf{3.14} and \rf{3.15}.
Because of the vanishing
\beq\label{3.21}
V'(\bz,\bg)=0,~~~~V''(\bz,\bg)=0
\eeq
in the point $(\bz,\bg)=(1,1/2)$,
expanding around $(\bz,\bg)$ according to \rf{3.14}, \rf{3.15}
leads automatically to a potential which 
is of order $a^2$ when expressed in terms
of the renormalized constants $(\cdtZ,\cdtL)$,
precisely like the square-root term when expressed in terms of 
$(\cdtZ,\cdtL)$. 

The point $(\bz,\bg )$ differs from the critical point 
$(z_c(g_s),g_c(g_s))$, as long as $g_s\not= 0$.
In fact, both $1/\bz$ and $\bg$ lie {\it beyond} 
the radii of convergence of 
$1/z$ and $g$, which are precisely $1/z_c(g_s)$ and $g_c(g_s)$. 
However, since the differences
are of order $a$ and $a^2$, respectively, they simply amount to {\it shifts}
in the renormalized variables, as
made explicit in eqs.\ \rf{3.14a} and \rf{3.15a}. Therefore, re-expressing 
$W(Z,\La,G_s)$ in \rf{3.17a} in terms of the variables 
$\cdtZ$ and $\cdtL$ simply leads to the expression 
$\cdtW(\cdtZ,\cdtL,G_s)$, first derived in \cite{alwz}.
Similarly, any geometric quantities defined with respect to $Z$ and $\La$ 
can equally well be expressed in terms of $\cdtZ$ and $\cdtL$.
For instance, the average continuum length of the boundary and the average 
continuum area of a triangulation are given by
\beq\label{3.22}
\la L \ra = \frac{\prt \ln W(Z,\La,G_s)}{\prt Z}= 
\frac{\prt \ln \cdtW(\cdtZ,\cdtL,G_s)}{\prt \cdtZ},
\eeq
\beq\label{3.23}
\la A \ra =\frac{\prt \ln W(Z,\La,G_s)}{\prt \La}= 
\frac{\prt \ln \cdtW(\cdtZ,\cdtL,G_s)}{\prt \cdtL}. 
\eeq
In the limit of $G_s \to 0$,
the variables $(Z,\La)$ and $(\cdtZ,\cdtL)$ become identical and the 
disk amplitude\index{disk amplitude} becomes the original CDT amplitude 
$(\cdtZ+\sqrt{2\cdtL})^{-1}$ alluded to in \rf{3.18}.

\section{Matrix model representation\index{matrix model} }\label{matrix}

Above we have solved the two-dimensional gravity models
by purely combinatorial techniques which emphasize
the geometric interpretation: the quantum theory as
a sum over geometries\index{sum over geometries}. The use of so-called 
matrix models allows one to perform the summation over the
piecewise linear geometries\index{piecewise linear geometries} 
in a relatively simple way.
Surprisingly, it turns out that the {\it scaling limit} of
the generalized CDT model\index{CDT} has itself a 
matrix model\index{matrix model}  
representation. We will here describe how matrix models 
can be used instead of the combinatorial methods and how 
one is led to the CDT matrix model\index{CDT matrix model}. 

\begin{figure}[t]
\centerline{\scalebox{0.7}{\rotatebox{0}{\includegraphics{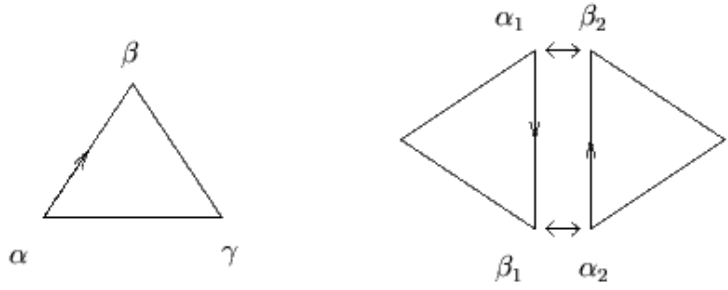}}}}
\caption[5_1]{The matrix representation of triangles which converts the
gluing along links to a Wick contraction.}
\label{5_1}
\end{figure}

Let $\phi$ be a Hermitian $N\times N$ matrix with 
matrix elements $\phi _{\a\b}$
and consider for $k=0,1,2,\ldots $ the integral
\beq\label{5.9}
\int d\phi \;e^{-\oh \Tr \phi^2 } \frac{1}{k!} \left(\frac{1}{3}
\Tr \phi^3 \right)^k,
\eeq
where
\beq\label{5.10}
d\phi =\prod_{\a \leq \b} d\, {\rm Re}\, \phi_{\a\b}
\prod_{\a < \b} d \,{\rm Im}\, \phi_{\a\b}.
\eeq
We can regard $\phi$ as a zero-dimensional matrix-valued field
so the integral can be evaluated in the standard way
by doing all possible Wick contractions of $(\Tr\phi^3)^k$ and using
\beq\label{5.11}
\lan \phi_{\a\b} \phi_{\a'\b'} \ran =C
\int d \phi \; e^{-\oh \sum_{\a\b}|\phi_{\a\b}|^2} \phi_{\a\b}
\phi_{\a'\b'}
= \del_{\a\b'} \del_{\b\a'},
\eeq
where $C$ is a normalization factor. The evaluation of the expression
\rf{5.9} can be
interpreted graphically by associating to each factor $\Tr\phi ^3$ an oriented
triangle and to each term $\phi _{\a\b}\phi_{\b\g}\phi_{\g\a}$ contributing
to the trace a labeling of its vertices by $\a,\b,\g$ in cyclic order,
such that the matrix element $\phi_{\a\b}$ is associated with the
oriented link whose endpoints are labeled by $\a$ and $\b$ in accordance
with the orientation.  eq.\ \rf{5.11} can then be interpreted as
a gluing of the link labeled by $(\a\b)$ to an oppositely
oriented copy of the same link, see Fig.\ \ref{5_1}.

In this way the integral \rf{5.9} can be represented as a sum over closed,
possibly disconnected, triangulations with $k$ triangles.
Triangulations with an arbitrary genus arise in this representation.
By summing over
$k$ in \rf{5.9} it follows by standard arguments that the (formal)
logarithm of the corresponding integral is represented as a sum over
all closed and {\em connected} triangulations. The contribution of a given
triangulation can be determined by observing that in the process of gluing
we pick up a factor of $N$ whenever a vertex becomes an internal vertex in
the triangulation. Thus, the weight of a triangulation $T$ is simply
$$
N^{N_v(T)}.
$$
If we make the substitution
\beq\label{5.12}
\Tr \phi^3 \to \frac{g}{\sqrt{N}} \Tr \phi^3,
\eeq
the weight of $T$ is replaced by
$$
g^k N^{N_v(T)-k/2}=g^k N^{\chi (T)},
$$
where $\chi (T)$ is the Euler characteristic of $T$.
Note also that the factor $(k!)^{-1}$ in \rf{5.9} is cancelled in the sum
over different triangulations because of the $k!$ possible permutations of
the triangles, except for trian\-gu\-la\-ti\-ons with non-trivial
automorphisms, in which case the
symmetry factor $C_T^{-1}$ survives. With the identifications
\beq\label{5.14a}
\frac{1}{G} = \log N,~~~~~ \m = -\log g
\eeq
we conclude that
\beq\label{5.14}
Z(\m,G)=\log \frac{Z(g,N)}{Z(0,N)}
\eeq
where $Z(\m,\k )$ is defined by eq.\ \rf{5x.11}--\rf{5x.new11} and
\beq\label{5.16}
Z(g,N)=\int d\phi \;\exp\left(-\oh \Tr \phi^2 +
\frac{g}{3\sqrt{N}}\Tr \phi^3\right).
\eeq
The integral \rf{5.16} is of course
divergent and should just be regarded as a shorthand
for the formal power series in the coupling constant $g$.

It is straightforward to generalize the preceding arguments to the case of
general unrestricted triangulations, where arbitrary polygons are allowed.
Instead of one coupling constant $g$ we have a set
$g_1,g_2,g_3,\ldots$, but we will still use the notation $g$ or $g_i$.
In this case eq.\ \rf{5.16} is replaced by
\beq\label{5.19a}
Z(g_i,N))= \int d \phi\; e^{-N \Tr V(\phi)},
\eeq
where the potential $V$, which depends on all the coupling constants
$g_i$, is given by
\beq\label{5.20}
V(\phi ) = \oh \phi^2-\sum_{j=1}^{\infty} \frac{g_j}{j}  \phi^j.
\eeq
In eq.\ \rf{5.20} we have scaled $\phi \to \sqrt{N}\phi$
for later convenience.
It is not difficult to check that the 
obvious generalization of eq.\ \rf{5.14} holds
and the weight of a triangulation $T$ is given by
$$
C_T^{-1}N^{\chi (T)}\prod _{j\geq 1}g_j^{N_j(T)},
$$
where $N_j(T)$ is the number of $j$-gons in $T$.
eq.\ \rf{5.19a} is of course a representation
of a formal power series which is obtained by expanding the exponential of
the non-quadratic terms as a power series in the coupling constants and
then performing the Gaussian integrations term by term.

Differentiating $\log Z(g_i,N)$
with respect to the coupling constants
$g_j$, one obtains the expectation values of products of traces
of powers of $\phi$. These expectation values have a straightforward
interpretation in terms of triangulations. Denoting the expectation with
respect to the measure
$$
 Z(g_i,N)^{-1} e^{-N\Tr V(\phi )}\,d\phi
$$
by $\lan\cdot\ran$ we see, for example, that $\lan
N^{-1}\Tr\phi ^n\ran$ is given by the sum over all connected triangulations
of arbitrary genus whose boundary is an $n$-gon with one marked link.
Similarly,
\beq\label{5.21}
\frac{1}{N^2} \bra \Tr \phi^n \Tr \phi^m \ket- \frac{1}{N^2}
\bra\Tr \phi^n\ket \bra\Tr \phi^m \ket
\eeq
is given by the sum over all 
connected triangulations\index{sum over geometries}
whose boundary consists of two components with $n$ and $m$ links.
More generally, the relation to the combinatorial problem is given by
\beq\label{5.23}
w(g_i,z_1,\ldots,z_b) = N^{b-2} \sum_{k_1,\ldots ,k_b}
\frac{ \bra \Tr \phi^{k_1} \cdots \Tr \phi^{k_b}\ket_{conn}}{z^{k_1+1}_1
\cdots z^{k_b +1}_b},
\eeq
where the subscript
{\it conn} indicates the connected part of the expectation
$\lan\cdot\ran$.  One can rewrite eq.\ \rf{5.23} as
\beq\label{5.24}
w(g_i,z_1,\ldots,z_b) =
N^{b-2} \bra \Tr \frac{1}{z_1-\phi}\;\cdots\;
\Tr \frac{1}{z_b-\phi}\ket_{conn}.
\eeq

The one-loop function\index{loop correlator} $w(g_i,z)$
is related to the density $\rho(\lam)$
of eigenvalues of $\phi$ defined by
\beq\label{*rho1}
\rho(\lam) = \bra \sum_{i=1}^N \del(\lam-\lam_i) \ket ,
\eeq
where $\lam_i$, $i=1,\ldots,N$ denote the $N$ eigenvalues of the matrix
$\phi$. With this definition we have
\beq\label{5.28a}
\frac{1}{N} \bra \Tr \phi^n \ket =
\int^\infty_{-\infty} d\lam\; \rho(\lam) \lam^n, ~~~ n \geq 0.
\eeq
Hence,
\beq\label{*cut}
w(g_i,z) = \int_{-\infty}^\infty d\lam\; \frac{\rho(\lam)}{z-\lam}.
\eeq
In the limit $N\to \infty$ the support of $\rho$ is confined to a finite
interval $[c_-,c_+]$ on the real axis. In this case $w(z)$ will be
an analytic function in the complex plane, except for a cut along the
interval $[c_-,c_+]$.  Note that $\rho(\lam)$ is determined from $w(z)$ by
\beq\label{*cut1}
2\pi i\rho(\lam) = \lim_{\ep \to 0} \left( w(\lam-i\ep)-w(\lam+i\ep)\right) .
\eeq

\subsection{The loop equations\index{loop equation}}

A standard method in quantum field theory is to derive identities by
a change of variables in functional integrals. Here we apply this method
to the matrix models\index{matrix model}  
and explore the invariance of the matrix integral
\rf{5.19a} under infinitesimal field redefinitions of the form
\beq\label{*4.29}
\phi \to \phi+\ep \phi^n,
\eeq
where $\ep$ is an infinitesimal parameter. One can show
that to first order in $\ep$ the measure $d \phi$ defined by eq.\ \rf{5.10}
transforms as
\beq\label{*4.30}
d\phi \to d\phi \left(1 + \ep \sum_{k=0}^n \Tr \phi^k
\; \Tr \phi^{n-k} \right).
\eeq
The action transforms according to
\beq\label{*4.31}
\Tr V (\phi) \to \Tr V(\phi) + \ep  \Tr \phi^{n} V'(\phi)
\eeq
to first order in $\ep$.
We can use these formulas to study the transformation of the measure
under more general field redefinitions of the form
\beq\label{*4.32}
\phi \to \phi +\ep \sum_{k=0}^\infty \frac{\phi^k}{z^{k+1}} =
\phi+ \ep \frac{1}{z-\phi}.
\eeq
This field redefinition only makes sense if $z$ is
on the real axis outside the support $\rho$.
In the limit $N\to\infty$ this is possible for $z$ outside the interval
$[c_-,c_+]$. Under the field redefinitions \rf{*4.32}
the transformations of the measure and the action are given by
\beq\label{*4.33}
d \phi \to d\phi \left( 1+ \ep\,  \Tr \frac{1}{z-\phi}\;\Tr \frac{1}{z-\phi}
\right),
\eeq
\beq\label{*4.34}
\Tr V(\phi) \to \Tr V(\phi) +
\ep\, \Tr \left ( \frac{1}{z-\phi} V'(\phi)\right).
\eeq
The integral \rf{5.19a} is of course
invariant under this change of the
integration variables. By use of eqs.\ \rf{*4.33}
and \rf{*4.34} we obtain the identity
\beq\label{*4.35}
\int d\phi \left\{ \left( \Tr \frac{1}{z-\phi} \right)^2 -
N\Tr \left( \frac{1}{z-\phi} V'(\phi) \right)\right\}\, e^{-N\Tr V(\phi)}=0.
\eeq
The contribution to the integral coming from the
first term in $\{\cdot\}$ in eq.\ \rf{*4.35} is, by definition,
\beq\label{*4.36}
N^2 w ^2(z) +  w(z,z).
\eeq
The contribution from the second term inside $\{ \cdot\}$ in eq.\ \rf{*4.35}
can be written as an integral over the  
one-loop function\index{loop correlator} as follows:
\beq\label{*4.37}
\frac{1}{N} \bra \Tr \frac{V'(\phi)}{z-\phi} \ket =
\int d\lam \,\rho(\lam)\, \frac{V'(\lam)}{z-\lam} =
\oint_C \frac{d \om}{2 \pi i}  \,\frac{V'(\om)}{z-\om} \; w(\om),
\eeq
where the second equality follows from eq.\ \rf{*cut1}.
The curve $C$ encloses the support of $\rho$ but not $z$.
It is essential for the existence of $C$ that $\rho$ have compact support.
We can then write eq.\ \rf{*4.35} in the
form
\beq\label{*loop}
\oint_C \frac{d \om}{2 \pi i} \frac{V'(\om)}{z-\om} \;w(\om) =
w^2(z) + \frac{1}{N^2} w(z,z),
\eeq
where $z$ is outside the interval $[c_-,c_+]$ on the real axis.
Since both sides of eq.\ \rf{*loop} can be analytically continued to
$C\setminus [c_-,c_+]$ the equation holds in this domain.

We recognize  eq.\ \rf{*loop} as the loop equation\index{loop equation}
already derived by combinatorial means, except that 
we here have an additional term involving $w(z,z)$ and with 
a coefficient $1/N^2$. In fact, \rf{*loop} is the starting 
point for a $1/N^2$-expansion, i.e.\ a higher-genus expansion.  
To leading order in $1/N^2$ we have (as already derived)
\beq\label{*sphere2}
w(z)=w_0(z)= \oh \left( V'(z) -M(z) \sqrt{(z-c_+)(z-c_-)} \right)
\eeq
and from eq.\ \rf{*cut1} the corresponding
eigenvalue density\index{eigenvalue density} is
\beq\label{*4.46}
\rho(\lam) =  \frac{1}{2\pi}M(\lam) \sqrt{(c_+-\lam)(\lam-c_-)}.
\eeq

Let us now discuss how the new scaling limit is described in the matrix 
formalism. Hermitian matrix models\index{matrix model}  
are often analyzed in terms of the
dynamics of their eigenvalues. Since the action in \rf{5.19a} is invariant 
under the transformation $\phi \to U \phi U^\dg$, with $U \in U(N)$
a unitary $N\times N$-matrix, 
one can integrate out the ``angular'' degrees of freedom. What is left is
an integration over the eigenvalues $\lam_i$ of $\phi$ only, 
\beq\label{3.1}
Z(g) \propto \int \prod_{i=1}^N d\lam_i \; 
e^{-N \sum_j V(\lam_j)} \; \prod_{k<l} |\lam_k-\lam_l|^2,
\eeq
where the last factor, the Vandermonde determinant, 
comes from integrating over the angular variables, and where
\beq\label{3.2}
{\rm tr}\ V(\phi) = \sum_{i=1}^N V(\lam_i).
\eeq
Naively one might expect that 
the large-$N$ limit is dominated by a saddle-point
with $V'(\lam)=0$. However, this is not the case since 
the Vandermonde determinant in \rf{3.1} contributes in the large-$N$ limit. 
The cut which appears in $w(z)$ is a direct
result of the presence of the Vandermonde determinant. In this way
one can say that the dynamics of the eigenvalues is ``non-classical'',
deviating from $V'(\lam) =0$, the size of the cut being a measure of this 
non-classicality.
To get to the generalized CDT model\index{CDT}\index{disk amplitude} 
we introduced a new coupling 
constant $g_s$ in the matrix model\index{matrix model} by substituting
\beq\label{3.3}
V(\phi) \to \frac{1}{g_s} \; V(\phi),
\eeq
and considered the limit $g_s \to 0$. As we have seen 
the coupling constant $g_s$ 
controls and reduces the size of the cut and thus brings 
the system closer to a ``classical'' behaviour. Thus 
the quantum fluctuations\index{quantum fluctuations} 
are reduced in the generalized CDT models\index{disk amplitude}.

Let us now consider the matrix potential \rf{3.7}, 
which formed the starting point
of our new scaling analysis. We are still free to perform a change 
of variables. Inspired by
relations \rf{3.14}--\rf{3.15a}, let us transform to new ``CDT''-variables
\beq\label{3.24}
\phi \to \bz \, \hI + a \Phi + O(a^2),
\eeq 
at the same time re-expressing $g$ as
\beq
g=\bg(1-a^2\cdtL+O(a^4)),
\eeq
following eq.\ \rf{3.14}.
Substituting the variable change into the matrix potential, and discarding
a $\phi$-independent constant term, one obtains
\beq\label{3.25}
V(\phi) = \bV(\Phi) \equiv \frac{\cdtL \Phi -\frac{1}{6} \Phi^3}{2G_s}
\eeq
in the limit $a \to 0$, from which it follows that
\beq\label{3.26}
Z(g,g_s) = a^{N^2} Z(\cdtL,G_s), ~~~Z(\cdtL,G_s) = \int d\Phi \; 
\e^{-N\tr \bV(\Phi)}.
\eeq
The disk amplitude\index{disk amplitude} 
for the potential $\bV(\Phi)$ is precisely 
$W(\cdtZ,\cdtL,G_s)$, and since by definition 
\beq\label{3.27}
\frac{1}{z-\phi}= \frac{1}{a}\, \frac{1}{\cdtZ-\Phi},
\eeq
the first equal sign in eq.\ \rf{3.17a} follows straightforwardly from
the simple algebraic equation \rf{3.25}. 
We conclude that the continuum generalized
CDT-theory\index{disk amplitude} 
is described by the matrix model\index{matrix model} 
with potential $\bV(\Phi)$.

The conclusion is that to leading order in $N$, the combinatorial
method which works with a regularized lattice theory with a geometric
interpretation and an explicit cut-off $a$ has a matrix model
representation, and even in the continuum limit\index{continuum limit} 
where $a \to 0$ there
exists a matrix model representation of the theory. Once this 
is proved, one can actually ``derive'' a number of the 
results known for the generalized CDT model\index{CDT} from the matrix model
by the formal manipulations of eqs.\ \rf{3.25}-\rf{3.26}, 
where $a$ appears merely as a parameter without any obvious 
geometric interpretation as a cut-off. Also, once the 
matrix model equivalence is established, it is clear that 
one has automatically a higher-genus expansion available.
This higher-genus expansion was first established working
with the combinatorial, generalized CDT theory, and based
on purely geometric arguments of splitting and joining 
of space as a function of proper time, as we (partly) described
above. Such a theory could be turned into a kind of string field theory,
imitating the work of Kawai and collaborators 
on Euclidean quantum gravity\index{Euclidean quantum gravity} \cite{sft}. 
Like in the Euclidean case of Kawai et al., 
the formulation of the CDT string field theory 
used entirely a continuum notation (i.e.\ to cut-off $a$ was 
already taken to zero). However, contrary to the situation
for the Euclidean string field theory, we now have a matrix
model representation even in the continuum, and many of the
string field theory results derived in \cite{alwz} follow
easily from the 
loop equations\index{loop equation} \rf{*loop} for the potential \rf{3.25}.

\subsection{Summation over all genera in the CDT matrix 
model\index{matrix model}\index{CDT matrix model}   }

One remarkable application of the CDT matrix model representation is 
that we can find the Hartle-Hawking 
wave function\index{Hartle-Hawking wave function} summed over all genera.

In \cite{alwz} the matrix model given by 
eqs.\ \rf{3.25} and \rf{3.26} 
was related to the CDT Dyson-Schwinger equations 
by (i) introducing into the latter an expansion parameter $\a$, 
which kept track 
of the genus of the two-dimensional spacetime, and (ii) identifying this 
parameter with $1/N^2$, where $N$ is the size of the matrix 
in the matrix integral.
The $1/N$-expansion of our matrix model therefore plays 
a role similar to the $1/N$-expansion originally introduced 
by 't Hooft: it reorganizes an asymptotic expansion in a 
coupling constant $t$
($t=G_s/(\sqrt{\cdtL})^{3/2}$ in our case) into convergent sub-summations in 
which the $k$th summand appears with a 
coefficient $N^{-2k}$. In QCD applications, 
the physically relevant value is $N=3$, 
to which the leading-order terms in the large 
$N$-expansion\index{large $N$ expansion} 
can under favourable circumstances give a reasonable approximation. 

As we will see, for the purposes of solving our string 
field-theoretic model non-perturbatively, an additional 
expansion in inverse powers of $N$ (and thus an identification
of the contributions at each particular genus) is neither 
essential nor does it provide any new insights. 
This means that we will consider the entire sum over 
topologies ``in one go", which simply amounts
to setting $N=1$, upon which
the matrix integral \rf{3.26} reduces to the 
ordinary integral\footnote{Starting 
from a matrix integral for $N\times N$-matrices 
like (\ref{3.26}), performing a {\it formal} expansion
in (matrix) powers commutes with setting $N=1$, 
as follows from the following property of
expectation values of products of traces, which holds for
any $n=1,2,3,\dots$ and any set of non-negative
integers $\{ n_k\} $, $k=1,\dots , 2n$, such that 
$\sum_{k=1}^{2n}n_k =2n$. For any particular choice of
such numbers, consider
\beq\label{foot1}
\langle \prod_{k=1}^{2n} \Big(\frac{1}{N}\tr M^{n_k}\Big) \rangle \equiv
\frac{\int \d M \, \e^{-\oh \tr M^2} 
\prod_{k=1}^{2n} \Big(\tr M^{n_k}/N\Big)}{\int 
\d M \, \e^{-\oh \tr M^2}}=
\sum_{m=-n}^{n} \omega_m N^{m},
\eeq
where the last equation {\it defines} the numbers 
$\omega_m$ as coefficients in the power expansion in 
$N$ of the expectation value. Now, we have that
\beq\label{foot2}
\sum_{m=-n}^{n}\omega_m = (2n-1)!!
\eeq
independent of the choice of partition $\{ n_k\}$.
The number $(2n-1)!!$ simply counts the 
``Wick contractions" of $x^{2n}$ which we
could have obtained directly as the expectation 
value $\langle x^{2n}\rangle$, evaluated
with a one-dimensional Gaussian measure.
In the model at hand, we will calculate sums of the form
$\sum_{m=-n}^{n}\omega_m$ directly, since we 
are summing over all genera {\it without} introducing an
additional coupling constant for the genus expansion. 
In other words, the dimensionless
coupling constant $t$ in this case already 
contains the information about the splitting
and joining of the surfaces, and the coefficient of $t^k$ 
contains contributions from 2d geometries 
whose genus ranges between 0 and $[k/2]$. We cannot 
disentangle these contributions further unless we 
introduce $N$ as an extra parameter.}
\beq\label{j3.1}
Z(G_s,\cdtL) = \int \d m \; 
\exp \left[ -\frac{1}{2G_s} 
\left( \cdtL m - \frac{1}{6}\; m^3\right)\right],
\eeq 
while the disk amplitude\index{disk amplitude} can be written as 
\beq\label{j3.2}
\cdtW(X) = \frac{1}{Z(G_s,\cdtL)} 
\int \d m\; \frac{\exp \left[ -\frac{1}{2G_s} 
\left( \cdtL m - \frac{1}{6}\; m^3\right)\right]}{X-m}.
\eeq
These integrals should be understood as formal power series
in the dimensionless variable $t=G_s/(\sqrt{\cdtL})^{3/2}$ 
appearing in eq.\ \rf{3.17}.
Any choice of an integration contour which makes the integral well 
defined and reproduces the formal power series is a potential
non-perturbative definition. However, different
contours might produce different non-perturbative contributions
(i.e.\ which cannot be expanded in powers of 
$t$), and there may even be non-perturbative contributions 
which are not captured by any choice of integration contour. 
As usual in such situations, additional
physics input is needed to fix these contributions.

To illustrate the point, let us start by evaluating the 
partition function given in 
\rf{j3.1}. We have to decide on an integration path in the 
complex plane in order to define the integral. One possibility is to take a 
path along the negative 
axis and then along either the positive or the negative imaginary 
axis. The corresponding integrals are 
\beq\label{j3.2a}
Z(g,\lam)= \sqrt{\cdtL}\; t^{1/3} F_{\pm} (t^{-2/3}),~~~
F_{\pm} (t^{-2/3}) =2\pi \; e^{\pm i\pi/6}{\rm Ai}(t^{-2/3}\e^{\pm 2\pi i/3}),
\eeq
where Ai denotes the Airy function. Both $F_\pm$ 
have the same asymptotic expansion
in $t$, with positive coefficients. Had we chosen the integration path 
entirely along the imaginary axis we would have obtained ($2\pi i$ times)
${\rm Ai}(t^{-2/3})$, but this has an asymptotic expansion 
in $t$ with coefficients of oscillating sign, which is at odds with its
explicit power expansion in $t$. We have (using the standard
notation of Airy functions)
\beq\label{j3.2b}
F_{\pm}(z) = \pi \Big({\rm Bi}(z) \pm i {\rm Ai}(z)\Big),
\eeq 
from which one deduces immediately 
that the functions $F_{\pm}(t^{-2/3})$ are not real.
However, since ${\rm Bi}(t^{-2/3})$ grows like 
$e^{\frac{2}{3t}}$ for small $t$ while ${\rm Ai}(t^{-2/3})$ 
falls off like $e^{-\frac{2}{3t}}$, 
their imaginary parts are exponentially small 
in $1/t$ compared to the real part, and therefore do not contribute to
the asymptotic expansion in $t$.
An obvious way to {\it define} a partition 
function which is real and shares the
same asymptotic expansion is by symmetrization,
\beq
\oh (F_+ +F_-)\equiv \pi {\rm Bi}.
\eeq
The situation parallels the one encountered in the 
double-scaling limit\index{double-scaling limit} of the 
``old'' matrix model but is less complicated.

Presently, let us collectively denote by $F(z)$ any of the functions 
$F_{\pm}(z)$ or $\pi {\rm Bi}(z)$, leading to the
tentative identification
\beq\label{j3.3}
Z(G_s,\cdtL) = 
\sqrt{\cdtL}\; t^{1/3} \, F\Big(t^{-2/3}\Big),~~~~F''(z) = z F(z),
\eeq 
where we have included the differential equation 
satisfied by the Airy functions for
later reference. Assuming $X > 0$, we can write
\beq\label{j3.5}
\frac{1}{X-m}
= \int_0^{\infty} \d L
\; \exp\left[-\left(X-m\right)L\right].
\eeq
We can use this identity 
in eq.\ \rf{j3.2} to obtain the integral representation 
\beq\label{j3.6}
\cdtW(X) =  
\int_0^{\infty}\d L
\; \e^{-X L}\; 
\frac{F\Big(t^{-2/3}-t^{1/3}\sqrt{\cdtL} L\Big)}{F\Big(t^{-2/3}\Big)}.
\eeq
From the explicit expression of the Laplace transform
we can now read off the 
Hartle-Hawking amplitude\index{Hartle-Hawking wave function} 
as function of the boundary length $L$,
\beq\label{j3.7}
\cdtW(L) = 
\frac{F(t^{-2/3}-t^{1/3}\sqrt{\cdtL}\,L)}{F(t^{-2/3})}.
\eeq
Before turning to a discussion of the non-perturbative
expression for $\cdtW(L)$ we have just derived, 
let us remark that the asymptotic
expansion in $t$ of course agrees with that obtained
by recursively solving the CDT Dyson-Schwinger equations.
Using the standard asymptotic expansion of the Airy function one obtains
\beq\label{j3.9}
\cdtW(L) = \e^{-\sqrt{\cdtL} \,L} \; \e^{t \, h(t,\sqrt{\cdtL} \,L)}\;\;
\frac{\sum_{k=0}^\infty  c_k \;t^k 
\;(1-t\sqrt{\cdtL}\,L)^{-\frac{3}{2} 
k-\frac{1}{4}}}{\sum_{k=0}^\infty c_k \;t^k},
\eeq
where the coefficients $c_k$ are given by $c_0=1$, 
$c_k=\frac{1}{k!} \left(\frac{3}{4}\right)^k
\left(\frac{1}{6}\right)_k \left(\frac{5}{6}\right)_k$, $k>0$.
In (\ref{j3.9}), we have rearranged the exponential factors to 
exhibit the exponential
fall-off in the length variable $L$, multiplied by a term 
containing the function 
\beq\label{j3.10}
h(t,\sqrt{\cdtL}\,L) = \frac{2}{3t^2} 
\left[(1-t\sqrt{\cdtL}\,L)^{3/2}-1+\frac{3}{2}t\sqrt{\cdtL}\,L\right],
\eeq
which has an expansion in positive powers of $t$.

$\cdtW(L)$ has the interpretation of the wave function of the 
spatial universe according to the hypothesis of 
Hartle and Hawking\index{Hartle-Hawking wave function}.
$L \in [0,\infty]$ and the probability of finding a spatial universe
with length between $L$ and $L+\d L$ is 
\beq\label{j3.11}
P(L) = \frac{|\cdtW (L)|^2}{L},
\eeq
since the integration measure is $\d L/L$. Thus the probability is 
not normalizable in a conventional way and peaked 
at $L = 0$ since $\cdtW(L=0)=1$.
However, for each term in the asymptotic expansion \rf{j3.9} we 
obtain a finite value $\la L \ra \sim 1/\sqrt{\cdtL}$ as one would 
naturally expect. Since termination of the series in \rf{j3.9}
at a finite $k$ also implies restricting the 2d spacetime to have 
a finite genus we can say that as long as we restrict spacetime 
to have finite genus we have $\la L \ra \sim 1/\sqrt{\cdtL}$. However, if
we allow spacetimes of arbitrarily large genus to appear, i.e.\ if 
topology fluctuations are unconstrained (that means 
at most suppressed by a coupling 
constant, but no upper limit on the genus imposed by hand), 
{\it a remarkable change
appears}: $\la L \ra = \infty$ because the full non-perturbative 
$\cdtW(L)$ does not fall off like $\e^{-\sqrt{\cdtL}\;L}$ but only 
as $L^{-1/4}$ (Note that
$\cdtW(L)$ is still integrable at infinity since the integration
measure is $\d L/L$). This dramatic change in large $L$ behaviour
($W(L)$ also becomes oscillatory for large $L$, despite the fact that 
each term in the asymptotic expansion \rf{j3.9} is positive) is clearly
to be attributed to surfaces of arbitrarily large genus, i.e.\ it is 
a genuinely non-perturbative result.

\section{Discussion and perspectives}\label{discuss}

The four-dimensional CDT model\index{CDT} 
of quantum gravity is extremely simple. 
It is the path integral\index{CDT path integral}
over the class of causal geometries with a global time foliation. In order
to perform the summation explicitly, we introduce a grid of piecewise linear 
geometries\index{piecewise linear geometries}, 
much in the same way as when defining the path integral in 
quantum mechanics. Next, we rotate each of these geometries to Euclidean 
signature and use as bare action the Einstein-Hilbert 
action\index{Einstein-Hilbert action}\footnote{Of course, 
the full, effective action, 
including measure contributions, will contain all higher-derivative 
terms.} in Regge form\index{Regge action}. That is all.

The resulting superposition exhibits a nontrivial scaling 
behaviour as function of the 
four-volume, and we observe the appearance of a well-defined average 
geometry, that of de Sitter space\index{de Sitter universe}, 
the maximally symmetric solution to the classical
Einstein equations in the presence of a positive cosmological constant.  
We are definitely in a quantum regime,
since the fluctuations of the three-volume around de 
Sitter space are sizable, as can be seen in 
Fig.\ \ref{fig1}. Both the average geometry and the 
quantum fluctuations\index{quantum fluctuations} 
are well described in terms of the mini\-superspace action \rf{n5}. 
A key feature to appreciate is that, 
unlike in standard (quantum-)cosmological treatments, 
this description is the {\it outcome} of a 
non-perturbative evaluation\index{non-perturbative} of 
the {\it full} path integral, 
with everything but the scale factor (equivalently, $V_3(t)$)
summed over. Measuring the correlations of the 
quantum fluctuations in the computer
simulations for a particular choice of bare coupling constants
enabled us to determine the continuum gravitational coupling
constant $G$ as $G\approx 0.42 a^2$, thereby introducing an absolute physical 
length scale into the dimensionless lattice setting. 
Within measuring accuracy, our 
de Sitter universes\index{de Sitter universe} (with volumes lying
in the range of 6.000-47.000 $\ell_{Pl}^4$) are seen to behave perfectly 
semi-classically with regard to their large-scale properties. 

We have also indicated how we may be able to penetrate  
into the sub-Planckian regime by suitably changing the bare coupling constants.
By ``sub-Planckian regime" we mean that  
the lattice spacing $a$ is (much) smaller than 
the Planck length\index{Planck length}. 
While we have not yet analyzed this region in
detail, we expect to eventually observe a breakdown of the 
semi-classical approximation.
This will hopefully allow us to make contact with
continuum attempts to define a theory of quantum gravity  
based on quantum field theory. One such attempt has been described
in the introduction and is based on the concept  
of asymptotic safety\index{asymptotic safety}.
It uses renormalization group techniques in the continuum
to study scaling violations in quantum gravity around 
a UV fixed point\index{ultraviolet fixed point} 
\cite{reuteretc}. Other recent continuum field theoretical
models of quantum gravity which are not 
in disagreement with our data are the so-called Lifshitz gravity 
model \cite{horava}
and the so-called scale-invariant gravity model \cite{shaposh1,shaposh2}.
In principle it is only a question of computer power to decide 
if any of the models agree with our CDT model of quantum gravity.

On the basis of these results two major issues suggest 
themselves for further research. First, we need to establish  
the relation of our effective gravitational coupling constant $G$ 
with a more conventional gravitational 
coupling constant, defined directly in terms of coupling 
matter to gravity. In the present work, we have defined $G$ as the coupling
constant in front of the effective action, 
but it would be desirable to verify
directly that a gravitational coupling 
defined via the coupling to matter agrees
with our $G$. In principle it is easy to couple matter to 
our model, but it is less straightforward to define in a simple way 
a set-up for extracting the semi-classical 
effect of gravity on the matter sector. 
Attempts in this direction were already undertaken in the ``old'' Euclidean 
approach \cite{js,newton}, and it is possible that similar ideas can be used
in CDT quantum gravity.

The second issue concerns the precise nature 
of the ``continuum limit''. Recall our discussion in the Introduction
about this in a conventional lattice-theoretic setting. The 
continuum limit\index{continuum limit} is usually linked to a 
divergent correlation length\index{correlation length}
at a critical point. It is unclear whether 
such a scenario is realized in our case.
In general, it is rather unclear how one could define at all 
the concept of a divergent length related to 
correlators\index{correlation function} in quantum gravity, 
since one is integrating over all geometries, and it is the geometries
which dynamically give rise to the notion of ``length".

This has been studied in detail in two-dimensional (Euclidean) 
quantum gravity\index{Euclidean quantum gravity} 
coupled to matter with central charge $c \le 1$ 
\cite{corr2d}. It led to the conclusion that one
could associate the critical behaviour of the 
matter fields (i.e.\ approaching the critical point of 
the Ising model) with a divergent 
correlation length\index{correlation length}, although 
the matter correlators themselves had to be defined
as non-local objects due to the requirement
of diffeomorphism invariance. On the other hand, the two-dimensional
studies do not give us a clue of how to treat the 
gravitational sector itself, since they do not possess 
gravitational field-theoretic degrees of freedom. 
As we have seen the two-dimensional lattice models
can be solved analytically and the  only fine-tuning needed
to approach the continuum limit\index{continuum limit} 
is an additive renormalization of the cosmological constant.
Thus, fixing the two-dimensional spacetime volume $N_2$ (the 
number of triangles), such that the cosmological constant plays no role,
there are no further coupling constants to adjust and the continuum limit is 
automatically obtained by the assignment $V_2 = N_2 a^2$ and
taking $N_2 \to \infty$. This situation can also occur in special 
circumstances in ordinary lattice field theory. A term like
\beq\label{k5}
\sum_i c_1 (\phi_{i+1}-\phi_i)^2 + c_2(\phi_{i+1}+\phi_{i-1}-2\phi_i)^2
\eeq
(or a higher-dimensional generalization) will also go to the continuum free 
field theory simply by increasing the lattice size and using
the identification $V_d = L^d a^d$ ($L$ denoting the linear size of 
the lattice in lattice units), the higher-derivative term being 
sub-dominant in the limit. It is not obvious that in quantum gravity 
one can obtain a continuum quantum field theory without fine-tuning in
a similar way, because the action in this case is multiplied by  
a dimensionful coupling constant. Nevertheless, it is certainly remarkable
that the infrared limit of our effective action apparently 
reproduces -- within the cosmological setting -- the 
Einstein-Hilbert action\index{Einstein-Hilbert action}, which is
the unique diffeomorphism-invariant generalization of the 
ordinary kinetic term, containing at most second derivatives
of the metric. A major question is whether and how far our theory can 
be pushed towards an ultraviolet limit. We have indicated how to obtain such 
a limit by varying the bare coupling constants of the theory, but 
the investigation of the limit $a \to 0$ with fixed $G$ has only just begun
and other scenarios than a conventional 
UV fixed point\index{ultraviolet fixed point} might 
be possible. One scenario, which has often been discussed as a 
possibility, but which is still missing an explicit implementation is the 
following: when one approaches sub-Planckian scales the theory effectively
becomes a topological quantum field theory where the metric plays no 
role. Also in our very explicit implementation of a quantum gravity model
it is unclear how such a scenario would look. 
  
\subsection*{Acknowledgments}
The material presented in this review is based on collaborations
with numerous colleagues. In particular, we would like to thank
Andrzej G\"{o}rlich, Willem Westra and Stefan Zohren who have 
contributed in an essential way to the more recent part of the material
presented. All authors acknowledge support by
ENRAGE (European Network on
Random Geometry), a Marie Curie Research Training Network, 
contract MRTN-CT-2004-005616, and JJ by COCOS 
(Correlations in Complex Systems), a Marie Curie Transfer
of Knowledge Project, contract MTKD-CT-2004-517186, both in the 
European Community's Sixth Framework Programme.
RL acknowledges support by the Netherlands
Organisation for Scientific Research (NWO) under their VICI program.
JJ acknowledges a partial support by the Polish Ministry of Science and
Information Technologies grant 1P03B04029 (2005-2008).

\printindex

\end{document}